\documentclass[10pt]{article}
\pdfoutput=1

\usepackage{bm,wrapfig,float,array,mathtools}
\usepackage{amsfonts}
\usepackage{amssymb}
\usepackage{cite}
\usepackage{amsmath}
\usepackage{color}
\usepackage{graphicx}
\usepackage{hyperref}
\usepackage{float}
\usepackage[T1]{fontenc}
\usepackage[utf8]{inputenc}
\usepackage{alphabeta}
\usepackage{fancyvrb}
\usepackage[dvipsnames]{xcolor}
\usepackage{bold-extra}
\usepackage{lmodern}
\usepackage{cleveref}
\usepackage[normalem]{ulem}
\usepackage{tikz-cd}
\usepackage{tensor}
\usepackage{multirow}

\hypersetup{colorlinks=true, citecolor= NavyBlue,
  linkcolor= Maroon, urlcolor=NavyBlue}

\usepackage{colortbl}

\graphicspath{ {figures/} }

\usepackage{tikz}
\usetikzlibrary{arrows}
\usetikzlibrary{shapes.geometric,calc,arrows, positioning,shapes.misc,decorations.markings}
\tikzset{
  big arrow/.style={
    decoration={markings,mark=at position 1 with {\arrow[scale=2,#1]{>}}},
    postaction={decorate},
    shorten >=0.4pt},
  big arrow/.default=black}
\usetikzlibrary{external}
\usetikzlibrary{positioning}
\usetikzlibrary{calc}
\tikzset{gauge-node/.style={shape=circle, draw, minimum width=.6cm}}
\tikzset{global-node/.style={shape=rectangle, draw, minimum width=.6cm}}

\tikzstyle{brane}=[draw]
\tikzset{D7/.style={circle, draw=black, inner sep=0pt, fill=white, minimum size=3mm}}
\tikzset{hasse/.style={circle, fill,inner sep=2pt}}
\tikzset{flavor/.style={regular polygon,regular polygon sides=4,inner sep=2.5pt, draw}}
\tikzset{gauge/.style={circle, draw,inner sep=2.5pt}}
\tikzset{gaugeb/.style={circle, draw,fill=black,inner sep=2.5pt}}
\tikzset{gaugecyan/.style={circle, draw,fill=cyan,inner sep=2.5pt}}
\tikzset{gaugegreen/.style={circle, draw,fill=green,inner sep=2.5pt}}
\tikzset{gaugeblue/.style={circle, draw,fill=blue,inner sep=2.5pt}}
\tikzset{gaugeorange/.style={circle, draw,fill=orange,inner sep=2.5pt}}
\tikzset{bd/.style={circle, draw=black, inner sep=0pt, fill=black, minimum size=2mm}}
\tikzset{wd/.style={circle, draw=black, inner sep=0pt, fill=white, minimum size=2mm}}
\tikzset{Dynkin/.style={circle, draw=black, inner sep=0pt, fill=white, minimum size=2mm}}
\tikzstyle{ligne}=[draw, thick] 
\tikzset{doublearrow/.style={ draw=black!75, color=black!75, thick, double distance=3pt, }}

\pgfdeclarelayer{edgelayer}
\pgfdeclarelayer{nodelayer}
\pgfsetlayers{edgelayer,nodelayer,main} 
\tikzstyle{none}=[inner sep=0pt] 

\tikzstyle{NodeCross}=[draw, shape=circle, cross out, inner sep=0pt, minimum size=6pt,line width=0.25mm]
\tikzstyle{Circle}=[draw, shape=circle, black, inner sep=0pt, minimum size=6pt]
\tikzstyle{rtriangle}=[fill=black, regular polygon, regular polygon sides=3, rotate=90, inner sep=0pt, minimum size=8pt]
\tikzstyle{ltriangle}=[fill=black, regular polygon, regular polygon sides=3, rotate=270, inner sep=0pt, minimum size=8pt]
\tikzstyle{rtriangleblue}=[fill={rgb,255: red,17; green,160; blue,255}, regular polygon, regular polygon sides=3, rotate=90, inner sep=0pt, minimum size=8pt]
\tikzstyle{ltriangleblue}=[fill={rgb,255: red,17; green,160; blue,255}, regular polygon, regular polygon sides=3, rotate=270, inner sep=0pt, minimum size=8pt]
\tikzstyle{rtrianglegreen}=[fill={rgb,255: red,69; green,255; blue,28}, regular polygon, regular polygon sides=3, rotate=90, inner sep=0pt, minimum size=8pt]
\tikzstyle{ltrianglegreen}=[fill={rgb,255: red,69; green,255; blue,28}, regular polygon, regular polygon sides=3, rotate=270, inner sep=0pt, minimum size=8pt]
\tikzstyle{Uprtriangle}=[fill=black, regular polygon, regular polygon sides=3, rotate=0, inner sep=0pt, minimum size=8pt]
\tikzstyle{Downltriangle}=[fill=black, regular polygon, regular polygon sides=3, rotate=180, inner sep=0pt, minimum size=8pt]
\tikzstyle{rtriangleAmber}=[fill={rgb,255: red, 191; green, 144; blue, 63}, regular polygon, regular polygon sides=3, rotate=90, inner sep=0pt, minimum size=8pt]
\tikzstyle{UprtriangleViolett}=[fill={rgb,255: red,255; green,0; blue,0}, regular polygon, regular polygon sides=3, rotate=0, inner sep=0pt, minimum size=8pt]

\tikzstyle{Downltriangle}=[fill=black, regular polygon, regular polygon sides=3, rotate=180, inner sep=0pt, minimum size=8pt]
\tikzstyle{UpRighttriangle}=[fill=black, regular polygon, regular polygon sides=3, rotate=45, inner sep=0pt, minimum size=8pt]
\tikzstyle{UpLefttriangle}=[fill=black, regular polygon, regular polygon sides=3, rotate=315, inner sep=0pt, minimum size=8pt]
\tikzstyle{DownRighttriangle}=[fill=black, regular polygon, regular polygon sides=3, rotate=135, inner sep=0pt, minimum size=8pt]
\tikzstyle{DownLighttriangle}=[fill=black, regular polygon, regular polygon sides=3, rotate=225, inner sep=0pt, minimum size=8pt]

\tikzstyle{Star}=[draw, shape=star, fill=black, star points=8, inner sep=0pt, minimum size=8pt]

\tikzstyle{DashedLine}=[-, densely dashed, line width=0.25mm]
\tikzstyle{DashedLineBrown}=[-, densely dashed, line width=0.25mm, draw={rgb,255: red,155; green,103; blue,51}]
\tikzstyle{DashedLineFall}=[-, densely dashed, line width=0.25mm, draw={rgb,255: red,195; green,0; blue,0}]
\tikzstyle{DashedLineViolett}=[-, densely dashed, line width=0.25mm, draw={rgb,255: red,139; green,41; blue,148}]
\tikzstyle{DottedLine}=[-, dotted, line width=0.25mm]
\tikzstyle{BlueLine}=[-, fill=none, draw={rgb,255: red,17; green,160; blue,255}, line width=0.25mm]
\tikzstyle{GreenLine}=[-, fill=none, draw={rgb,255: red,69; green,255; blue,28}, line width=0.25mm]
\tikzstyle{RedLine}=[-, draw={rgb,255: red,191; green,0; blue,0}, fill=none, line width=0.25mm]
\tikzstyle{DashedLineRed}=[-, densely dashed, fill=none, draw={rgb,255: red,191; green,0; blue,0}, line width=0.25mm]
\tikzstyle{ThickLine}=[-, line width=0.25mm]
\tikzstyle{ViolettLine}=[-, draw={rgb,255: red,132; green,60; blue,191}, fill=none, line width=0.25mm]
\tikzstyle{ViolettDashedLine}=[-, densely dashed, draw={rgb,255: red,132; green,60; blue,191}, fill=none, line width=0.25mm]
\tikzstyle{AmberLine}=[-, draw={rgb,255: red,191; green,144; blue,63}, fill=none, line width=0.25mm]
\tikzstyle{DashedRedThick}=[-, densely dashed, fill=none, draw={rgb,255: red,191; green,0; blue,0}, line width=0.40mm]
\tikzstyle{DashedBlueThick}=[-, densely dashed, fill=none, black, line width=0.40mm]

\newcommand{\bea}{\begin{eqnarray}}
\newcommand{\eea}{\end{eqnarray}}
\newcommand{\be}{\begin{equation}}
\newcommand{\ee}{\end{equation}}
\newcommand{\ba}{\begin{aligned}}
\newcommand{\ea}{\end{aligned}}
\newcommand{\bit}{\begin{itemize}}
\newcommand{\eit}{\end{itemize}}
\newcommand{\ben}{\begin{enumerate}}
\newcommand{\een}{\end{enumerate}}
\newcommand{\nn}{\nonumber}

\newcommand{\beq}{\begin{equation}}
\newcommand{\eeq}{\end{equation}}
\def\be {\begin{equation}}
\def\ee {\end{equation}}
\def\bs#1\es{\begin{split}#1\end{split}}
\def\bg#1\eg{\begin{gathered}#1\end{gathered}}
\def\bea{\begin{eqnarray}}
\def\eea{\end{eqnarray}}

\newcommand{\Sym}{S_{\text{Sym}}}

\usepackage{customprelude}
\newcommand{\diff}{\breve}

\newcommand{\bd}{\star}

\begin{document}

\baselineskip=18pt  
\numberwithin{equation}{section}  
\allowdisplaybreaks  






\thispagestyle{empty}


\vspace*{0.8cm} 
\begin{center}
{{\huge 
Symmetry TFTs from String Theory
%

}}

 \vspace*{1.5cm}
Fabio Apruzzi$\,^{\sharp\,\flat}$,\  Federico Bonetti\,$^{\sharp}$, \ I\~naki Garc\'ia Etxebarria\,$^*$, \\
\smallskip
Saghar S. Hosseini\,$^*$,\  and  Sakura Sch\"afer-Nameki\,$^{\sharp}$\\

 \vspace*{.7cm} 
{\it  $^\sharp$ Mathematical Institute, University of Oxford, \\
Andrew-Wiles Building,  Woodstock Road, Oxford, OX2 6GG, UK}\\
\smallskip

{\it $^\flat$Albert Einstein Center for Fundamental Physics, Institute for Theoretical Physics,\\
University of Bern, Sidlerstrasse 5, CH-3012 Bern, Switzerland}\\

\smallskip

{\it $^*$ Department of Mathematical Sciences\\
Durham University, Durham, DH1 3LE, United Kingdom}

\end{center}

\vspace*{0.8cm}

\noindent
We determine the $d+1$ dimensional topological field theory, which
encodes the higher-form symmetries and their 't Hooft anomalies for
$d$-dimensional QFTs obtained by compactifying M-theory on a
non-compact space $X$. The resulting theory, which we call the
Symmetry TFT, or SymTFT for short,  is derived by reducing the
topological sector of 11d supergravity on the boundary $\partial X$ of
the space $X$.  Central to this endeavour is a reformulation of
supergravity in terms of differential cohomology, which allows the
inclusion of torsion in cohomology of the space $\partial X$, which in
turn gives rise to the background fields for discrete (in particular
higher-form) symmetries.  We apply this framework to 7d super-Yang
Mills, where $X= \mathbb{C}^2/\Gamma_{ADE}$, as well as the
Sasaki-Einstein links of Calabi-Yau three-fold cones that give rise to
5d superconformal field theories. This M-theory analysis is
complemented with a IIB 5-brane web approach, where we derive the
SymTFTs from the asymptotics of the 5-brane webs. Our methods apply to 
both Lagrangian and non-Lagrangian theories, and allow for many generalisations.
\newpage

\tableofcontents

 \section{Introduction}

\subsection{Symmetry TFTs for QFTs from Supergravity}

Quantum Field Theories (QFTs) have a rich structure of symmetries: in
addition to the familiar symmetry groups acting on point operators one
encounters in textbooks, we can also have more general kinds of
symmetries: higher-form symmetries \cite{Gaiotto:2014kfa}, higher
group symmetries \cite{Sharpe:2015mja, Tachikawa:2017gyf,
  Cordova:2018cvg,Benini:2018reh,Cordova:2020tij}, non-invertible
symmetries \cite{Bhardwaj:2017xup,Chang:2018iay,Rudelius:2020orz,Heidenreich:2021xpr, Kaidi:2021xfk, Choi:2021kmx},
and more generally symmetries described by abstract categorical
structures \cite{Huang:2021nvb}. Furthermore, theories with identical
local dynamics can have different symmetries in this generalised sense
\cite{Gaiotto:2010be,Aharony:2013hda}, and different symmetry
structures for a given set of local dynamics can sometimes be related
by gauging \cite{Kapustin:2014gua}, which in the presence of mixed 't
Hooft anomalies can relate theories with more conventional symmetry
structures to theories with less familiar ones
\cite{Bhardwaj:2017xup,Tachikawa:2017gyf}.

A very useful way of organising these structures, that as we will see
seems to arise naturally in string theory, is in terms of the
following construction (which we learned from Dan Freed
\cite{FreedTalk}, here we are only sketching an outline of a more
detailed construction): if the original theory $\mathfrak{T}$ is
formulated on $d$ dimensional spacetimes $\cM_d$, we introduce a
generically non-invertible topological $(d+1)$ dimensional quantum
field theory (which in this paper we will call the {\it symmetry
  theory}, or {\it symmetry TFT (SymTFT)} when we want to emphasize
that it is topological\footnote{In this paper we will be considering
  cases with $d\in 2\bZ+1$, so there are no local anomalies.}) with
the property that it admits a non-topological theory
$\tilde{\mathfrak{T}}$ as the theory of edge modes on manifolds with
boundary (a relative theory, in the framework of \cite{Freed:2012bs}),
and also a gapped interface $\rho$ to the anomaly theory $\cA$ of the
theory $\mathfrak{T}$:
\begin{center}
  \vspace{0.3cm}
  \includegraphics[height=4cm]{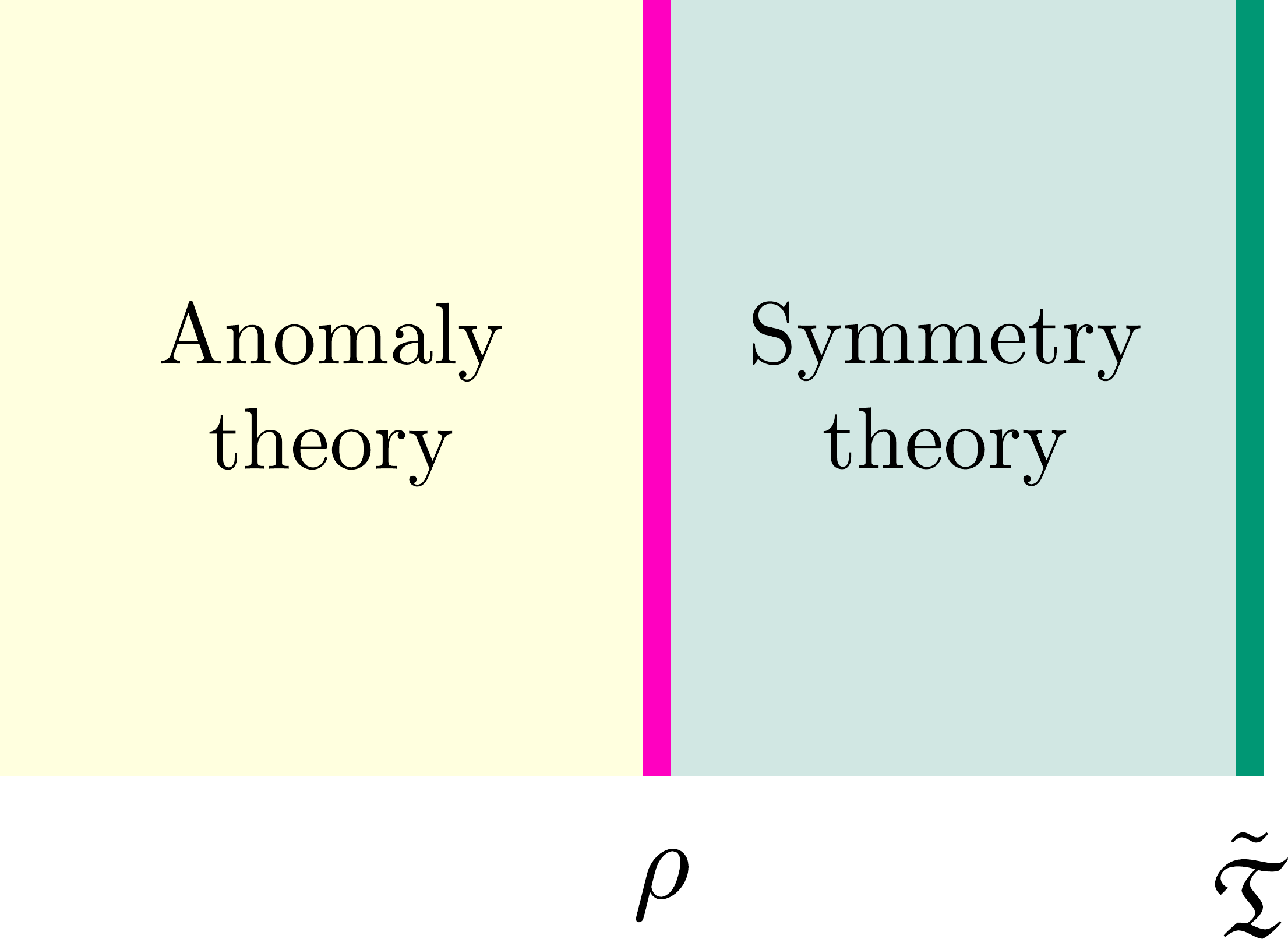}
\end{center}
The anomaly theory is a well understood object (see
\cite{Freed:2014iua,Monnier:2019ytc} for reviews): it is an invertible
theory that gives us a way of defining the phase of the partition
function of $\mathfrak{T}$ by evaluating the partition function of
$\cA$ on a $d+1$ manifold with boundary (see \cite{Dai:1994kq} for the
original discussion in the case of anomalies of fermions). The theory
$\mathfrak{T}$, attached to its anomaly theory $\cA$, arises when we
collide $\rho$ with $\tilde{\mathfrak{T}}$.

We will argue in this paper that the picture that arises in string
theory is the complementary one, in which we focus on the symmetry
theory by sending $\rho$ to infinity. More concretely, in this paper
we will consider singular string configurations, where we have a set
of local degrees of freedom (often strongly coupled) living at the
singular point of some non-compact cone $X$. We identify these local
degrees of freedom with $\tilde{\mathfrak{T}}$. The choice of the
actual symmetries of $\mathfrak{T}$ (which in our picture above would
be associated with a choice of $\rho$), has been previously argued to
live ``at the boundary of $X$''
\cite{GarciaEtxebarria:2019caf,Morrison:2020ool,Albertini:2020mdx}, a
behaviour that is also familiar in the context of holography
\cite{Witten:1998wy}. The goal of this paper is to sharpen this
picture by giving a direct derivation of the symmetry theory from the
string construction: we will see that we can obtain in a natural way a
non-invertible topological theory encoding both the choices of
symmetries for $\mathfrak{T}$ and their anomalies.

Our methods do not require knowledge of a holographic dual, or of a
weakly coupled description of the QFT. We find our results
particularly illuminating in the case that the local degrees of
freedom $\tilde{\mathfrak{T}}$ are those of a strongly coupled CFT
without a Lagrangian description (generically we know little about
such theories, so any additional information is useful), but we do not
require conformality of $\mathfrak{T}$ either.

\begin{figure}
\begin{center}
\includegraphics[width=8cm]{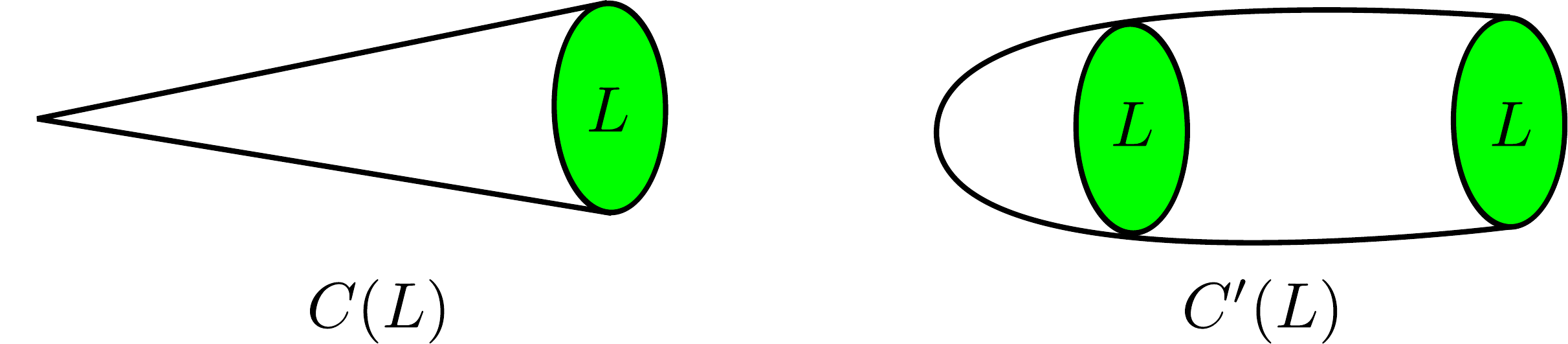}
\end{center}\caption{The cone $C(L)$ over the link $L$, and the deformation, shown on the right, to a long cylinder where the singularity is at the far end. \label{fig:Cone}}
\end{figure}

The main idea is as follows. In string theory we can construct the
$d$-dimensional theories $\mathfrak{T}$ by introducing defects or
singularities extending along $d$-dimensional submanifolds of the
11-dimensional spacetime. For concreteness we will focus in this paper
on M-theory on singular spaces with a single isolated singularity. The
non-trivial local dynamics arise from massless M2 branes wrapping
vanishing cycles at the singularity. Close to the singular point the
geometry will look like a real cone over some manifold $L$ with
$\dim(L)=10-d$. We can deform the cone into an infinitely long cigar,
with the singularity at the tip, and $L$ as the base of the cylinder
along the cigar, see figure \ref{fig:Cone}. The information that we
are after is topological, so it is reasonable to expect that we can
still obtain it from this deformed background (our results will
support this expectation). If we now dimensionally reduce the M-theory
action on $L$ we will obtain a theory on the remaining $d+1$
dimensions, which look like $\cM_d\times \bR_{\geq 0}$. We claim that
the topological sector -- i.e. couplings that are metric independent
-- arising from this reduction on $L$ is precisely the symmetry theory
for $\tilde{\mathfrak{T}}$.

In our specific context of M-theory we will obtain this topological
sector by ``reducing'' the Chern-Simons sector of M-theory on $L$,
additionally including the effect of flux-noncommutativity
\cite{Moore:2004jv,Freed:2006ya,Freed:2006yc}. As we will see,
flux-noncommutativity leads to choices of higher form symmetries which
appear in a way familiar from holography. For example, in the
$\text{AdS}_5 \times S^5$ case studied in \cite{Witten:1998wy,
  Maldacena:2001ss, Belov:2004ht} the 5d supergravity contains the
coupling
\begin{equation}
\Sym= {N} \int B_2 \wedge dC_2 \,,
\end{equation}
which upon imposing boundary conditions on $(B_2, C_2)$ yields
different global forms of the gauge group of 4d $\mathcal{N}=4$ SYM.
(Similar couplings have been studied in other holographic setups such
as ABJM in \cite{Bergman:2020ifi} and the non-conformal dual to
confinement in Klebanov-Strassler in \cite{Apruzzi:2021phx}.) We will
obtain analogous couplings from compactification. The topological
reduction that we consider also generates in a natural way the
anomalies that are expected in cases where the answer from field
theory is known.


We expect the general idea that we are putting forward to be much more
general and applicable in a wide range of setups.\footnote{One
  complication in the general case is that topological terms in the
  parent string theory are not the only source of topological terms in
  the lower dimensional theory. For instance, in the presence of
  chiral fermions the lower dimensional theory will generically
  include couplings proportional to the $\eta$ invariant evaluated in
  the compactification space.} In order to illustrate this point, in
section~\ref{sec:5branes} we will apply the same topological reduction
prescription for the same 5d SCFTs we analyse from the M-theory
viewpoint, but now in terms of their realization from $(p,q)$ 5-branes
\cite{Aharony:1997bh}. We will show how 1-form symmetries for 5d SCFTs
are encoded in the brane-web and compute from a supergravity point of
view, expanding on the boundary of the IIB spacetime. The latter is
inspired by the back-reacted holographic solutions
\cite{DHoker:2017mds}. These are AdS$_6 \times M_4$ solutions, which
are near-horizon limit of $(p,q)$ 5-branes webs. What will be
important for our analysis is the topology of $M_4$ of the
near-horizon geometry, which we will be using to dimensionally reduce
the topological coupling of IIB supergravity. The topology of $M_4$ is
based on isometries as well as asymptotic charges of the semi-infinite
5-branes and it is not affected by any large $(p,q)$ charge
holographic limit, therefore will be valid in any IIB $(p,q)$
5-brane-setup. Our focus here is on the dimensional reduction of the
IIB 10-dimensional topological coupling on $M_4$ where we also need to
include contributions coming from $(p,q)$ 5-brane source to determine
the anomalies for 5d SCFTs for both Lagrangian and non-Lagrangian
theories. This beautifully complements the geometric analysis in
M-theory. At last, as another application we also derive a mixed
anomaly between continuous \cite{Cordova:2020tij} and discrete 1-form
symmetries of the $(1,1)$ little string theory (LST) engineered by
NS5-branes in IIB \cite{Intriligator:1997dh} from the holographic
linear dilaton background \cite{Maldacena:1997cg}.

\subsection{A Differential Cohomology Refinement of Dimensional Reduction}

M-theory compactification on singular Calabi-Yau 2- and 3-folds gives
rise to 7d super-Yang Mills (SYM) and 5d superconformal field theories
(SCFTs), respectively.  These theories have 1-form symmetries, and in
the 5d case also 0-form symmetries. The 1-form symmetry in all these
cases is discrete and is characterized in terms of the relative
homology quotient of the Calabi-Yau $X$, with respect to its boundary
$\partial X$ \cite{GarciaEtxebarria:2019caf,Morrison:2020ool,
  Albertini:2020mdx, Bhardwaj:2020phs}
\begin{equation}
\Gamma^{(1)}= {H_2 (X,
  \partial X; \mathbb{Z})\over H_2 (X; \mathbb{Z})}\,.
\end{equation}
To derive SymTFTs for global 1-form symmetries, one will have to
incorporate their backgrounds $B_2 \in H^2 (M_5; \Gamma^{(1)})$ into
the supergravity formalism. The torsional nature of these fields
introduces various subtleties in the process. We will use differential
cohomology to address these subtleties.\footnote{In supergravity
  theories different prescriptions to incorporate torsion have been
  put forward \cite{Camara:2011jg, Berasaluce-Gonzalez:2012abm}, but
  none that are mathematically entirely satisfactory or unambiguous --
  we will comment on a detailed comparison to these approaches
  shortly.}

Differential cohomology has seen numerous applications within quantum
field theory and string/M-theory.  Some of the earlier works on the
subject include \cite{Dijkgraaf:1989pz, Diaconescu:2003bm, Bauer:2004nh, Moore:2004jv, Freed:2006yc}.
For more recent examples of differential cohomology applications in
formal high-energy physics that are of some relevance to this work see
\cite{Kapustin:2014gua,Kapustin:2014zva, Monnier:2016jlo,
  Monnier:2017klz, Monnier:2018nfs, Cordova:2019jnf,
  Cordova:2019uob,Hsieh:2020jpj, Bah:2020uev,Davighi:2020vcm,
  Debray:2021vob}. In this paper differential cohomology will be used
to refine the notion of dimensional reduction (or KK-reduction) of
supergravity theories, with the goal of providing a precise treatment
of the effect of torsion cohomology classes in the compactification
manifold.

To illustrate, without going into the mathematical intricacies of
differential cohomology yet, we now give a very concrete example of
how the concept of dimensional reduction is reformulated using this
approach.  We begin with M-theory on a 5d space $L_5$ (which in this
paper will be a manifold linking the singular point of a non-compat
Calabi-Yau three-fold) which has
$H^2 (L_5 ; \mathbb{Z}) = \mathbb{Z}_n \oplus \mathbb{Z} = \langle t_2
\rangle \oplus \langle v_2\rangle$, where $t_2$ is a torsional
generator of the degree two cohomology group and $v_2$ is a free
generator. The reduction for the latter is the standard
KK-reduction. It is the torsion part that will most benefit from the
uplift to differential cohomology. We denote the differential
cohomological uplifts of $t_2$ and $v_2$ by $\diff t_2$ and
$\diff v_2$, the precise meaning of these will be explained below.

In the case of M-theory we need to introduce a differential refinement
$\diff G_4$ of $G_4$ as well. As in the standard KK-expansion, this
has a decomposition in terms of differential cohomology classes along
the internal space $L_5$ (torsion and free), as well as external
spacetime $M_6$:
\begin{equation} \label{eq:g4fb2}
\diff{G}_4 = \diff{F} \bd \diff{v}_2  + \diff{B}_2 \bd \diff{t}_2 \,.
\end{equation}
There is an extra term, discussed below, that we are ignoring here
for simplicity. The meaning and properties of the product ``$\bd$''
will also be explained below.  The CS-term in the M-theory action is
\begin{equation}
  \frac{S_{\text{top}}}{2\pi}=-\frac{1}{6} \int_{M_{11}} \diff G_4 \bd \diff G_4 \bd \diff G_4  \,,
\end{equation}
which upon inserting the decomposition~\eqref{eq:g4fb2} of $\diff G_4$
we integrate over $L_5 \times \mathcal{W}_6$. The integration over the
internal space $L_5$ results in the SymTFT on $\cW_6$. In ordinary
cohomology the integral on $L_5$ would pick out only the forms of
degree 5, however that would mean the purely torsional part
$ \int_{L_5} \diff t_2 \bd \diff t_2 \bd \diff t_2$ would naively not
contribute. Differential cohomology works differently, and as reviewed
below $ \int_{L_5} \diff t_2 \bd \diff t_2 \bd \diff t_2$ can be
non-vanishing.



Reformulating the problem in terms of differential cohomology on the
link of the singularity involves some additional technical
complications, but the effort pays off in a number of ways:

\begin{itemize}
\item Geometric engineering of QFTs corresponds to
  ``compactification'' of string and M-theory on non-compact spaces
  $Y^d$.  This can be mathematically challenging, in particular it is
  difficult to define in a precise mathematical sense what one means
  by reducing the Chern-Simons action on a non-compact space $Y^d$. In
  our approach we are instead reducing the Chern-Simons action on the
  closed manifold $X^{d-1}= \partial Y^d$, which is the boundary or
  link of the non-compact space $Y^d$.  This is a much better defined
  mathematical question, that can be clearly analysed using the
  formalism of differential cohomology.
  
\item The effective field theory in $11-d$ dimensions is most
  interesting when $Y^{d}$ is singular, so it becomes a non-abelian
  Yang-Mills theory in 7d (for $d=4$) or a non-trivial interacting 5d
  SCFT (for $d=6$). But it is precisely in this singular geometric
  regime that it is most difficult to pin down what one means by doing
  a geometric reduction of the effective action. By contract, our
  formalism is entirely agnostic about the singular structure of
  $Y^d$, and can be applied without issues even when $Y^d$ is singular
  (in fact, it is arguably at the singular cone point in moduli space
  where it is most natural to apply our techniques!).
  
\item Often, the analysis of reduction on singular spaces is done by
  removing these singularities, e.g.  in 5d going to the Coulomb
  branch. It is well-documented, e.g. in the set of canonical
  singularities realizing 5d SCFTs from isolated hypersurface
  singularities (see \cite{Xie:2017pfl, Closset:2020scj,
    Closset:2020afy, Closset:2021lwy} for a discussion in the context
  of 5d SCFTs), that we can have terminal singularities that do not
  admit a Calabi-Yau (crepant) resolution.  This is obviously a class
  of theories where many standard methods will fail. Another setting
  where field theory-inspired arguments (including those employed in
  \cite{Cvetic:2021sxm}) are not extendable, is when the 5d SCFT may
  have a Coulomb branch, but does not admit a non-abelian gauge theory
  description.

In contrast, our approach of deriving the SymTFT and thereby the anomaly of the QFT in terms of the reduction on the boundary is applicable in all those instances, and we will provide some examples of non-Lagrangian 5d SCFTs and their SymTFTs below. 
\item The approach uniformly encodes the entire SymTFT, for all
  symmetries arising from the compactification.  E.g. in 5d we derive
  both the mixed 0-1-form symmetry anomalies as well as the 1-form
  symmetry $(B_2)^3$ anomaly \cite{Gukov:2020btk} (see
  section~\ref{sec:5d}).

\item This approach lends itself to applications also in the context
  of string and M-theory compactifications with torsional cycles are
  in the geometry, such as those studied in \cite{Camara:2011jg,
    Berasaluce-Gonzalez:2012abm}, which give rise to theories with
  discrete gauge symmetries in string theory.

\end{itemize}

\subsection{Comparison to other Approaches}

Continuous global symmetries are usually rather manifest in terms of
the geometry, e.g. the R-symmetry of a 4d $\mathcal{N}=1$ SCFT is
often encoded in some geometric isometry (such as in the setup of
D3-branes probing Calabi-Yau cones), or the flavor symmetry in terms
of non-compact divisors in a Calabi-Yau compactification -- as will be
the case in this paper.  Discrete and continuous higher-form
symmetries are encoded also in the topology of the compactification
space -- usually in terms of relative homology, that captures the
defect operators, modulo screening \cite{DelZotto:2015isa}.  In the
reduction on the link, the anomalies arise in terms of topological
couplings for the background fields of global symmetries.  Systematic
tools for treating symmetries associated to isometries and
non-torsional cohomology classes in brane constructions in M-theory
and Type IIB are studied in \cite{Bah:2019rgq,Bah:2020jas} in
connection to equivariant cohomology, see also
\cite{Hosseini:2020vgl}.

In this paper we focus instead mostly on the torsional sector. Dealing
with torsion cycles in supergravity has of course a history. One
particularly promising framework was put forward in
\cite{Camara:2011jg, Berasaluce-Gonzalez:2012abm}.  In supergravity,
form-fields are usually expanded in harmonic forms, which however do
not capture the torsion parts ${\rm Tor }\, H^p(X_d;\mathbb Z)$.  The
key idea of these papers is to use non-harmonic forms to model classes
in ${\rm Tor }\, H^p(X_d;\mathbb Z)$.  More precisely, a non-trivial
class in ${\rm Tor }\, H^p(X_d,\mathbb Z)$ of order $k$ is modeled by
a $(p-1)$-form $\beta_{p-1}$ and a $p$-form $\alpha_{p}$ subject to
the condition \be\label{form_relation} k \alpha_p = d\beta_{p-1}\,.
\ee Moreover, motivated by the above general remarks on the
KK-expansion, $\beta_{p-1}$ is required to be a co-exact eigenfunction
of the Laplacian with a non-zero eigenvalue.  (It then follows
automatically that $\alpha_p$ is also an eigenform with the same
eigenvalue.)

To illustrate the physical picture underlying this proposal,
let us consider  for example the terms in the expansion of $C_3$ associated
to a pair $(\beta_1, \alpha_2)$ satisfying \eqref{form_relation}. We have schematically
\beq
C_3 \supset  \widetilde B_2 \wedge \beta_1 +  \widetilde A_1 \wedge \alpha_2 \ , \qquad
G_4 \supset (d\widetilde A_1 + k \widetilde B_2) \wedge \alpha_2 + d\widetilde B_2 \wedge \beta_1 \ .
\eeq
The 11d kinetic term for $G_4$ induces lower-dimensional kinetic terms
of the schematic form
\beq
S \supset \int - \frac 12 \, g_{\widetilde A_1 \widetilde A_1} \, (d\widetilde A_1 + k \widetilde B_2) \wedge * (d\widetilde A_1 + k \widetilde B_2)
 - \frac 12 \, g_{\widetilde B_2 \widetilde B_2} \, d \widetilde B_2  \wedge * d \widetilde B_2  \ ,
\eeq
where the Hodge star and the integration are now over external spacetime
in $11-d$ dimensions. The quantities $g_{\widetilde A_1 \widetilde A_1}$,
$g_{\widetilde B_2 \widetilde B_2}$ depend on the details of the compactification
and on the eigenvalue of the forms $\beta_1$, $\alpha_2$ under the action of the internal
Laplacian. We do not need to discuss them in detail for this argument.
We observe that the lower-dimensional fields $\widetilde A_1$, $\widetilde B_2$
participate in a St\"uckelberg mechanism: 
the 2-form field $\widetilde B_2$ ``eats'' the 1-form field $\widetilde A_1$
and gets massive (as is expected, since the eigenforms
$\beta_1$, $\alpha_2$
have a non-zero eigenvalue
of the internal Laplacian).
The St\"uckelberg mechanism
leaves behind an unbroken $\mathbb Z_k$ gauge symmetry: in the IR, 
the field $\widetilde B_2$ is a continuum description of a $\mathbb Z_k$ discrete 2-form gauge field.
We identify the latter with the discrete gauge field
that originates from a formal expansion of $G_4$
of the form $G_4 \supset \widetilde B_2 \cup x$,
where  $x\in {\rm Tor} \, H^2(X_d;\mathbb Z)$ is the torsion class 
modelled by the pair of forms
$(\beta_1,\alpha_2)$.

This approach has been applied successfully in various
compactification scenarios with torsion (co)-cycles.  However,
clearly, this prescription leaves various mathematical questions open,
and equally importantly, in more complicated compactification
settings, carrying out this approach consistently can be
difficult. The differential cohomology approach that we propose here
has the advantage of providing a sound mathematical framework, which
unambiguously lets us implement torsion in a supergravity setting.  It
would be interesting to provide a precise map to the above
prescription using non-closed forms.

We should also comment that the M-theory approach often has a IIA-avatar, e.g. when there is a circle-fibration in the geometry. We will see that this can be a useful complementary check of our proposal. However, obviously this only applies in a very limited set of geometric situations. To contrast and compare to the IIA setting, we discuss the IIA counterparts to the M-theory computations in this paper in appendix \ref{app:IIA}.

In the context of 5d SCFTs and 7d gauge theories based on elliptic
Calabi-Yau singularities, the paper \cite{Cvetic:2021sxm} has
discussed in specific instances\footnote{The cases discussed do not
  have discriminant components of the elliptic fibration intersecting
  the boundary of the Calabi-Yau.} the anomalies of 1-form
symmetries. The analysis is based on a Coulomb branch (CB) computation
in the resolved geometry. What is implemented is a combination of a
field theoretic analysis, with information from the geometry about the
structure of the CB. The point of view in that paper is roughly
complementary to ours: they integrate the Chern-Simons term on the
singular Calabi-Yau $X_d$ to obtain an anomalous coupling in the
$(11-d)$-dimensional field theory, while we integrate the Chern-Simons
coupling on the $(d-1)$-dimensional link of the singular point in
$X_d$ to obtain the $(10-d)$-dimensional anomaly theory (together
with the non-invertible sectors, which are not considered in
\cite{Cvetic:2021sxm}). As emphasised above, this change in perspective
has many benefits, for instance it allows us to approach problems
where we do not have a gauge theory description to guide us.

We also resolve a puzzle in the computation in \cite{Cvetic:2021sxm}
of the mixed anomaly between the 1-form symmetry and the instanton
symmetry. It was observed in that paper that the results of the
computation did not always agree with the field theory expectation
\cite{BenettiGenolini:2020doj}. In the present paper, we obtain the
expected field theory mixed anomaly,\footnote{Our understanding is
  that the results in \cite{Cvetic:2021sxm} differ because their
  choice of non-compact divisor dual to the instanton generator is not
  always the same as in the field theory, but can differ from that by
  torsion generators.} and in addition also compute the $B^3$ 1-form
symmetry anomaly, where we also find agreement with
\cite{Gukov:2020btk}.  Reproducing these very involved results gives
us confidence that our approach is both sound and fruitful.  More
generally we expect to be able to apply our approach to
non-geometric, and also not-resolvable geometries, which would provide
a substantial extension compared to both field theory and Coulomb
branch approaches.

\paragraph{Plan.} The plan of this paper is as follows: We start in
section \ref{sec:DC} with an overview of differential cohomology,
providing a (hopefully physics-friendly) summary of its salient
features. We then apply this to the M-theory topological couplings --
the CS and $C_3\wedge X_8$ terms, and give the general KK-expansion in
terms of differential cohomology.  This framework is put to work in
the context of $\mathbb{C}^2/\Gamma_{ADE}$ compactifications in
M-theory in section \ref{sec:7d} and in section \ref{sec:5d} to 5d
SCFTs realized on canonical singularities in Calabi-Yau
three-folds. The anomalies of the 1-form symmetry are determined in
these cases, including the mixed 0-1-form symmetry anomaly in 5d. For
5d we derive the anomalies from a complementary point of view in
section \ref{sec:5branes} using the 5-brane webs in Type IIB -- again
from the boundary of the spacetime.  Finally, we determine the anomaly
of the little string theory (LST) from an edge mode approach in
section \ref{sec:LST}.  In appendix \ref{app:DC} we address a
technical point regarding $G_4$-flux quantisation and in appendix
\ref{app:IIA} we provide a countercheck to the M-theory results, using
more conventional IIA reductions, which are available in some setups.

\section{Differential Cohomology and M-theory}
\label{sec:DC}

In this section we discuss how to apply the language of differential
cohomology to describe the topological couplings in the M-theory low-energy effective action.
We then describe the dimensional reduction
of these couplings on a generic internal space $L$ 
including contributions originating from both free and torsion
elements in the cohomology of~$L$.

\subsection{Aspects of Differential Cohomology}

Differential cohomology provides a mathematical framework to
describe $U(1)$ $(p-1)$-form gauge fields 
on arbitrary spacetimes. This formalism is particularly
useful to keep track of subtler 
aspects that emerge when we consider spacetimes with non-trivial topology
(and in particular with torsional cycles).
We refer the reader to
\cite{Freed:2000ta,Hopkins:2002rd,Freed:2006ya,Freed:2006yc,Freed:2006mx,Hsieh:2020jpj}
for reviews aimed at physicists. We also highly recommend the textbook
by Bär and Becker \cite{BarBecker} for a pedagogical discussion of
most of the results below.

\paragraph{Characteristic class and field strength.}

Let us consider a  closed, connected, oriented manifold $\cM$.
The $p$-th differential cohomology group $\diff H^p(\cM)$ of $\cM$ 
is an Abelian group  furnishing a differential refinement of the
ordinary cohomology group $H^p(\cM;\mathbb Z)$.
The group $\diff H^p(\cM)$ sits at the center of the following commutative
diagram:
\begin{equation}    \label{eq:SS-}
\begin{tikzcd}[row sep={1.732050808cm,between origins},column sep={2cm,between origins}]
{} & {} & {} & \text{Tor} H^p(\cM; \mathbb Z) \arrow[dr] & {} & {} & {} \\
{} & {} & H^{p-1}(\cM;\mathbb R / \mathbb Z) \arrow[rr,"-\beta"] \arrow[dr, hookrightarrow, "i" ] \arrow[ur] & {} & H^{p}(\cM;\mathbb Z) \arrow[dr , "\varrho"] & {} & {} \\ 
{} & \frac{H^{p-1}(\cM;\mathbb R)}{H^{p-1}_{\text{Free}}(\cM;\mathbb Z)} \arrow[ur] \arrow[dr  ] & {} &  \diff  H ^{p}(\cM) \arrow[ur, twoheadrightarrow, "I" ] \arrow[dr, twoheadrightarrow, "R"] & {} & H^{p}_{\text{Free}}(\cM ;\mathbb Z) & {} \\
{} & {} & \displaystyle \frac{ \Omega^{p-1}(\cM)}{\Omega_{\mathbb Z}^{p-1}(\cM)} \arrow[rr,"d_{\mathbb Z}" ] \arrow[ur, hookrightarrow, "\tau"  ] \arrow[rd, "d"] & {} & \Omega_{\mathbb Z}^{p}(\cM) \arrow[ur, "r"] & {} & {}\\
{} & {} & {} & d\Omega^{p-1}(\cM) \arrow[ur] & {} & {} & {} 
\end{tikzcd}
\end{equation}
where all the diagonals are short exact sequences.

The maps $i$, $I$, $\tau$, $R$ are natural
(that is, given a
smooth map $f\colon \cM \to \cM'$ they commute with the pullback $f^*$ of
$f$).
Let us now proceed to unpack the relevant information contained in the above diagram,
and to provide some physical interpretation: 

\begin{itemize}

\item The symbol $\Omega_\mathbb Z^q(\cM)$ denotes closed differential $q$-forms
on $\cM$ with integral periods.
The surjective map $R$
associates to each element $\diff a \in \diff H^p(\cM)$ a   $p$-form
 $R(\diff a) \in \Omega^p_{\mathbb Z}(\cM)$, which we refer to as 
the field strength of $\diff a$. Physically, $\diff a$ models 
a $U(1)$ $(p-1)$-form field, up to gauge equivalences,
and the $p$-form $R(\diff a)$ is identified with the physical
field strength of the $(p-1)$-form gauge field.
(The fact that $R(\diff a)$ has integral periods encodes the fact that the gauge group
is $U(1)$ and not $\mathbb R$.)

\item An element $\diff a \in \diff H^p(\cM)$ with $R(\diff a) = 0$ is called flat.
Exactness of the central NW-SE diagonal in the  diagram \eqref{eq:SS-}
demonstrates that flat elements of $\diff H^p(\cM)$ can be identified 
with elements in $H^{p-1}(\cM ; \mathbb R/\mathbb Z)$.
Physically, the gauge-invariant information about a flat $(p-1)$-form
gauge field is encoded in its holonomies
around non-trivial $(p-1)$-cycles, which take values in $U(1) \cong \mathbb R/\mathbb Z$
and can be encoded in an element of  $H^{p-1}(\cM ; \mathbb R/\mathbb Z)$.

\item The surjective map
  $I: \diff H^p(\cM) \rightarrow H^p(\cM ; \mathbb Z)$ is the map that
  ``forgets'' the differential refinement, yielding back ordinary
  cohomology with coefficients in $\mathbb Z$.  Given an element
  $\diff a \in \diff H^p(\cM)$, we refer to
  $I (\diff a) \in H^p(\cM;\mathbb Z)$ as the characteristic class of
  $\diff a$.

\item An element $\diff a \in \diff H^p(\cM)$ with $I(\diff a) = 0$ is
  called topologically trivial.  Exactness of the central SW-NE diagonal in
  the  diagram \eqref{eq:SS-} implies that topologically
  trivial elements of $\diff H^p(\cM)$ can be identified with globally
  defined $(p-1)$-forms on $\cM$, up to additive shifts by closed
  $(p-1)$-forms with integral periods.  In physics term, a
  topologically trivial $(p-1)$-form gauge field can be described
  globally by specifying a $(p-1)$-form. The shift by closed
  $(p-1)$-forms with integral periods is interpreted as a gauge
  transformation (a ``large gauge transformation'' if the $(p-1)$-form
  is closed but not exact).

\item Commutativity of the square on the RHS of the diagram \eqref{eq:SS-}
is the statement that, for any $\diff a \in \diff H^p(\cM)$,
\beq \label{eq:compatibility}
r\big( R(\diff a) \big)  = \varrho \big(  I(\diff a) \big) \ .
\eeq
The short exact sequence in the lower NW-SE diagonal of \eqref{eq:SS-} comes from the isomorphism $ \Omega^{p}_\mathbb{Z}(\cM)/d\Omega^{p-1}(\cM)\cong  H_{\text{free}}^p(\cM;\mathbb{Z})$ which is a by-product of de Rham's theorem.
From the physics perspective, it is well-known that
information about the topological aspects of a $(p-1)$-form gauge field
configuration  can be extracted from  its field strength
(for example, the integer charge of a monopole configuration for a $U(1)$ 1-form
gauge field on $\cM = S^2$ is extracted integrating the 2-form field strength on $S^2$).
Crucially, however, the field strength encodes only $\varrho (I(\diff a))$ and not
necessarily
$I(\diff a)$. To see this, let $I(\diff a)=[a]\in H^p(\cM; \mathbb{Z})$ and embed $\mathbb Z$ into $\mathbb R$ to get $[a]_{\mathbb R}\in H^p(\cM; \mathbb R)$. Then, for a de Rham cohomology class $[F]_{\text{dR}}\in H_{\text{Free}}^p(\cM; \mathbb Z)\otimes \mathbb R$ of $F\in \Omega^p_{\mathbb Z}(\cM)$ we have $[F]_{\text{dR}}=[a]_{\mathbb R}$. Thus, $[a]$ contains more information than $[F]_{\text{dR}}$ at the differential level since $[a]_{\mathbb R}$ can be obtained from $[a]$ but the converse is not true. 
In particular, information about torsional components
in $I( \diff a )$ is lost in passing to $\varrho (I(\diff a))$.

\item A flat element in $\diff H^p(\cM)$ is not necessarily topologically trivial.
Suppose $\diff a \in \diff H^p(\cM)$ is flat; we aim to compute its characteristic class
$I(\diff a)$.
 From exactness of the
NE-SW diagonal we  know that
$\diff a = i(u)$ for some $u \in H^{p-1}(\cM ; \mathbb R/\mathbb Z)$.
Commutativity of the upper triangle in the  diagram \eqref{eq:SS-} gives us
\beq
I(\diff a ) = I (i(u)) = - \beta (u)  \ .
\eeq
Here $\beta: H^{p-1}(\cM; \mathbb R/\mathbb Z) \rightarrow H^p(\cM ; \mathbb Z)$
is the Bockstein homomorphism associated to the short exact sequence
$0 \rightarrow \mathbb Z \rightarrow \mathbb R \rightarrow \mathbb R / \mathbb Z \rightarrow 0$,
\beq
  \label{eq:exponential-ses}
  \ldots \to H^{p-1}(\cM ; \mathbb Z)  \xrightarrow{\varrho} H^{p-1}(\cM ;\bR) \to H^{p-1}(\cM ; \mathbb R /\mathbb Z)
  \xrightarrow{\beta} H^p(\cM ; \mathbb Z ) \xrightarrow{\varrho} H^p(\cM ; \mathbb R )  \to \ldots
  \eeq
  which is in general non-vanishing.

\item A topologically trivial element in $\diff H^p(\cM)$ is not necessarily flat.
Suppose $\diff a \in \diff H^p(\cM)$ is topologically trivial; we aim to compute its field strength  
$R(\diff a)$.
 From exactness of the central
SW-NE diagonal we  know that
$\diff a = \tau([\omega])$ for some 
class $[\omega]$ in the quotient $\Omega^{p-1}(\cM) / \Omega^{p-1}_\mathbb Z(\cM)$.
Commutativity of the lower triangle in the  diagram \eqref{eq:SS-} gives us
\beq \label{eq:rtd}
R(\diff a ) = R (\tau([\omega])) = d_{\mathbb Z}[\omega]  \ .
\eeq
The symbol $d_{\mathbb Z}$ in the  diagram denotes the standard de Rham differential
on forms, which passes to the quotient of $\Omega^{p-1}(\cM)$ by $\Omega^{p-1}_\mathbb Z(\cM)$.
The relation \eqref{eq:rtd} is familiar in physics:
if we have a topologically trivial $(p-1)$-form gauge field,
described by the globally defined form $\omega$ in some gauge,
its field strength is simply $d\omega$.

\item An element $\diff a \in \diff H^p(\cM)$ can be both flat and topologically trivial.
Such elements in $\diff H^p(\cM)$ are usually referred to as Wilson lines.
A Wilson line in $\diff H^p(\cM)$ can be identified with 
an element in the quotient $H^{p-1}(\cM;\mathbb R)/H^{p-1}_{\text{Free}}(\cM;\mathbb Z)\cong H^{p-1}(\cM;\mathbb Z)\otimes \mathbb{R}/ \mathbb{Z}$.
The latter is in turn isomorphic to
\be\label{eq:wilsonline}
 \frac{H^{p-1}(\cM;\mathbb R)}{H^{p-1}_{\text{Free}}(\cM;\mathbb Z)} \cong \frac{\Omega_{\rm closed}^{p-1}(\cM)}{ \Omega^{p-1}_\mathbb Z(\cM)} \,,
\ee
which is a torus of dimension $b^{p-1} = \dim H^{p-1}(\cM;\mathbb R)$.

\end{itemize}

Two differential cohomology
classes $\diff a,\diff b\in \diff H^p(\cM)$ with $I(\diff a)=I(\diff b)$
necessarily differ by a topologically trivial class.
Exactness of the central NW-SE exact sequence in \eqref{eq:SS-}
then implies that $\diff a - \diff b$ can be represented by an element
in $\Omega^{p-1}(\cM)/\Omega^{p-1}_\bZ(\cM)$.
We conclude that we can view 
$\diff H^p(\cM)$ as a fibration with basis the set of points in
$H^p(\cM;\mathbb{Z})$, and fiber isomorphic to
$\Omega^{p-1}(\cM)/\Omega^{p-1}_\bZ(\cM)$:
\be
\begin{tikzpicture}
\node (v1) at (1,0) {$\Omega^{p-1}(\cM)/\Omega^{p-1}_\bZ(\cM)$};
\node (v2) at (5,0)  {$\diff H^p(\cM)$};;
\node(v3) at (5,-1.5) {$H^p(\cM;\mathbb{Z})$};
\draw[->] (v1) --(v2);
\draw[->] (v2) -- (v3);
\end{tikzpicture} \,.
\ee
Concretely, if we pick some origin $\diff \Phi$ for the
fiber on top of $I(\diff \Phi)$, we can write
the most general element $\diff a$ of the fiber as
\beq \label{eq:choice_of_origin}
  \diff a = \diff \Phi + \tau([\omega]) \ ,
\eeq
where $\omega \in \Omega^{p-1}(X)$
is a differential form representing a class $[\omega]$
in the quotient of  $\Omega^{p-1}(\cM)$ by $\Omega^{p-1}_\bZ(\cM)$.
As pointed out above, a different choice for $\omega$ in the same class
$[\omega]$ is simply a gauge transformation.

\paragraph{Torsion Classes.}
Let us consider a torsion cohomology class $t\in H^p(\cM ; \mathbb Z)$.
It will be useful for us to choose a convenient origin $\diff \Phi$ for the fiber
on top of $t$.
By exactness of the long exact
sequence~\eqref{eq:exponential-ses}, we have that if
$t\in H^{p}(\cM;\mathbb{Z})$ is torsion then there is some (not
necessarily unique) $u\in H^{p-1}(\cM; \mathbb R/\mathbb Z)$ such that
$t = - \beta(u)$. 
Our choice for the origin of the fiber above $t$ is 
$\diff \Phi = i(u)$.
Commutativity of \eqref{eq:SS-} ensures
$I(\diff \Phi) = t$, confirming indeed that $\diff \Phi$ lies in the fiber on top of
$t$.
Moreover, the differential cohomology class $\diff \Phi$ is flat, $R(\diff \Phi) = 0$,
as follows from exactness
of the central NW-SE diagonal in \eqref{eq:SS-}.

\paragraph{Product structure in differential cohomology.}

There exists a bilinear product operation on differential
cohomology classes,  
\beq
  \bd \colon \quad \diff H^p(\cM )\times \diff H^q(\cM )\to \diff H^{p+q}(\cM) \ .
\eeq
The product $\bd$ is natural
and satisfies the following identities:
for any  $\diff a \in \diff H^p(\cM)$, $\diff b \in \diff H^q(\cM)$,
\begin{align} \label{eq:star_ids}
\diff a \bd \diff b   = (-)^{pq} \, \diff b \bd \diff a \ , \qquad
 I(\diff a \bd \diff b)  = I(\diff a) \smile I(\diff b) \ , \qquad
 R(\diff a \bd \diff b)   = R(\diff a) \wedge R(\diff b)   \ .
\end{align}
In the above relations, $\wedge$ is the standard wedge product of differential
forms and $\smile$ is the standard cup product of cohomology classes.

The product of a topologically trivial (respectively flat)
element in $\diff H^p(\cM)$ with any element in 
$\diff H^q(\cM)$ is again  topologically trivial (respectively flat).
More precisely, we have the identities
\beq \label{eq:flat_star}
\tau([\omega]) \bd \diff b = \tau ([ \omega \wedge R(\diff b) ]) \ , \qquad
i(u) \bd \diff b = i ( u \smile I(\diff b)) \ ,   
\eeq
for any
$\omega \in \Omega^{p-1}(\cM)$,
   $u\in H^{p-1}(\cM ; \mathbb R /\mathbb Z)$, and 
 $\diff b \in \diff H^q(\cM)$.\footnote{Using the
fact that $\bd$, $\wedge$, $\smile$ are graded commutative,
these identities can also be written in the form
\begin{equation}
  \diff b \bd \tau([\omega])   = (-1)^q \, \tau( [ R(\diff b) \wedge \omega ] )  \ , \qquad
     \diff b \bd i(u)   = (-1)^q \, i(I(\diff b) \smile u) \ ,  \nn
\end{equation}
for $\diff b \in \diff H^q(\cM)$, $\omega \in \Omega^{p-1}(\cM)$,
   $u\in H^{p-1}(\cM ; \mathbb R /\mathbb Z)$. 
}
Recall that $[\omega]$ denotes the equivalence class of $\omega$ in    
$\Omega^{p-1}(\cM)/\Omega_\mathbb Z^{p-1}(\cM)$.

\paragraph{Fiber integration in differential cohomology.}
Given a locally trivial fiber bundle $\cM$ with base $\cB$ and closed
fiber $\mathcal F$, we can define an integration over the fiber
\begin{equation}
  \int_{\mathcal F} \colon \qquad \diff H^p(\cM) \to \diff H^{p-\dim(\mathcal F)}(\cB)\, ,
\end{equation}
which we can characterize axiomatically. First, it is a natural group
homomorphism that is compatible with taking the curvature and taking
the characteristic class:
\begin{align}
  \int_{\mathcal F} R(\diff a)   = R\left(\int_{\mathcal F} \diff a\right)\,  , \qquad 
  \int_{\mathcal F} I(\diff a)   = I\left(\int_{\mathcal F} \diff a\right)\, .
\end{align}
(On the left hand side of these expressions we are using the usual
notions of fiber integration of differential forms and cohomology
classes.) It is also compatible with the maps $i$ and $\tau$:
\begin{align}
  \label{eq:i-integration}
  \int_{\mathcal F} i(u)   = i\left(\int_{\mathcal F} u\right)\, , \qquad 
  \int_{\mathcal F} \tau([\omega])   = \tau\left( \left[ \int_{\mathcal F} \omega \right] \right)\, .
\end{align}

An important special case is when we take $\cB = {\rm pt}$ and
we identify the fiber  $\mathcal F$ with $\cM$ itself.
One has $\diff H^0({\rm pt}) \cong \mathbb Z$,
$\diff H^1({\rm pt}) \cong \mathbb R / \mathbb Z$,
while $\diff H^p({\rm pt})$ is trivial for $p \neq 0,1$.
We then have two non-trivial integration maps.
The first is integer-values and yields the so-called
\emph{primary invariant} of a differential cohomology class
of degree $\dim (\cM)$,
\beq
\int_\cM \diff a = \int_\cM I(\diff a) = \int_\cM R(\diff a) \in \mathbb Z \ , \qquad
\diff a \in \diff H^{\dim(\cM)}(\cM) \ .
\eeq
The second integration operator
is valued in $\mathbb R /\mathbb Z$ and yields the
so-called \emph{secondary invariant}
 of a differential cohomology class
of degree $\dim (\cM)+1$,
\beq
\int_\cM \diff a = \int_\cM u \in \mathbb R /\mathbb Z \ , \quad
\diff a \in \diff H^{\dim(\cM) + 1}(\cM) \ , \quad u \in H^{\dim(\cM)}(\cM; \mathbb R/\mathbb Z) \ , \quad
\diff a = i(u) \ .
\eeq
We have used the fact that any element $\diff a \in \diff H^{\dim(\cM) +1}(\cM)$
is necessarily flat for dimensional reasons, and therefore 
can be written as $\diff a  = i(u)$ for some $ u \in H^{\dim(\cM)}(\cM; \mathbb R/\mathbb Z)$.

\subsection{Differential Cohomological Formulation of M-theory}

The topological terms in the M-theory low-energy effective action
can be written schematically in the form 
\beq \label{eq:action_naive}
e^{i S_{\rm top}} = \exp \, 2\pi i \, \int_{\cM_{11}} \bigg[ - \frac 16 \, C_3 \wedge G_4 \wedge G_4 
- C_3 \wedge X_8 \bigg] \ ,
\eeq 
where $\cM_{11}$ is 11d spacetime, $C_3$ is the M-theory 3-form gauge field,
$G_4$ is its field strength, and $X_8$ is an 8-form  characteristic 
class constructed from the Pontryagin classes $p_i(T\cM_{11})$, $i=1,2$,
of the tangent bundle to $\cM_{11}$,
\beq
X_8 = \frac{1}{192} \, \bigg[ p_1(T\cM_{11})^2 - 4 \, p_2(T\cM_{11}) \bigg] \ .
\eeq
The expression \eqref{eq:action_naive} for the topological couplings
can only be taken literally 
if the 3-form is topologically trivial, in which case $C_3$ is a globally
defined 3-form on $\cM_{11}$, and the integral in \eqref{eq:action_naive} 
can be understood as the standard integral of an 11-form. 
In topologically non-trivial situations,
such as those studied in this work,
greater care is needed to 
make sense of the formal expression \eqref{eq:action_naive}.

For our purposes, it will be enough to model the M-theory 3-form gauge
field as a class $\diff G_4 \in \diff H^4(\cM_{11})$ in (ordinary)
differential cohomology.\footnote{It should be noted that in the
  mathematical literature this element of $\diff H^4(\cM_{11})$ is
  sometimes denoted $\diff C_3$. We prefer the notation $\diff G_4$ to
  make manifest the degree of this differential cohomology class.}{}$^{,}$\footnote{More subtle questions require working on some
  \emph{generalised} differential cohomology theory, of which there
  are many types. The generalised cohomology theory appropriate for
  M-theory has been postulated in \cite{Fiorenza:2019usl}. It would be
  interesting to see if this more refined picture leads to any
  interesting consequences in field theory.}  In particular, we are
implicitly restricting ourselves to situations in which the periods of
the M-theory 4-form field strength are integrally quantized.  As
explained in \cite{Witten:1996md}, on certain spacetimes the periods
must be half-integrally quantized.  We argue in appendix \ref{app:DC}
that this does not occur in the setups discussed in this
work.\footnote{We refer the reader to \cite{Monnier:2013rpa} for a
  model for the M-theory 3-form in terms of a shifted differential
  cohomology class, which can accommodate both integral and
  half-integral periods.}

The topological action  \eqref{eq:action_naive}  is interpreted
as the $\mathbb R/\mathbb Z$-valued secondary invariant of a differential cohomology class 
$\diff I_{12} \in \diff H^{12}(\cM_{11})$, 
\beq \label{eq:Stop}
\frac{S_{\rm top}}{2\pi} = \int_{\cM_{11}} \diff I_{12} \mod 1 \ ,
\eeq
where $\diff I_{12}$ is given by
\beq \label{eq:I12}
\diff I_{12} = - \frac 16 \, \diff G_4 \bd \diff G_4 \bd \diff G_4
- \frac{1}{192} \, \diff G_4 \bd \diff p_1(T\cM_{11}) \bd \diff p_1(T \cM_{11})
+ \frac{1}{48} \, \diff G_4 \bd \diff p_2(T\cM_{11}) \ .
\eeq
In the previous expression, 
$\diff p_i(T\cM_{11}) \in \diff H^{4i}(\cM_{11})$ denotes a differential
refinement of the Pontryagin classes $p_i(T\cM_{11}) \in H^{4i}(\cM_{11};\mathbb Z)$
\cite{10.1007/BFb0075216, myBunke, Hopkins:2002rd}. 

Within the formalism of differential cohomology we are allowed to
consider products and $\mathbb Z$-linear combinations of differential
cohomology classes, but multiplying by rational coefficients --- such
as the factor of 1/6 in front of the
$\diff G_4 \bd \diff G_4 \bd \diff G_4$ term in \eqref{eq:I12} ---
leads to a quantity which is not well defined in general.  The fact
that the particular combination $\diff I_{12}$ is nonetheless
well-defined stems from the analysis of
\cite{deAlwis:1996hr,Witten:1996md}, which demonstrates that the total
topological action $e^{i S_{\rm CS}}$ is well-defined up to a sign,
which cancels a potential sign problem in the definition of the
Rarita-Schwinger determinant.  This sign ambiguity arises if and only
if the periods of the $G_4$ field strength are half-integrally
quantized.  As mentioned above, this does not occur for the setups
discussed in this work, meaning that $e^{i S_{\rm CS}}$ is
well-defined by itself.

\subsection{Kaluza-Klein Reduction in Differential Cohomology} \label{sec:KK}

Let us consider an 11d spacetime $\cM_{11}$ that is the direct product
of an ``internal'' manifold $L_n$ of dimension $n$, and an
``external'' spacetime $\cW_{11-n}$ of dimension $11-n$,
\beq \label{eq:direct_product} \cM_{11} = \cW_{11-n} \times L_n \ .
\eeq It is standard to consider the expansion of the M-theory 3-form
onto harmonic forms on $L_n$, to obtain massless $U(1)$ gauge fields
on $\cW_{11-n}$ of various $p$-form degrees.  Our goal is to
generalise this picture, by expanding the M-theory 3-form onto
\emph{all} cohomology classes of $L_n$, both free and torsional.

On a factorized spacetime such as \eqref{eq:direct_product}, it is natural to 
start from objects (differential forms, cohomology classes, differential cohomology classes)
defined on the two factors, and combine them into objects on the total space.
Let 
\beq
p_\cW : \cM_{11} \rightarrow \cW_{11-n} \ , \qquad
p_L : \cM_{11} \rightarrow L_n 
\eeq
be the projection maps onto the two factors of $\cM_{11}$. 
For notational simplicity,
we henceforth omit the pullback maps $p_\cW^*$
and $p_L^*$ from various factorized expressions.
For example,
\begin{subequations}
  \begin{align}
    &\text{if $\lambda \in \Omega^r(\cW_{11-n})$ and $\omega \in \Omega^s(L_n)$,}   & 
                                                                                      \lambda \wedge \omega &\text{ is shorthand for } 
                                                                                                              p_\cW^*(\lambda) \wedge p_L^*(\omega)  \ ,
    \\
    &\text{if $a \in H^r(\cW_{11-n};\mathbb Z)$ and $b \in H^s(L_n;\mathbb Z)$, }   & 
                                                                                      a \smile b  &\text{ is shorthand for }  p_\cW^*(a) \smile p_L^*(b) \ ,
    \\
    &\text{if $\diff a \in \diff H^r(\cW_{11-n})$ and $\diff b \in \diff H^s(L_n)$,}  & 
                                                                                        \diff a \bd \diff b   &\text{ is shorthand for }   p_\cW^*(\diff a) \bd p_L^*(\diff b)  \label{eq:omit-star-pullbacks}\ .
  \end{align}
\end{subequations}
We observe that the naturality of the products $\smile$ and $\bd$, together with \eqref{eq:star_ids}, implies  
\beq  \label{eq:product_ids}
R(\diff a \bd \diff b) = R(\diff a) \wedge R(\diff b) \ , \quad
I(\diff a \bd \diff b) = I(\diff a) \smile I(\diff b) \ , \quad \text{
for $\diff a \in \diff H^r(\cW_{11-n})$, $\diff b \in \diff H^s(L_n)$
} \ .
\eeq

For each $p=0,\dots,n$, $H^p(L_n ; \mathbb Z)$ is a finitely generated
Abelian group.  We take the generators of $H^p(L_n ; \mathbb Z)$ to be
\be\ba
\text{free generators of $H^p(L_n ;\mathbb Z)$:}&\quad   v_{p(\alpha)} \,, \quad  \alpha  \in \{ 1 , \dots b^p \}
\cr 
\text{torsion generators of $H^p(L_n ;\mathbb Z)$:}&  \quad t_{p(i)}\,,\quad  i \in \mathcal I_p \,.
\ea
\ee
The subscript $p$ is a reminder that these are classes of degree $p$,
while $(\alpha)$, $(i)$ are labels that enumerate the generators.  We
define $b^p \df \dim H^p(L_n;\mathbb R)$, the $p$-th Betti number of
$L_n$.  For the torsion generators, the index set $\cI_p$ is some
finite set of labels, which can be specified more explicitly in
concrete examples.  Each torsional generator has a definite torsional
order: the minimal positive integer $n_{(i)}$ such that
$n_{(i)} \, t_{p(i)} = 0$ (no sum on $i$).

For simplicity we take
\begin{equation}
{\rm Tor} \, H^*(\cW_{11-n} ; \mathbb Z) = 0 \ .
\end{equation}
By the K\"unneth formula, we may then expand a generic cohomology
class $a_{4} \in H^4(\cM_{11};\mathbb Z)$ as
\beq \label{eq:a4_expansion} a_4 = \sum_{p=0}^4 \, \sum_{\alpha_p
  =1}^{b^p} \sigma_{4-p}^{(\alpha_p)} \smile v_{p(\alpha_p)} +
\sum_{p=0}^4 \, \sum_{i_p \in \cI_p } \rho_{4-p}^{(i_p)} \smile
t_{p(i_p)} \ .  \eeq In the above expression,
$\sigma^{(\alpha_p)}_{4-p} , \rho^{(i_p)}_{4-p} \in
H^{4-p}(\cW_{11-d}; \mathbb Z)$.

Recall that the map $I$ in \eqref{eq:SS-}
is surjective. This applies both for elements in $\diff H^*(\cW_{11-n})$
and $\diff H^*(L_n)$\footnote{We are implicitly considering a Wick-rotated
version of the theory, and we are taking external spacetime $\cW_{11-n}$
to be a closed, connected, oriented manifold.}.
It follows that there exist differential cohomology classes
$\diff F_{4-p}^{(\alpha_p)}, \diff B_{4-p}^{(i_p)} \in \diff H_{4-p}(\cW_{11-n})$
and  $\diff v_{p(\alpha_p)}, \diff t_{p(i_p)} \in \diff H_p(L_n)$ such that
\beq \label{eq:nice_objects}
\sigma_{4-p}^{(\alpha_p)} = I (  \diff F_{4-p}^{(\alpha_p)} ) \ , \qquad
\rho_{4-p}^{(i_p)} = I (  \diff B_{4-p}^{(\alpha_p)} ) \ , \qquad
  v_{p(\alpha_p)} = I (\diff v_{p(\alpha_p)}) \ , \qquad
   t_{p(i_p)}   = I(  \diff t_{p(i_p)}   ) \ .
\eeq
With the objects 
$\diff F_{4-p}^{(\alpha_p)}, \diff B_{4-p}^{(i_p)} \in \diff H_{4-p}(\cW_{11-n})$
and  $\diff v_{p(\alpha_p)}, \diff t_{p(i_p)} \in \diff H_p(L_n)$
we can construct the following differential cohomology class
\beq \label{eq:diff_a4}
\diff a_4 = \sum_{p=0}^4   \, \sum_{\alpha_p =1}^{b^p}  \diff  F_{4-p}^{(\alpha_p)}  \bd \diff v_{p(\alpha_p)}
+ \sum_{p=0}^4   \, \sum_{i_p \in \cI_p }  \diff  B_{4-p}^{(i_p)}  \bd \diff t_{p(i_p)}  \ .
\eeq
The salient property of $\diff a_4$ in \eqref{eq:diff_a4} is that
it represents a possible lift of $a_4$ in \eqref{eq:a4_expansion}, in the sense that
\beq
I(\diff a_4) = a_4 \ .
\eeq
This is verified using \eqref{eq:nice_objects}, the naturality of the differential
cohomology product $\bd$, and the second identity in \eqref{eq:star_ids}.

The differential cohomology class $\diff a_4$
is not the most general class that reduces to $a_4$ under the action of $I$.
From the discussion around \eqref{eq:choice_of_origin},
however, we know that any other class that reduces to $a_4$
must differ from $\diff a_4$ by a topologically trivial
element of $\diff H^4(\cM_{11})$, which can be 
represented by a globally defined 3-form.
These considerations lead us to 
 the following final form for the Ansatz 
for $\diff G_4$,
\begin{equation} \label{eq:diffG4_ansatz}
\diff G_4 = \sum_{p=0}^4   \, \sum_{\alpha_p =1}^{b^p}  \diff  F_{4-p}^{(\alpha_p)}  \bd \diff v_{p(\alpha_p)}
+ \sum_{p=0}^4   \, \sum_{i_p \in \cI_p }  \diff  B_{4-p}^{(i_p)}  \bd \diff t_{p(i_p)} 
+ \tau([\omega_3]) \ , \qquad \omega_3 \in \Omega^3(\cM_{11}) \ .
\end{equation}
The first two sums in \eqref{eq:diffG4_ansatz} encode all topological
information about $\diff G_4$, while the last term collects the
topologically trivial part of $\diff G_4$.

The differential cohomology classes
$\diff F_{4-p}^{(\alpha_p)}, \diff B_{4-p}^{(i_p)} \in \diff
H_{4-p}(\cW_{11-n})$ encode external gauge fields. More precisely, we
have:
\begin{itemize}
\item The class $\diff F_{4-p}^{(\alpha_p)}$, of degree $(4-p)$,
  represents a $(3-p)$-form gauge field with gauge group $U(1)$, which
  restricts to a background field for a $U(1)$ $(2-p)$-form symmetry on the
  boundary;
\item The class $\diff B_{4-p}^{(i_p)}$, of degree $(4-p)$, represents
  a discrete $(4-p)$-form gauge field with gauge group
  $\mathbb Z_{n_{(i_p)}}$, where $n_{(i_p)}$ is the torsion order of
  $t_{p(i_p)}$, which restricts to a background field for a
  $\bZ_{n_{(i_p)}}$ $(3-p)$-form symmetry on the boundary.
\end{itemize}
The first case is familiar, but the second one requires some
additional explanation. Notice in particular the difference in the
relation between the differential cohomology class degree and the
degree of the higher form symmetry on the boundary. Consider two
classes $\diff B,\diff B'\in \diff H^{4-p}(\cW_{11-d})$, such that
$I(\diff B)=I(\diff B')$. Then $I(\diff B - \diff B')=0$, so by
exactness of~\eqref{eq:SS-} there is some globally defined
differential form $\ssb$ of degree $3-p$ such that
$\diff B' = \diff B + \tau(\ssb)$. By~\eqref{eq:flat_star} and
naturality of $\tau$ and $R$ we then have that
$\tau(\ssb)\bd \diff t_{p(i_p)}=\tau(\ssb\wedge R(\diff
t_{p(i_p)}))=0$, since we have chosen $\diff t$ to be flat. This
implies $\diff B\star\diff t_{p(i_p)}=\diff B'\star\diff t_{p(i_p)}$,
so $\diff B_{4-p}^{(i_p)}\star \diff t_{p(i_p)}$ is fully determined
by its cohomology class $I(\diff B_{4-p}^{(i_p)})\smile I(\diff t_{p(i_p)})$
(given our canonical choice of $\diff t_{p(i_p)}$). This is an element of
$H^{4-p}(\cW_{11-d};\mathbb{Z})\otimes \Tor H^{p}(L_p;\mathbb{Z})$, which by the universal
coefficient theorem is isomorphic (since we are assuming
$\Tor H^{4-p}(\cW_{11-d};\mathbb{Z})=0$) to
$H^{4-p}(\cW_{11-d}; \Tor H^{p}(L_p;\mathbb{Z}))$. So by this isomorphism, we can
reinterpret $I(\diff B_{4-p}^{(i_p)})\smile I(t_{p(i_p)})$ as a class
in $H^{4-p}(\cW_{11-d}; \bZ_{n_{(i_p)}})$. But such a cohomology class
is a map (up to homotopy) from $\cW_{11-d}$ to the classifying space
$K(\bZ_{n_{(i_p)}}, 4-p)=B^{4-p}\bZ_{n_{(i_p)}}$, which is the data
that defines a principal bundle for a $(3-p)$-form symmetry. For
instance, when $p=3$ we have an ordinary (0-form) discrete symmetry,
and the backgrounds for such symmetries are maps from $\cW_{11-d}$ to
$B\bZ_{n_{(i_p)}}$, or equivalently elements of
$H^1(\cW_{11-d};\bZ_{n_{(i_p)}})$.

\paragraph{Integration on products.}

Finally, we want to integrate differential cohomology
classes on product spaces. Assume that
$\diff a\in \diff H^p(X)$, $\diff b\in \diff H^q(Y)$. Then
\begin{equation}
  \label{eq:external-product-general}
  \int_{X\times Y} \diff a \bd \diff b = (-1)^{(q-\dim(Y))\dim(X)}\left(\int_X\diff a\right)\bd \left(\int_Y\diff b\right)\, .
\end{equation}
For the Chern-Simons coupling~\eqref{eq:Stop} we have
$p+q=\dim(X)+\dim(Y)+1$. In this case:
\begin{equation}
  \label{eq:external-product-integral}
  \int_{X\times Y} \diff a \bd \diff b = \begin{cases}
    \left(\int_X u\right) \left(\int_Y R(\diff b)\right) & \text{if } p=\dim(X)+1\, ,\\
    (-1)^p  \left(\int_X R(\diff a)\right) \left(\int_Y v\right) & \text{if } p=\dim(X)\, ,\\
    0 & \text{otherwise}\, ,
  \end{cases}
\end{equation}
simply by taking into account that the integrals on the right hand
side of~\eqref{eq:external-product-general} are only non-vanishing for
very specific values of $p$ and $q$, as explained above. Here we have
used that in the first case $\diff a$ is flat for degree reasons, so
there is some $u\in H^{\dim(X)}(X; \bR/\bZ)$ such that
$\diff a = i(u)$, and similarly $\diff b = i(v)$ in the second
case.

In this paper we are particularly interested in those integrals over
the internal space involving torsional elements $\diff t$ (we omit the
subindices here for notational simplicity). First note that since we
have chosen these torsional generators to be flat, $R(\diff t)=0$, we
have
$\tau([\omega_3])\bd \diff t=\tau([\omega_3\wedge R(\diff t)])=0$, due
to~\eqref{eq:flat_star}. So any integral involving the $\diff t$
generators will be a topological invariant (including invariant under
deformations of the connection), by virtue of being independent of
$\tau([\omega_3])$.

This implies that when expanding~$\diff G_4^3$
using~\eqref{eq:diffG4_ansatz} we have
\begin{equation}
  \label{eq:G_4^3-expansion}
  \diff G_4^3 = \sum \text{monomials involving $\diff t$ and $\diff v$} + \sum \text{monomials involving $\diff v$ and $\tau([\omega_3])$}\, ,
\end{equation}
with no monomials involving both $\tau([\omega_3])$ and the torsional
classes. The second class of monomials are accessible using the
ordinary formalism based on differential forms, so we will not discuss
them further; both because they are well understood and because our
interest in this paper is on discrete higher form symmetries, which
arise from the torsional sector.

Now, regarding the first class of terms in~\eqref{eq:G_4^3-expansion},
by~\eqref{eq:star_ids} we have for all $\diff a=\diff t\bd \diff b$
that $R(\diff a)=0$, for any $\diff b$. (Note that
by~\eqref{eq:star_ids} $I(\diff a)$ is automatically torsion if
$I(\diff t)$ is.) This implies that when doing the integration over
the internal space $L_n$, torsional elements
$\diff a = \diff t\bd \diff b$ only contribute if
$\diff a\in \diff
H^{n+1}(L_n)$. By~\eqref{eq:external-product-integral} this leads to
effective actions on $\cW_{11-d}$ which are primary invariants, not
secondary ones. (Said more plainly: reducing the Chern-Simons term in
11d on the torsional sector leaves us with an ordinary integral of
characteristic classes in $\cW_{11-d}$.)

\medskip

We now turn to deriving these theories, which will be the SymTFTs,  in various geometric engineering settings.


\section{Symmetry TFTs from M-theory on $S^3/\Gamma$}
\label{sec:7d}

Consider first the case where the theory $\mathfrak{T}$ that we are
engineering is a seven dimensional \mbox{$\mathcal{N}=1$} theory with
Lie algebra $\fg_\Gamma$, obtained by putting M-theory on
$\mathcal{M}_{11}=\mathcal{M}_7\times\mathbb{C}^2/\Gamma$, where
$\Gamma$ is a discrete subgroup of $SU(2)$, and $\cM_7$ is a (closed
and torsion-free, for simplicity) manifold where $\mathfrak{T}$ lives. The
resulting seven dimensional theory has defect group
\cite{DelZotto:2015isa} $Z(G_\Gamma)^{(1)}\oplus Z(G_\Gamma)^{(4)}$
with $Z(G_\Gamma)$ the center of universal cover group $G_\Gamma$. For
instance, the field theory with gauge group $G_\Gamma$ has electric
1-form symmetry $Z(G_\Gamma)$, while the theory with gauge group
$G_\Gamma/Z(G_\Gamma)$ has magnetic 4-form symmetry
$Z(G_\Gamma)$. Other global forms for the gauge group are often
possible, depending on the choice of $\Gamma$, the analysis of the
possibilities is identical to the one in \cite{Aharony:2013hda}. The
seven dimensional theory has additionally a 2-form $U(1)$ instanton
symmetry, with generator the integral of the instanton density on a
closed 4-surface.

In the rest of this section we will argue that reducing M-theory on
$S^3/\Gamma=\partial(\bC^2/\Gamma)$ leads to an eight dimensional TFT
encoding both this choice of global form for the seven dimensional
theory (equivalently, the choice of its higher form symmetries) and
the anomalies of these higher form symmetries.

We expect these two sectors of the eight dimensional TFT to interact
in interesting ways: recall that choosing a global form for the gauge
group (which will be able to rephrase as a choice of boundary
behaviour in the BF theory~\eqref{eq:BF}) can also be understood as a
gauging of the higher form symmetries \cite{Kapustin:2014gua}. In the
presence of 't Hooft anomalies this gauging procedure might be
obstructed, or lead to less conventional symmetry structures (see for
instance \cite{Bhardwaj:2017xup,Tachikawa:2017gyf} for systematic
discussions). It would be very interesting to analyse this problem
from our higher dimensional vantage point, where it requires study of
gapped boundary conditions of the TFTs we construct, but we will not
do so in this paper.

\subsection{Choice of Global Structure from 8d}

We start by discussing how to see the choice of global form in terms
of a choice of boundary conditions for a gapped eight dimensional
theory. The basic idea is discussed in
\cite{GarciaEtxebarria:2019caf,Morrison:2020ool,Albertini:2020mdx},
where it was found that the geometric origin of the 1-form and 4-form
symmetries, and the fact that there is a choice to be made, can be
traced back to the non-commutativity of the boundary values of $F_4$
and $F_7$ fluxes in the presence of torsion in the asymptotic boundary
$\partial\mathcal{M}_{11}$ at infinity \cite{Freed:2006yc,
  Freed:2006ya}.

Consider M-theory compactified on $\bR\times X_{10}$, where we take the
first factor to be the time direction. For the moment we ignore the
Chern-Simons terms in the M-theory action, and view M-theory as a
generalised Maxwell theory. We will discuss the effect of the
Chern-Simons terms extensively below. We focus on the operators
$\Phi(\cT_3)$ and $\Phi(\cT_6)$ measuring the periods of $C_3$ and
$C_6$ on torsional cycles $\cT_3$ and $\cT_6$ inside $X_{10}$ (our
discussion is only sensitive to the homology class of the cycles). The
authors of \cite{Freed:2006yc, Freed:2006ya} found that these two
operators do not always commute, but rather there is a phase in their
commutation relation:
\be
  \label{eq:flux-noncommutativity}
  \Phi(\cT_3)\Phi(\cT_6) = e^{2\pi i\, \sL(\cT_3,\cT_6)} \Phi(\cT_6)\Phi(\cT_3)\, .
\ee
Here, $L$ is the linking pairing
\be
  \sL: \Tor H_3(X_{10};\bZ) \times \Tor H_6(X_{10};\bZ) \to \bQ/\bZ\, .
\ee
This pairing can be defined as follows. Assume that
$\cT_6\in\Tor H_6(X_{10};\bZ)$ is of order $n\in\bZ$. That is, there is some
chain $C_7$ such that $\partial C_7 = n \cT_6$. Then
\be
  \sL(\cT_3, \cT_6) \df \frac{1}{n} \cT_3 \cdot C_7 \mod 1\, .
\ee
One can analogously define a linking pairing
$\Tor H^k(\cM_d;\bZ)\times \Tor H^{d+1-k}(\cM_d;\bZ)\to \bQ/\bZ$ in
cohomology, which will appear below.

In our case we are putting M-theory on $\cM_7\times \bC^2/\Gamma$,
which is not quite of the form $\bR\times X_{10}$, but
\cite{GarciaEtxebarria:2019caf,Morrison:2020ool,Albertini:2020mdx}
argue that taking the radial direction of $\bC^2/\Gamma$ as time (that
is, taking $X_{10}=\cM_7\times S^3/\Gamma$) reproduces the answers one
expects from field theory. Our task is to modify the analysis in those
papers to adapt it to the viewpoint that we are advocating here. In
this section we will focus on the fate of the $\Phi(\cT_3)$ and
$\Phi(\cT_6)$ operators upon dimensional reduction. The Chern-Simons
part of the M-theory action gives additional contributions, which we
study in detail below. We will find that reducing M-theory on
$S^3/\Gamma$ leads to a non-invertible theory in eight dimensions
whose states naturally correspond to the possible choices of global
structure for the seven dimensional theory.

\begin{table}
  \def\arraystretch{1.6}
  \centering
  \begin{tabular}{|c|c|c|c|}\hline
    $\Gamma$ & $G_\Gamma$ & $\Gamma^{\text{ab}}$ & $\sL_{S^3/\Gamma}$ \\
    \hline
    \hline
    $A_{n-1}$ & $SU(n)$ & $\bZ_n$ 
    & $\frac{1}{n}$ \\ 
    $\mathrm{D}_{2n}$ & $\Spin(4n)$ & $\mathbb{Z}_2\oplus \mathbb{Z}_2$ & ${1\over 2}\left(\begin{matrix} n & n-1 \\ n-1  & n \end{matrix}\right) $\\ 
    $ \mathrm{D}_{2n+1}$ &   $\Spin(4n+2)$ &   $\bZ_4$
    & $ \frac{2n-1}{4} $ \\ 
    $2T$ & $E_6$ & $\bZ_3$ & 
    $\frac{2}{3}$ \\ 
    $2O$ & $E_7$ & $\bZ_2$ & 
    $\frac{1}{2}$ \\ 
    $2I$ & $E_8$ & $0$ & $0$\\\hline
  \end{tabular}  
  \caption{The abelian group $H_1(S^3/\Gamma;\bZ)=\Gamma^{\text{ab}}$ and
    the linking pairing
    $\sL\colon \Gamma^{\text{ab}}\times\Gamma^{\text{ab}}\to\bQ/\bZ$,
    from \cite{GarciaEtxebarria:2019caf}. We have altered the
    presentation of the results with respect to that paper to bring
    the table closer to its spin Chern-Simons refinement presented in
    table~\ref{table:7d-CS} below.}
  \label{table:7d-linking}
\end{table}

The analysis goes as follows. We place M-theory on a manifold of the
form $\bR\times \cM_7\times S^3/\Gamma$, where we view the $\bR$
factor as time when quantising. We are assuming that $\Tor^k(\cM_7)=0$
for all $k$, so by the Künneth formula
\be
  \Tor H_p(X_{10};\bZ) = H_{p-1}(\cM_7;\bZ)\otimes H_1(S^3/\Gamma;\bZ)\, ,
\ee
where we are using the fact that the torsion in the homology of
$S^3/\Gamma$ is localised on degree 1:
\be
  \label{eq:lens-(co)homology}
  H_q(S^3/\Gamma;\bZ) = H^{3-q}(S^3/\Gamma;\bZ) = \begin{cases}
    \mathbb{Z} & \text{for } q=0,3\, ,\\
    \Gamma^{\text{ab}} \df \frac{\Gamma}{[\Gamma, \Gamma]} & \text{for }
    q=1\, ,\\
    0 & \text{for } q=2\, .
  \end{cases}
\ee
The values for $\Gamma^{\text{ab}}$ when $\Gamma\subset SU(2)$ are given in table \ref{table:7d-linking}. Looking to this
table, one observes
\cite{Acharya:2001hq,DelZotto:2015isa,GarciaEtxebarria:2019caf} that
$\Gamma^{\text{ab}}=Z(G_\Gamma)$, which is a key fact necessary for
the whole picture to be consistent.

We find, in particular, that the torsional cycles $\cT_3$ and $\cT_6$
are necessarily of the form $\cT_3=\Sigma_2\times \cT_1$ and
$\cT_6=\Sigma_5\times \cT_2$, with $\cT_1$ and $\cT_2$ generators of
the torsional group $H_1(S^3/\Gamma;\bZ)$, and $[\Sigma_p]\in
H_p(\cM_7;\bZ)$. Assume for simplicity that $\Gamma^{\text{ab}}=\bZ_k$,
for some $k$. (The $D_{2n}$ cases can be treated similarly, at the
cost of introducing more fields.) Then we have that from the point of
view of the eight dimensional theory on $\bR\times \cM_7$ we have two
kinds of operators $\Psi(\Sigma_2)$ and $\Psi(\Sigma_5)$, parametrised
by two-surfaces $\Sigma_2$ and five-surfaces $\Sigma_5$. These
operators are the eight dimensional push-forwards of the operators
$\Phi(\Sigma_2\times\cT_1)$ and $\Phi(\Sigma_2\times\cT_1)$ in the
eleven dimensional theory, with $\cT_1$ the generator of
$H_1(S^3/\Gamma;\bZ)$. The commutation relation of these operators on a
spatial slice $\cM_7$ of the eight dimensional theory is determined
from~\eqref{eq:flux-noncommutativity}, using that
$\sL(\Sigma_2\times\cT_1, \Sigma_2\times\cT_1) =
(\Sigma_1\cdot\Sigma_2) \sL_{S^3/\Gamma}(\cT_1, \cT_1)$
\be
  \Psi(\Sigma_2)\Psi(\Sigma_5) = e^{2\pi i \ell^{-1} \, \Sigma_2\cdot\Sigma_5} \Psi(\Sigma_5)\Psi(\Sigma_2)\, ,
\ee
where we have defined $\ell^{-1}\df \sL_{S^3/\Gamma}(\cT_1,\cT_1)$.\footnote{Recall
  that the linking pairing is defined modulo 1, so in all cases ---
  except $D_{2n}$, where $\ell$ is not defined --- $\ell$ can be taken
  to be an integer.} This is precisely the operator content and
commutation relations of a BF theory with the following Lagrangian
(see \cite{Witten:1998wy} for a derivation)
\be
  \label{eq:BF}
 \Sym = \ell \int B_2 \wedge dC_5\, .
\ee
For $|\ell| > 1$ this is a non-invertible theory, whose state space
reproduces the choices of global structure expected from the field
theory side
\cite{Witten:1998wy,Aharony:2013hda,Tachikawa:2013hya,GarciaEtxebarria:2019caf}.

\subsection{Anomaly Theory from Link Reduction}

We will now describe how to obtain the eight dimensional anomaly
theory encoding the mixed 't Hooft anomaly between 1-form center
symmetries and the 2-form $U(1)$ instanton symmetry. We will find that
the anomaly theory on any (closed and torsion-free, for simplicity)
$\cW_8$ can be derived by taking the eleven dimensional Chern-Simons
part of the M-theory action on $\cW_8\times S^3/\Gamma$, and
integrating over $S^3/\Gamma$.

We are assuming $\Tor(H^3(\mathcal{W}_8;\bZ))=0$, so by the universal
coefficient theorem
$H^2(\mathcal{W}_8; \Gamma^{\text{ab}})=H^2(\mathcal{W}_8;\bZ)\otimes
H^2(S^3/\Gamma;\bZ)$, and we can parametrize a generic element
$a_4\in H^4(M^{11};\bZ)$ by
\begin{equation}
  \label{eq:torsional-expansion}
  a_4 = \sigma_4\smile 1 + \sum_{i} \rho_2^{(i)} \smile t_{2(i)} + \sigma_1\smile \text{vol}(S^3/\Gamma)\, .
\end{equation}
where $\text{vol}(S^3/\Gamma)$ is the generator of $H^3(S^3/\Gamma;\bZ)=\mathbb{Z}$,
$t_{2(i)}$ are the (torsional) generators of
$H^{2}(S^3/\Gamma;\bZ)=\Gamma^{\text{ab}}$, and ``$1$'' is the generator
of $H^0(S^3/\Gamma;\bZ)=\mathbb{Z}$.

Given an element $a_4\in H^4(\cM^{11};\bZ)$ written
as~\eqref{eq:torsional-expansion} we can uplift to a differential
cohomology class by using (\ref{eq:diffG4_ansatz}) to write
\begin{equation}\label{eq:G4for7d}
    \diff G_4 = \diff \gamma_4 \bd \diff 1 + \sum_i \diff B_2^{(i)}\bd
    \diff t_{2 (i)} + \diff \xi_1\bd \diff v_3 + \tau([\omega_3])\, ,
\end{equation}
where $\omega_3\in \Omega^3(\cM^{11})$,
$\diff \xi_1\in \diff H^1(\mathcal{W}_8;\bZ)$,
$\diff v_3 \in \diff H^3(S^3/\Gamma;\bZ)$, $I(\diff \xi_1)=\sigma_1$ and
$I(\diff v)=\text{vol}(S^3/\Gamma)$. We have
$I(\diff \gamma_4)=\sigma_4$ and we choose $\diff t_{2(i)}$ such that
$R(\diff t_{2(i)})=0$.

For the problem at hand, the terms in $\diff I_{12}$ in \eqref{eq:I12}
originating from $X_8$ do not contribute for degree reasons. (They
will play an important role in the case of 5d SCFTs below.) For
simplicity of exposition, let us assume first that $\Gamma^{\rm ab}$
has a single generator. Substituting (\ref{eq:G4for7d}) into the
Chern-Simons coupling (\ref{eq:Stop}) we obtain
\begin{equation}\label{eq:CSfor7d}
\begin{split}
    \frac{S_{\text{top}}}{2\pi}&=-\frac{1}{6} \int_{M_{11}} \diff G_4 \bd \diff G_4 \bd \diff G_4 \\ 
    &= 
    \frac{1}{2} \int_{\mathcal{W}_8}  \diff \gamma_4^2 \bd \diff \xi_1 \int_{S^3/\Gamma} \diff v-
    \frac{1}{2} \int_{\mathcal{W}_8} \diff \gamma_4 \bd \diff B_2^2 \int_{S^3/\Gamma} \diff t_2^2- \frac{1}{6} \int \diff
    \tau(w_3)\, .    
\end{split}
\end{equation}
The first term in this expression encodes a potential mixed anomaly
between a $(-1)$-form symmetry and the 2-form instanton symmetry. The
second term in (\ref{eq:CSfor7d}) corresponds to a mixed 't Hooft
anomaly between the center 1-form symmetry
$Z(G_\Gamma)=\Gamma^{\text{ab}}$ and the instanton 2-form symmetry. In
what follows we will concentrate on this last term. 
To find the coefficient of this
anomaly, we must evaluate the integral
\begin{equation}\label{eq:g4in7d}
{\rm CS}[S^3/\Gamma , \diff t_2] =   \frac 12  \int_{S^3/\Gamma} \diff t_2\bd \diff t_2\, .
\end{equation}
This $\mathbb R/\mathbb Z$-valued quantity is the spin Chern-Simons
invariant, evaluated for the 3-manifold $S^3/\Gamma$ and the flat
connection $\diff t_2 \in \diff H^2(S^3/\Gamma;\bZ)$. In general such
Chern-Simons invariants also depend on the spin structure on the
manifold. In our case, by construction, we have the spin connection
induced on the boundary of the supersymmetric compactification of
M-theory on $\bC^2/\Gamma$.

Using $\int_{S^3/\Gamma} \diff v = 1$, and neglecting the $\tau$ term
according to the general discussion of section \ref{sec:KK}, we
compute the SymTFT:
\be\label{7dSym}
\boxed{\Sym=  \frac 12 \, \int_{\cW_8}  \diff \gamma_4 \bd \diff \gamma_4 \bd \diff \xi_1  - {\rm CS}[S^3/\Gamma , \diff t_2] \, \int_{\cW_8} \diff \gamma_4 \bd \diff B_2 \bd \diff B_2  \,.}
\ee
The cases in which $\Gamma^{\rm ab}$ has two generators
can be treated in a completely analogous way, yielding
\be
  \label{eq:7d-anomaly-theory}
  \Sym =  \frac 12 \, \int_{\cW_8}  \diff \gamma_4 \bd \diff \gamma_4 \bd \diff \xi_1  - \sum_{i,j}{\rm CS}[S^3/\Gamma]_{ij} \, \int_{\cW_8} \diff \gamma_4 \bd \diff B_2^{(i)} \bd \diff B_2^{(j)}\ .
\ee
Note that formally one would be tempted to write
\be
  {\rm CS}[S^3/\Gamma]_{ij} = \frac{1}{2}\int_{S^3/\Gamma} \diff t_{2(i)} \bd
  \diff t_{2(j)}\, .
\ee
The factor of $\frac{1}{2}$ makes the right hand side not well
defined. Luckily (but unsurprisingly, given that our starting
Chern-Simons coupling in M-theory action is well-defined
\cite{Witten:1996md}), due to the symmetry
properties~\eqref{eq:star_ids} of the Cheeger-Simons product is only
the sum ${\rm CS}[S^3/\Gamma]_{ij}+{\rm CS}[S^3/\Gamma]_{ji}$ that
enters in the anomaly theory~\eqref{eq:7d-anomaly-theory}, and this
sum is well defined:
\begin{equation}
  {\rm CS}[S^3/\Gamma]_{ij} + {\rm CS}[S^3/\Gamma]_{ji} =
  \int_{S^3/\Gamma} \diff t_{2(i)} \bd \diff t_{2(j)}\, .
\end{equation}
Similar remarks apply to the off-diagonal entries in the $D_{2n}$ case
in table~\ref{table:7d-CS} below. We provide a more systematic
discussion of this issue at the end of this section.

\subsection{Evaluation of the Chern-Simons Invariant} \label{sec:CS3}

Let us now discuss a convenient formalism to evaluate the CS invariant
\eqref{eq:g4in7d} (including the 1/2 prefactor), obtained by a
straightforward generalisation of a discussion in the three
dimensional case by Gordon and Litherland
\cite{Gordon1978OnTS}.

\subsubsection{Cohomology of the Link and Bulk Compact Divisors} \label{sec:compact_cycles}

Let $L_{d-1}$ be a closed, connected, oriented $(d-1)$-manifold, and suppose that $L_{d-1}$
can be realised as boundary of a $d$-manifold $X_{d}$.
The long exact sequence in relative homology yields
\beq \label{eq:long_rel_hom}
\dots \rightarrow H_{d-2}(X_d;\bZ) \rightarrow H_{d-2}(X_d,L_{d-1};\bZ) \rightarrow H_{d-3}(L_{d-1};\bZ) \rightarrow 
H_{d-3}(X_d;\bZ) \rightarrow \dots  \ .
\eeq
We now make the assumption
\beq \label{eq:assumption1}
H_{d-3}(X_d;\bZ) = 0 \ .
\eeq
Using Poincar\'e duality in $L_{d-1}$ we have
$H_{d-3}(L_{d-1};\bZ) \cong H^2(L_{d-2};\bZ)$, and from \eqref{eq:long_rel_hom} we get the exact sequence
\beq \label{eq:exact_seq1}
H_{d-2}(X_d;\bZ) \xrightarrow{A} H_{d-2}(X_d, L_{d-1};\bZ) \xrightarrow{f} H^2(L_{d-1};\bZ) \rightarrow 0 \ .
\eeq
Notice in particular that the homomorphism $f$ is surjective:
any class in $H^2(L_{d-1};\bZ)$ can be lifted to an element in $H_{d-2}(X_d, L_{d-1};\bZ)$.
Let us now consider a torsional class  $a_2 \in H^2(L_{d-1};\bZ)$, satisfying
$n  a_2 = 0$ for some positive integer $n$.
We know that there exists an element
$\kappa \in H_{d-2}(X_d, L_{d-1};\bZ)$ such that $f(\kappa) = a_2$.
Since $f$ is a homomorphism, $0 = n a_2 = nf(\kappa) = f(n \kappa)$,
i.e.~$n \kappa \in \ker f$. Exactness of the sequence \eqref{eq:exact_seq1}
implies that there exists an element $Z\in H_{d-2}(X_d;\bZ)$ such that
$A (Z) = n  \kappa$.

The manifolds $L_{d-1}$ that we want to study are the link of a
canonical Calabi-Yau singularity $X^{\text{singular}}_d$, so in our
case there is a very natural family of choices of $X_d$: we can take
any crepant resolution of $X^{\text{singular}}_d$. Relatedly, we refer
to elements of $H_p(X_d, L_{d-1};\bZ)$ as non-compact $p$-cycles in $X_d$,
and to elements of $H_p(X_d;\bZ)$ as compact $p$-cycles in $X_d$.  The
observations made so far can be summarised as follows:
\begin{itemize}
\item To every class $a_2 \in H^2(L_{d-1};\bZ)$ we can associate a 
non-compact 
$(d-2)$-cycle $\kappa$ in $X_d$.
\item To every torsional class $a_2 \in H^2(L_{d-1};\bZ)$ we can associate a
compact
$(d-2)$-cycle $Z$ in $X_d$ via the following relations,
\beq
n a_2 = 0 \ , \qquad a_2 = f(\kappa) \ , \qquad A (Z) = n \kappa \ ,
\eeq
where $\kappa$ is a  non-compact $(d-2)$-cycle in $X_d$.
\end{itemize}

\begin{figure}\centering
\includegraphics[width = 10cm]{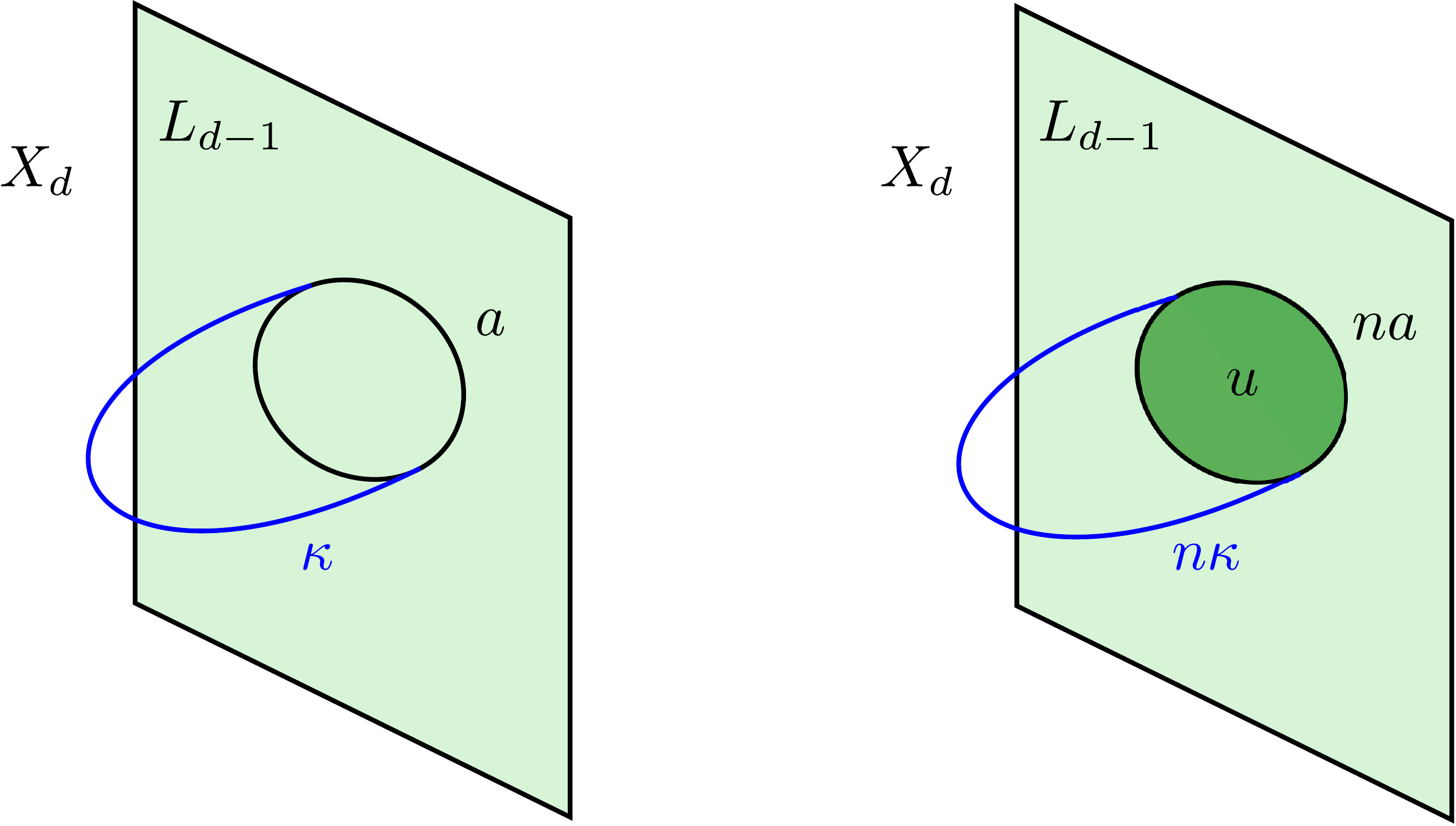}
\caption{On the left: 
Under the assumption $H_{d-3}(X_d;\bZ) =0$,
any $(d-3)$-cycle $a \in Z_{d-3}(L_{d-1})$ in the link  
can be realised as boundary of a $(d-2)$-chain $\kappa \in C_{d-2}(X_d)$ in the bulk $X_d$.
On the right: If $a$ represents a torsional homology class, $na = \partial u$ for some
$(d-2)$-chain $u\in C_{d-2}(L_{d-1})$ in the link, which can also be naturally
  regarded as an element in $C_{d-2}(X_d)$.
Combining the chains $u$ and $n\kappa$ we get a cycle,
$\partial(n\kappa - u) = 0$. This cycle can now be smoothly
retracted to the interior of $X_d$, and can therefore be thought of as a compact cycle.
Its homology class $[n\kappa -u]$   represents 
$Z \in H_{d-2}(X_d;\bZ)$.
\label{fig:gordon}}
\end{figure}

Let us now analyse the map $A$ in \eqref{eq:exact_seq1} in greater detail.
By Lefschetz duality,
\beq
H_{d-2}(X_d, L_{d-1};\bZ) \cong H^2(X_d;\bZ) \ .
\eeq
To proceed, we make the further assumption (that holds in all the cases in this paper)
\beq \label{eq:assumption2}
{\rm Tor} \, H_1(X_d;\bZ) = 0 \ .
\eeq
The universal coefficient theorem then guarantees that
\beq
H^2(X_d;\bZ) \cong {\rm Hom} (H_{2}(X_d;\bZ) , \mathbb Z) \ .
\eeq
We may then recast \eqref{eq:exact_seq1} in the form
\beq \label{eq:exact_seq2}
H_{d-2}(X_d;\bZ) \xrightarrow{A} {\rm Hom} (H_{2}(X_d;\bZ) , \mathbb Z) \xrightarrow{f} H^2(L_{d-1};\bZ) \rightarrow 0 \ .
\eeq
The homomorphism $A$ can be equivalently regarded as a bilinear $\mathbb Z$-valued
pairing between $H_{2}(X_d;\bZ)$ and $H_{d-2}(X_d;\bZ)$,
\beq
A : H_{d-2}(X_d;\bZ) \otimes H_2(X_d;\bZ) \rightarrow \mathbb Z \ .
\eeq
Indeed, $A$ is identified with the intersection pairing of
compact $(d-2)$-cycles  and compact 2-cycles in $X_d$.
Once we choose a basis for $H_{2}(X_d;\bZ)$ and $H_{d-2}(X_d;\bZ)$,
the map $A$ is represented by the intersection matrix $\cM_{d-2,2}$.

\subsubsection{Evaluation for $S^3/\Gamma$} \label{GL_formalism}

To compute the SymTFT for the 7d theory (\ref{7dSym}) we need to evaluate the CS invariant. 
Let us now specialise to a 3d link $L_3$, and fix a class $t_2 \in {\rm Tor}  \,H^2(L_3;\bZ)$
such that $n t_2 =0$.
Let $Z \in H_2(X_4;\bZ)$ be the compact 2-cycle in $X_4$ associated to $t_2$.
By the discussion above, the linking pairing of (the Poincar\'e dual to) $t_2$ with itself
can be computed as
\be
\int_{L_3} \diff t_2 \bd \diff t_2 = \mathsf L_{L_3} ( {\rm PD}[t_2]  ,{\rm PD}[t_2]) 
= \bigg[ \frac{Z \cdot Z}{n^2} \bigg]_\text{mod 1}
\ .
\ee
In the above expression, $\cdot$ denotes the intersection pairing among
compact 2-cycles in $X_4$.
Our task is actually to compute a CS invariant  of the form \eqref{eq:g4in7d}.
In the Gordon-Litherland approach, this quantity  is given by
\begin{equation}\label{eq:g4in7dd}
{\rm CS}[L_3 , \diff t_2] =   \frac 12  \int_{L_3} \diff t_2\bd \diff t_2 
=  \bigg[ \frac{Z \cdot Z}{2\,n^2} \bigg]_\text{mod 1} \, .
\end{equation}

\begin{table}[t]
    \centering
    \def\arraystretch{2}
    \begin{tabular}{|c|c|c|}
    \hline
    $\Gamma$ & Dynkin diagram & $Z$ \\ \hline\hline

    $A_{n-1}$ &
    
    \newcommand{\ctikz}[1]{$\vcenter{\hbox{#1}}$}
    \ctikz{%
    \begin{tikzpicture}[scale=1]
    \draw[fill=black]
    (0,0) node (v1) {} circle [radius=.1]
    (1,0) node (v3) {} circle [radius=.1]
    (2,0) node (v4) {} circle [radius=.1]
    (4,0) node (v5) {} circle [radius=.1];
    \node (v2) at (3,0) {$\cdots$};
    \draw  (v1) edge (v3);
    \draw  (v3) edge (v4);
    \draw  (v4) edge (v2);
    \draw  (v2) edge (v5);
    \node at (0,-0.4) {{$1$}};
    \node at (1,-0.4) {{$2$}};
    \node at (2,-0.4) {{$3$}};
    \node at (4,-0.4) {{$n-1$}};
    \end{tikzpicture}}

    &  $\sum_{i=1}^{n-1} i S_i$ \\ \hline
    
    $\mathrm{D}_{2n}$ &
    
    \newcommand{\ctikz}[1]{$\vcenter{\hbox{#1}}$}
    \ctikz{%
    \begin{tikzpicture}[scale=1]
    \draw[fill=black]
    (-0.4,0) node (v1) {} circle [radius=.1]
    (0.8,0) node (v3) {} circle [radius=.1]
    (2,0) node (v4) {} circle [radius=.1]
    (4,0) node (v5) {} circle [radius=.1]
    (5,0) node (v6) {} circle [radius=.1]
    (0.8,1.2) node (v7) {} circle [radius=.1];
    \node (v2) at (3,0) {$\cdots$};
    \draw  (v1) edge (v3);
    \draw  (v3) edge (v4);
    \draw  (v4) edge (v2);
    \draw  (v2) edge (v5);
    \node at (-0.4,-0.4) {{$2n+1$}};
    \node at (0.8,-0.4) {{$2n-1$}};
    \node at (2,-0.4) {{$2n-2$}};
    \node at (4,-0.4) {{$2$}};
    \node at (5,-0.4) {{$1$}};
    \draw  (v5) edge (v6);
    \draw  (v7) edge (v3);
    \node at (0.8,1.6) {{$2n$}};
    \end{tikzpicture}}

    & $\sum_{i=1}^{2n-1}(1-(-1)^i)S_i$\\ \hline
    $\mathrm{D}_{2n+1}$ &
    
    \newcommand{\ctikz}[1]{$\vcenter{\hbox{#1}}$}
    \ctikz{%
    \begin{tikzpicture}[scale=1]
    \draw[fill=black]
    (-0.4,0) node (v1) {} circle [radius=.1]
    (0.8,0) node (v3) {} circle [radius=.1]
    (2,0) node (v4) {} circle [radius=.1]
    (4,0) node (v5) {} circle [radius=.1]
    (5,0) node (v6) {} circle [radius=.1]
    (0.8,1.2) node (v7) {} circle [radius=.1];
    \node (v2) at (3,0) {$\cdots$};
    \draw  (v1) edge (v3);
    \draw  (v3) edge (v4);
    \draw  (v4) edge (v2);
    \draw  (v2) edge (v5);
    \node at (-0.4,-0.4) {{$2n-1$}};
    \node at (0.8,-0.4) {{$2n-2$}};
    \node at (2,-0.4) {{$2n-3$}};
    \node at (4,-0.4) {{$2$}};
    \node at (5,-0.4) {{$1$}};
    \draw  (v5) edge (v6);
    \draw  (v7) edge (v3);
    \node at (0.8,1.6) {{$2n$}};
    \end{tikzpicture}}
    
    &  
    $\begin{matrix}
    {1\over 2} \sum_{i=1}^{2n-1}(1-(-1)^i) S_i\, ,\\ 
    {1\over 2} \sum_{i=1}^{2n-2}\left({1-(-1)^i}\right)S_i+S_{2n}
    \end{matrix}$
    \\ \hline
    
    $E_6$ & 
    
    \newcommand{\ctikz}[1]{$\vcenter{\hbox{#1}}$}
    \ctikz{%
    \begin{tikzpicture}[scale=1]
    \draw[fill=black]
    (1,0) node (v1) {} circle [radius=.1]
    (2,0) node (v3) {} circle [radius=.1]
    (3,0) node (v4) {} circle [radius=.1]
    (4,0) node (v5) {} circle [radius=.1]
    (5,0) node (v6) {} circle [radius=.1]
    (3,1) node (v7) {} circle [radius=.1];
    \node at (1,-0.4) {{$5$}};
    \node at (2,-0.4) {{$4$}};
    \node at (3,-0.4) {{$3$}};
    \node at (4,-0.4) {{$2$}};
    \node at (5,-0.4) {{$1$}};
    \draw  (v5) edge (v6);
    \node at (3,1.3) {{$6$}};
    \draw  (v4) edge (v5);
    \draw  (v3) edge (v4);
    \draw  (v1) edge (v3);
    \draw  (v7) edge (v4);
    \end{tikzpicture}}

    & $\sum_{i=1}^5 iS_i+S_{2n}+3S_{2n+1}$ 
    \\ \hline
    
    $E_7$ & 
    
    \newcommand{\ctikz}[1]{$\vcenter{\hbox{#1}}$}
    \ctikz{%
    \begin{tikzpicture}[scale=1]
    \draw[fill=black]
    (1,0) node (v1) {} circle [radius=.1]
    (2,0) node (v3) {} circle [radius=.1]
    (3,0) node (v4) {} circle [radius=.1]
    (4,0) node (v5) {} circle [radius=.1]
    (5,0) node (v6) {} circle [radius=.1]
    (6,0) node (v8) {} circle [radius=.1]
    (3,1) node (v7) {} circle [radius=.1];
    \node at (1,-0.4) {{$6$}};
    \node at (2,-0.4) {{$5$}};
    \node at (3,-0.4) {{$4$}};
    \node at (4,-0.4) {{$3$}};
    \node at (5,-0.4) {{$2$}};
    \node at (6,-0.4) {{$1$}};
    \draw  (v5) edge (v6);
    \node at (3,1.3) {{$7$}};
    \draw  (v4) edge (v5);
    \draw  (v3) edge (v4);
    \draw  (v1) edge (v3);
    \draw  (v7) edge (v4);
    \draw  (v6) edge (v8);
    \end{tikzpicture}}

    & $S_1+S_3+S_7$ 
    \\ \hline
    
    

    
    \end{tabular}
    \caption{The center divisors $Z$ written in terms of the compact curves associated to the simple roots for ADE-singularities. We omit the $E_8$ case as it has trivial center symmetry.}
    \label{tab:divisors}
\end{table}

In particular, we apply this formalism to the case of interest
$L_3 = S^3/\Gamma$.  The bulk $X_4$ can be chosen to be the resolved
ALE space $\mathbb C^2/\Gamma$.  We notice that the assumptions
\eqref{eq:assumption1} and \eqref{eq:assumption2} are indeed
satisfied.  The intersection matrix $\cM_{2,2}$ representing the map
$A$ equals minus the Cartan matrix of the Lie algebra
$\mathfrak g_\Gamma$. For the ADE-singularities, the choice of central
divisors which gives the center of the gauge group has been identified
in \cite{Bhardwaj:2020phs}. We list their results in table
\ref{tab:divisors}. Using these compact divisors as representatives of
the torsional generators in order to be able to compare easily with
known field theory
results,\footnote{\label{fn:torsional-generator-ambiguity}Note that
  there is a genuine ambiguity here: if $t$ is a generator of $\bZ_n$,
  so is $kt$ for any $k$ such that $\gcd(k,n)=1$. We have
  $CS[S^3/\Gamma,k\diff t_2]=k^2CS[S^3/\Gamma,\diff t_2]$ mod 1, so
  this rescaling can potentially change the coefficient of the
  anomaly. This is why it is important to choose the right generator
  of the torsional group when comparing with field theory results,
  even though any choice of $k$ is in principle physically allowed.}
we obtain the results given in \ref{table:7d-CS} for the spin
Chern-Simons invariant ${\rm CS}[S^3/\Gamma , \diff t_2]$. It is a
nice check of our formalism that the resulting coefficients in the
anomaly theory perfectly reproduce the answer one gets from a purely
field theory analysis \cite{Cordova:2019uob}. (The $A_{n-1}$ answer
was also recently obtained in \cite{Cvetic:2021sxm} from a related
viewpoint.)

\begin{table}
  \def\arraystretch{1.6}
  \centering
  \begin{tabular}{|c|c|c|c|}\hline
    $\Gamma$ & $G_\Gamma$ & $\Gamma^{\text{ab}}$ & $-{\rm CS}[S^3/\Gamma , \diff t_2] $ \\
    \hline
    \hline
    $A_{n-1}$ & $SU(n)$ & $\bZ_n$ 
    & $\frac{n-1}{2n}$ \\ 
    $\mathrm{D}_{2n}$ & $\Spin(4n)$ & $\mathbb{Z}_2\oplus \mathbb{Z}_2$ & ${1\over 4}\left(\begin{matrix} n & n-1 \\ n-1  & n \end{matrix}\right) $\\ 
    $ \mathrm{D}_{2n+1}$ &   $\Spin(4n+2)$ &   $\bZ_4$
    & $ \frac{2n+1}{8} $ \\ 
    $2T$ & $E_6$ & $\bZ_3$ & 
    $\frac{5}{3}$ \\ 
    $2O$ & $E_7$ & $\bZ_2$ & 
    $\frac{3}{4}$ \\ 
    $2I$ & $E_8$ & $0$ & $0$\\\hline
  \end{tabular}  
  \caption{$\Gamma$ of ADE type. $G_\Gamma$ denotes the simply-connected gauge group in 7d, $\Gamma^{ab}$ the abelianization of $\Gamma$. Finally ${\rm CS}[S^3/\Gamma , \diff t_2] $ is the Chern-Simons invariant (\ref{eq:g4in7d}), which is closely related to the linking pairing, as explained in the main text. With these expressions one can then evaluate the SymTFT for 7d SYM in (\ref{7dSym}).}
  \label{table:7d-CS}
\end{table}

\paragraph{Relation to the Linking Pairing.}

Finally, we want to comment on the relation between the Chern-Simons
invariant $\text{CS}[S^3/\Gamma, \diff t_2]$ and the linking pairing
$\sL(t_2, t_2)$. The relation is that
$\text{CS}[S^3/\Gamma, \diff t_2]$ provides a quadratic refinement
\cite{Witten:1996md, Hopkins:2002rd} of the linking pairing:
\begin{equation}
  \sL_{S^3/\Gamma} ( {\rm PD}[s_2] ,{\rm PD}[t_2]) = \int_{S^3/\Gamma} \diff s_2 \bd \diff
  t_2 = \text{CS}[S^3/\Gamma, \diff s_2+\diff t_2] - \text{CS}[S^3/\Gamma, \diff s_2] - \text{CS}[S^3/\Gamma, \diff t_2] \mod 1\, ,
\end{equation}
where $\diff s_2,\diff t_2\in \diff H^2(S^3/\Gamma)$ are chosen to be
flat: $R(\diff s_2)=R(\diff t_2)=0$. The equality on the left can be
proven as follows.

Since $\diff s_2$ and $\diff t_2$ are flat, we can use the
commutativity of (\ref{eq:SS-}), and (\ref{eq:flat_star}) to write
\begin{equation}\label{eq:linkingp7d}
  \diff s_2\bd \diff t_2 = -\diff s_2 \bd i(\beta^{-1}(I(\diff t_2))) =
  - i\left(I(\diff s_2) \smile \beta^{-1}(I(\diff t_2))\right) \, ,
\end{equation}
where we have also used that
$H^1(S^3/\Gamma; \bR)=H^2(S^3/\Gamma;\bR)=0$, so the Bockstein map
$\beta$ in~\eqref{eq:SS-} is an isomorphism. For the integral of
\eqref{eq:linkingp7d} we then have
\begin{equation}\label{eq:linkp}
    \int_{S^3/\Gamma} i\left(I(\diff s_2) \smile \beta^{-1}(I(\diff t_2))\right)
    = i\left( \int_{S^3/\Gamma} I(\diff s_2) \smile \beta^{-1}(I(\diff t_2))\right)
    = \int_{S^3/\Gamma} I(\diff s_2) \smile \beta^{-1}(I(\diff t_2)) \,,
\end{equation}
since $i: H^0(\text{pt}; U(1)) \to \diff{H}^1(\text{pt})$ is an
isomorphism. The final expression in (\ref{eq:linkp}) is just the
linking pairing $\sL$ on $S^3/\Gamma$ \cite{conway2018linking} (up to
a sign convention)
\begin{equation}\label{eq:Lmod2}
  \int_{S^3/\Gamma} \diff s_2\bd \diff t_2
  = -\int_{S^3/\Gamma} I(\diff s_2) \smile \beta^{-1}(I(\diff t_2))
  = \sL_{S^3/\Gamma}({\rm PD}[s_2] , {\rm PD}[t_2]) \mod 1\, .
\end{equation}

This refinement of the linking pairing extends, in particular, the
observation in \cite{GarciaEtxebarria:2019caf} that the fractional
instanton number for an instanton bundle in the presence of background
1-form flux is half of the linking pairing in $S^3/\Gamma$ for the
torsional class $t_2$ representing the 1-form flux background. The
more refined statement that follows from our M-theory construction is
instead:
\begin{equation}
  \label{eq:n-inst-refined}
  n_{\text{inst}} = -\text{CS}[S^3/\Gamma, \diff t_2] \mod 1\, .
\end{equation}
The discussion in \cite{GarciaEtxebarria:2019caf} was specific to four
dimensional theories on Spin manifolds, and the two statements agree
on that class of manifolds (up to an overall sign that was chosen
oppositely in \cite{GarciaEtxebarria:2019caf}),
but~\eqref{eq:n-inst-refined} gives the correct answer on non-Spin
manifolds too.


\section{Symmetry TFTs for 5d SCFTs}
\label{sec:5d}

Placing M-theory on toric Calabi-Yau threefold singularities leads to
a rich and interesting class of five dimensional SCFTs. This is a very
active area of investigation, started by
\cite{Seiberg:1996bd,Morrison:1996xf,Intriligator:1997pq,Aharony:1997ju,Leung:1997tw}. These
theories have an intricate set of global symmetries, starting with
0-form flavor symmetries, which are enhanced at the UV fixed point, as
well as discrete higher-form symmetries -- both 1-form (or 2-form)
symmetries \cite{Morrison:2020ool, Albertini:2020mdx,
  Bhardwaj:2020phs} and 3-form symmetries \cite{Closset:2020scj,
  Closset:2020afy}. These symmetries can have 't Hooft anomalies, and
by gauging some of the symmetries one obtains theories with 2-group
structure \cite{Tachikawa:2017gyf,Apruzzi:2021vcu}.

In this section we will determine a subset of the six dimensional
symmetry TFT by reducing M-theory on the five-dimensional
Sasaki-Einstein manifold $L_5$ linking the singular point. In
section~\ref{sec:5branes} below we will present an alternative
derivation based on reducing the Chern-Simons terms on the worldvolume
of the dual system on $(p,q)$ 5-branes down to six dimensions. One
important omission from our analysis is the sector leading to 2-group
structures, which we leave for future work.

\subsection{5d SCFTs from M-theory and Higher Form Symmetries}

We will start by giving a summary of the salient features of 5d SCFT
engineering from M-theory that will play a role in this paper. We are
interested in the dynamics of M-theory on a singular toric Calabi-Yau
threefold $X$. The global 0-form flavor symmetries are encoded in the
non-compact divisors $\{D_i\}$ of $X$ and their intersections with the
compact divisors $S_a$, $a= 1, \cdots, r$, which furnish the Cartan
generators of the gauge group on the Coulomb branch (CB), of rank
$r$. The general flavor symmetry on the CB contains an instanton
$U(1)_I$, whose current is $j_I= * \Tr(F\wedge F)$, and which often in
the UV enhances the flavor symmetry compared to the gauge theory
description in the IR. Many geometric methods of computing the UV
flavor symmetry from the non-compact divisors have been developed 
\cite{Apruzzi:2018nre, Apruzzi:2019vpe, Apruzzi:2019opn,Apruzzi:2019enx,
  Apruzzi:2019kgb, Bhardwaj:2020ruf, Bhardwaj:2020avz, Tian:2021cif}, which enable
computing also the global form of the flavor symmetry groups
\cite{Apruzzi:2021vcu}.

The higher form symmetries arise from the homology groups of
$L_5\df \partial X$. The 2-form symmetry $\Gamma^{(2)}$ under which
the 't Hooft surface operators are charged, is determined by the group
\cite{GarciaEtxebarria:2019caf,Morrison:2020ool, Albertini:2020mdx}
\be
  \Gamma^{(2)} = {H_2(X, L_5; \mathbb{Z} ) \over H_2 (X; \mathbb{Z})}
  \cong H_1 (L_5; \mathbb{Z}) \,,
\ee
where the last isomorphism is true if all 1-cycles in $L_5$ trivialise
in the bulk (this will be true for $X$ simply because
$H_1(X;\bZ)=0$). In the cases of interest to us this group will be of
the form $\bZ_k$ for some $k$ that depends on $X$.

As noted in \cite{GarciaEtxebarria:2019caf,Morrison:2020ool,
  Albertini:2020mdx}, this group can be computed in any smooth
resolution of $X$ using the intersection of compact divisors and
compact curves in the Calabi-Yau by
\be
  \Gamma^{(1)} \cong \mathbb{Z}^{b_4}/\mathcal{M}_{4,2}
  \mathbb{Z}^{b_2} \,,
\ee
where the Betti numbers are related to the rank of the CB gauge group
$r$ and flavor rank $f$ by $b_2 =r+f$ and $b_4=r$.
$\mathcal{M}_{4,2} = (S_i \cdot_X C_k)$ is the intersection matrix of
compact divisors $S_i$ with compact curves $C_k$. Alternatively, it
can be computed without needing to resolve $X$ by looking to the
structure of the external points in the toric diagram
\cite{Garcia-Etxebarria:2016bpb,Albertini:2020mdx}.

Similarly, as explained in section~\ref{sec:KK}, the torsional part of
$H_3 (L_5;\mathbb{Z})$ will lead to a finite abelian 1-form symmetry
under which  line operators are charged, while the free projection of
$H_3 (L_5;\mathbb{Z})$ leads to a continuous 0-form symmetry, which includes the
 instanton symmetry or $U(1)_I$.

These higher form symmetries are not all realised simultaneously in a
given field theory. As in the seven dimensional case studied above, we
will obtain a generically non-invertible BF sector in the symmetry
theory when reducing on $L_5$, and different choices of boundary
conditions for the symmetry theory will determine which higher form
symmetries are actually realised. The derivation of the BF sector is
very similar to the one in that case, so we will be brief. Consider
the operators $\Phi(\cT_3)$ and $\Phi(\cT_6)$ wrapped on generators
$t_1$ and $t_3$ of $H_1(L_5;\bZ)$ and $\Tor H_3(L_5;\bZ)$. They will
lead to operators $\Psi(\Sigma_2)$ and $\Psi(\Sigma_3)$ in the
effective six dimensional symmetry theory, with a commutation relation
\be
  \Psi(\Sigma_2)\Psi(\Sigma_3) = e^{2\pi i \ell^{-1} \Sigma_2\cdot\Sigma_3} \Psi(\Sigma_3)\Psi(\Sigma_2)
\ee
on a spatial slice, where $\ell^{-1}\df \sL_{L_5}(t_1, t_3)$. This is
the content of a BF theory with action\footnote{We are abusing
  notation slightly here, and denoting by $B_2$ both the continuous
  field that we use in writing the BF action and the discrete
  background for the 1-form symmetry, since they are identified at low
  energies.}
\be
  S_{\text{BF}} = \ell \int B_2\wedge dC_3\, .
\ee

The 1-form symmetries also participate in 't Hooft anomalies. Denote
the background fields for the 1-form symmetry by
$B_2 \in H^2 (M_5; \Gamma^{(1)})$.  From general field theory
considerations, obtained by studying the Coulomb branch, there are two
types of anomalies: the purely 1-form symmetry cubic anomaly ($B^3 $)
\cite{Gukov:2020btk}, and the mixed $U(1)_I$ and 1-form symmetry
anomaly ($B^2 F_I$) \cite{BenettiGenolini:2020doj}.  The cubic 1-form
symmetry anomaly was derived from field theory in the context of the
SCFTs that have a Coulomb branch description as $SU(p)_q$
\cite{Gukov:2020btk}
\be
  \label{Bcubed}
\mathcal{A}_{B^3} = {q p (p-1) (p-2) \over 6 \gcd (p,q)^3} B_2^3 \,.
\ee
We are using conventions where the periods of $B_2$ are integrally
quantised (as opposed to $2\pi/\gcd(p,q)$ quantised) and the 1-form
symmetry in this case is $\Gamma^{(1)} = \mathbb{Z}_{\gcd(p,q)}$. This coupling corresponds to a 't Hooft anomaly for the 1-form symmetry, and therefore field theoretically obstructs its gauging. This will imply that (potentially) some asymptotic flux choice might be obstructed, and not all the global form of the gauge group are allowed, unless a more complicated structure arises, which mixes the 1-form symmetry with other symmetries present in the theory. We plan to explore deeper consequences of this coupling by using our methods in the future.

There is also a field theoretic mixed anomaly between the instanton
$U(1)_I$ and 1-form symmetry, determined in
\cite{BenettiGenolini:2020doj} for the $SU(2)_0$ theory using field
theory arguments. In the IR for $SU(p)_q$ it takes the form \cite{Gukov:2020btk, BenettiGenolini:2020doj}
\be
  \label{PL}
  \mathcal{A}_{FB^2} = {p (p-1)\over 2 \gcd(p,q)^2} F_I B_2^2  \, .
\ee
(This contribution to the anomaly was also analysed in
\cite{Cvetic:2021sxm} using string theory methods, reaching a
different conclusion. We believe that the discrepancy between their
result and ours might be due to a different choice of torsional
representative, see footnote~\ref{fn:torsional-generator-ambiguity}
above.)

We will now derive these anomalies from first principles using the
differential cohomology approach developed in this paper, being
agnostic about whether this is a UV or IR computation. We will see
that an essential contribution for these anomalies comes from the
$C_3 \wedge X_8$ term in the M-theory effective action.

\subsection{Link Reduction Using Differential Cohomology}

The integral cohomology of $L_5$, the base of the toric Calabi-Yau
cone $X$, takes the form
\be
  H^*(L_5;\mathbb Z) = \big\{ \mathbb Z ,
0 , \mathbb Z^{b^2} \oplus {\rm Tor}\, H^2(L_5;\mathbb Z) , \mathbb
Z^{b^2} \oplus {\rm Tor}\, H^3(L_5;\mathbb Z) , {\rm Tor}\,
H^2(L_5;\mathbb Z) , \mathbb Z \big\} \ .
\ee
For simplicity, we assume that the Betti number $b^1$ of $L_5$ is
zero. (This is true in all examples we study.)  The expansion of
$\diff G_4$ then reads
\begin{align} \label{G4_ansatz_L5}
    \diff G_4 &= \diff \gamma_4 \bd \diff 1
    +  \sum_{\alpha =1}^{b^2} \diff F_2^{(\alpha)} \bd \diff v_{2(\alpha)}
        +  \sum_{\alpha = 1}^{b^2} \diff \xi_{1(\alpha)} \bd \diff v_3 ^{(\alpha)}
        \nn \\
&    + \sum_i \diff B_2^{(i)} \bd \diff t_{2(i)} 
    + \sum_m \diff b_1^{(m)} \bd \diff t_{3(m)}
    + \sum_i \diff \psi_{0(i)} \bd \diff t_4^{(i)}  +  
    \tau([\omega_3])\, .
\end{align}
The label $\alpha$ runs over generators of the free part of $H^2(L_5;\mathbb Z)$,
the label $i$ runs over generators of ${\rm  Tor} \, H^2(L_5 ;\mathbb Z_5)$,
while the label $m$ runs over generators of ${\rm Tor} \, H^3(L_5;\mathbb Z)$.

We can now consider the reduction of the $G_4^3$ coupling in M-theory.
Using \eqref{G4_ansatz_L5} and collecting all relevant terms, we arrive at
\be
\ba
\label{eq:expand_Gcube}
& - \frac 16 \, \int_{\cM_{11}} \diff G_4 \bd \diff G_4 \bd \diff G_4 \cr 
& = - \sum_\alpha \int_{\cW_6} \diff \gamma_4 \bd \diff F_2^{(\alpha)} \bd \xi_{1(\alpha)}
-  \sum_{i,j,k}\bigg[ \frac 16 \, \int_{L_5} \diff t_{2(i)} \bd  \diff t_{2(j)} \bd  \diff t_{2(k)} \bigg]
\, \int_{\cW_6} \diff B_2^{(i)} \bd \diff B_2^{(j)} \bd \diff B_2^{(k)}
 \\
& -  \sum_{i,j,\alpha}  \bigg[\frac 12 \,  \int_{L_5} \diff t_{2(i)} \bd  \diff t_{2(j)} \bd  \diff v_{2(\alpha)} \bigg]
\, \int_{\cW_6} \diff B_2^{(i)} \bd \diff B_2^{(j)} \bd \diff F_2^{(\alpha)}\cr 
& -   \sum_{i,\alpha,\beta} \bigg[ \frac 12 \, \int_{L_5}  \diff t_{2(i)} \bd  \diff v_{2(\alpha)} \bd  \diff v_{2(\beta)} \bigg]
\, \int_{\cW_6} \diff B_2^{(i)} \bd \diff F_2^{(\alpha)} \bd \diff F_2^{(\beta)}
 \\
& +  \sum_{m,n}\bigg[ \frac 12 \, \int_{L_5} \diff t_{3(m)} \bd \diff t_{3(n)} \bigg] \, \int_{\cW_6} 
\diff \gamma_4 \bd  \diff b_1^{(m)} \bd  \diff b_1^{(n)}
+ \sum_{m,\alpha} \bigg[ \int_{L_5} \diff t_{3(m)} \bd  \diff v_3^{(\alpha)}  \bigg] \, \int_{\cW_6} \diff \gamma_4
\bd \diff b_1^{(m)} \bd \diff \xi_{1(\alpha)}
 \\
& - \sum_{i,j} \bigg[ \int_{L_5} \diff t_{2(i)} \bd \diff t_4^{(j)} \bigg] \,
\int_{\cW_6} \diff   \gamma_4 \bd \diff B_2^{(i)} \bd \diff \psi_{0(j)}
- \sum_{\alpha,j}\bigg[ \int_{L_5} \diff v_{2(\alpha)} \bd \diff t_4^{(j)} \bigg] \,
\int_{\cW_6} \diff \gamma_4 \bd \diff F_2^{(\alpha)} \bd \diff \psi_{0(j)} \ .
\ea
\ee
In the first term, we have used $\int_{L_5} v_{2(\alpha)} \bd v_3^{(\beta)} = \delta^\beta_\alpha$.

Next, let us consider the terms that originate from the $G_4 X_8$
coupling in M-theory.  Recall that 11d spacetime is taken to be the
direct product $\cM_{11} = \cW_6 \times L_5$.  As a result, at the
level of cohomology classes with integer coefficients, one
has\footnote{The Whitney sum formula for Pontryagin classes in integral cohomology is
\cite{brown1982cohomology}
\beq
p_q(A \oplus B) = \sum r_{2q-j}(A) \smile r_j(B) \ , \quad
r_{2s} = p_s \ , \quad
r_{2s+1} = {\rm Bock}(w_{2s}) \smile  {\rm Bock}(w_{2s})  + p_s \smile {\rm Bock}(w_1) \ ,
\eeq
where $A$, $B$ are $O$ bundles, the $w_i$'s are 
Stiefel-Whitney classes,
and Bock
is the Bockstein homomorphism associated to the short exact
sequence $0 \rightarrow \mathbb Z \xrightarrow{2} \mathbb Z \rightarrow \mathbb Z_2 \rightarrow 0$. In appendix~\ref{app:DC} we show $w_1(TL_5)=w_2(TL_5)=0$, which implies that \eqref{eq:pontryagin} holds at the level of cohomology with integral coefficients.
}
\be\ba \label{eq:pontryagin}
p_1(T\cM_{11}) & = p_1(T\cW_6) + p_1(TL_5)  \cr 
p_2(T\cM_{11}) &=  p_2(T\cW_6) + p_2(TL_5) +  p_1(T\cW_6) \smile p_1(TL_5) \,.
\ea
\ee
These relations imply
\beq
X_8 = - \frac {1}{96} \, p_1(T\cW_6) \smile p_1(TL_5) \ .
\eeq
Promoting integral cohomology classes to differential cohomology classes (the precise representative of $p_1$ one chooses is not important \cite{Freed:2019sco})
we can write the $G_4 X_8$ coupling in the form
\be \label{eq:G4X8_on_L5}
- \int_{\cM_{11}} \diff G_4 \bd \diff X_8 = \frac{1}{96} \, \int_{\cM_{11}} \diff G_4 \bd \diff p_1(T\cW_6)
\bd \diff p_1(TL_5)
= \frac{1}{96} \, \sum_i  \int_{L_5} \diff t_{2(i)} \bd \diff p_1(TL_5)   \,
\int_{\cW_6} \diff B_2^{(i)} \bd \diff p_1(T\cW_6) \ .
\ee
In the second step we used \eqref{G4_ansatz_L5} and we 
observed that the only internal differential cohomology classes that can have a non-trivial
pairing with $\diff p_1(TL_5)$ are the degree-2 torsional classes $\diff t_{2(i)}$.

To proceed, we make use of the following congruence for integral
cohomology classes \cite{wall}
\beq \label{eq:congruence}
p_1(T\cW_6) \smile a_2 = 4 \, a_2 \smile a_2 \smile a_2 \mod 24 \qquad \text{for any 
$a_2 \in H^2(\cW_6 ; \mathbb Z)$} \ .
\eeq
This congruence
can be derived using the Atiyah-Singer index theorem as follows \cite{Witten:1996qb}.
We take external spacetime $\cW_6$ to be a Spin manifold.
Consider an arbitrary
$a_2 \in H^2(\cW_6 ; \mathbb Z)$. There exists a line bundle with connection $A$
on $\cW_6$ such that its first Chern class equals $a_2$. Consider the Dirac
operator on $\cW_6$ twisted by this line bundle. The Atiyah-Singer theorem implies
\beq
{\rm Ind}(\slashed D_A) = \int_{\cW_6} \bigg[ \frac 16 \, F \wedge F \wedge F - \frac{1}{24} \, F \wedge p_1(T\cW_6) \bigg] 
=  \int_{\cW_6} \bigg[ \frac 16 \, a_2 \smile a_2 \smile a_2 - \frac{1}{24} \, a_2 \smile p_1(T\cW_6) \bigg] \ ,
\eeq 
where $F$ is the curvature 2-form of the connection $A$, satisfying
$[F]_{\rm dR} = \varrho(a_2)$.
We conclude that
$\int_{\cW_6} [ 4\,a_2 \smile a_2 \smile a_2 - a_2 \smile p_1(T\cW_6) ] \in 24\,\mathbb Z$,
which is equivalent to \eqref{eq:congruence}.

Relation~\eqref{eq:congruence} then implies
\be\ba
  \label{eq:congruenceBIS}
\int_{\cW_6} \diff B_2^{(i)} \bd \diff p_1(T\cW_6) &= \int_{\cW_6}   B_2^{(i)} \smile   p_1(T\cW_6) 
= 24\, M^{(i)} + 4\,  \int_{\cW_6} B_2^{(i)} \smile B_2^{(i)} \smile B_2^{(i)} \\
& = 24 \, M^{(i)} + 4 \, \int_{\cW_6} \diff B_2^{(i)} \bd  \diff B_2^{(i)} \bd  \diff B_2^{(i)}  \ ,
\ea\ee
where we have used the fact that $\int_{\cW_6} \diff a_6 = \int_{\cW_6} I(\diff a_6)$
for any $\diff a_6 \in \diff H^6(\cW_6)$, together with
\eqref{eq:star_ids}, $I(\diff B_2^{(i)}) =   B_2^{(i)}$,
and $I( \diff p_1(T\cW_6)) =   \diff p_1(T\cW_6)$. 
The quantities $M^{(i)}$ are unspecified integers, encoding the ambiguity
 in the mod 24 congruence \eqref{eq:congruence}. 
There is no summation on the repeated label $i$ in \eqref{eq:congruenceBIS}. 
Inserting \eqref{eq:congruenceBIS} into \eqref{eq:G4X8_on_L5}, we arrive at
\beq \label{eq:G4X8_result}
- \int_{\cM_{11}} \diff G_4 \bd \diff X_8 = 
\frac{1}{24} \, \sum_i  \int_{L_5} \diff t_{2(i)} \bd \diff p_1(TL_5)   \,
\int_{\cW_6} \diff B_2^{(i)} \bd \diff B_2^{(i)} \bd  \diff B_2^{(i)}  
+ \sum_i M^{(i)} \, \int_{L_5} \diff t_{2(i)} \bd \frac{\diff p_1(TL_5)}{4} \ .
\eeq
In appendix \ref{app:DC} we show that $p_1(TL_5)/4$ is an integral class. Although we have no general proof, we also find that in all of our examples the quantity multiplying $M^{(i)}$ is an integer, so we will drop the last term in what follows, and focus on the terms in the symmetry theory that
contain only the  fields $\diff B_2^{(i)}$ and $\diff F_2^{(\alpha)}$.

Notice that there is an ambiguity in the definition of the 
differential cohomology classes $\diff v_{2(\alpha)}$ associated to the free part
of $H^2(L_5;\mathbb Z)$, which can be shifted
by integral multiples of the differential cohomology classes $\diff t_{2(i)}$
associated to ${\rm Tor}\,H^2(L_5;\mathbb Z)$,  $\diff v_{2(\alpha)} \rightarrow 
\diff v_{2(\alpha)} + m_{(\alpha)}{}^{(i)} \, \diff t_{2(i)}$, with
$m_{(\alpha)}{}^{(i)} \in \mathbb Z$. Our choices are such that for the examples in the paper we have
\beq \label{eq:fix_good_basis}
\int_{L_5} \diff t_{2(i)} \bd \diff v_{2(\alpha)} \bd \diff v_{2(\beta)} = 0 \ .
\eeq
Combining \eqref{eq:expand_Gcube} and \eqref{eq:G4X8_result}, we
obtain the following anomaly couplings in the symmetry TFT:
\be
\boxed{
\Sym =  \sum_{i,j,k} \Omega_{ijk} \,  \int_{\cW_6} \diff B_2^{(i)} \bd \diff B_2^{(j)} \bd \diff B_2^{(k)} 
+ \sum_{i,j,\alpha} \Omega_{ij\alpha} \,  \int_{\cW_6} \diff B_2^{(i)} \bd \diff B_2^{(j)} \bd \diff F_2^{(\alpha)} 
\,,}
\ee
where the $\mathbb R/\mathbb Z$-valued
quantities $\Omega_{ijk}$, $\Omega_{ij\alpha}$ are CS invariants defined
by
\be \label{eq:CS5}
\ba
\Omega_{ijk} &= - \frac 16 \, \int_{L_5} \diff t_{2(i)} \bd  \diff t_{2(j)} \bd  \diff t_{2(k)}
+ \frac{1}{24} \, \delta_{i,j} \, \delta_{i,k} \, \int_{L_5} \diff t_{2(i)} \bd
\diff p_1(TL_5)  \cr 
\Omega_{ij\alpha} &= - \frac 12 \, \int_{L_5}\diff t_{2(i)} \bd  \diff t_{2(j)} \bd  \diff v_{2(\alpha)} \ .
\ea
\ee
As we demonstrate in appendix \ref{app:DC}, for the setups of interest in this work
$G_4$ is integrally quantised, and therefore the 11d couplings in the M-theory effective
action are guaranteed to be well-defined. It follows that the CS invariants \eqref{eq:CS5}
are also well-defined. 
Let us emphasize, however,  that 
the two terms 
in $\Omega_{ijk}$ with $i=j=k$
are not separately well-defined, in general.

The CS invariants $\Omega_{ijk}$, $\Omega_{ij\alpha}$ are defined
purely in terms of the link geometry $L_5$. In order to 
evaluate them for a given $L_5$, however, it can be convenient
to resort to a computation in the bulk of the Calabi-Yau   $X_6$, 
using an extension of the  Gordon-Litherland formalism discussed in section \ref{sec:CS3}.
 Let $n_{(i)}$ denote the torsional degree
of $t_{2(i)} \in H^2(L_5;\bZ)$, and let $Z_{(i)}$ be the compact divisor in the bulk 
associated to $t_{2(i)}$. We also associate a non-compact divisor $D_{(\alpha)}$
to the non-torsional classes $v_{2(\alpha)} \in H^2(L_5;\bZ)$, which correspond to
flavor symmetries.
With this notation, the invariants  \eqref{eq:CS5}  can be computed as
\be \label{eq:Omega_formulae}
\ba
\Omega_{ijk} &= \bigg[ - \frac 16 \,  \frac{Z_{(i)} \cdot Z_{(j)}  \cdot Z_{(k)} }{n_{(i)} \, n_{(j)} \, n_{(k)} }  
+ \frac{1}{24} \, \delta_{i,j} \, \delta_{i,k} \,  \frac{Z_{(i)}  \cdot p_1(TX_6)  }{n_{(i)}} 
 \bigg]_\text{mod 1}  \ , \cr 
\Omega_{ij\alpha} &= \bigg[
 - \frac 12 \,    \frac{Z_{(i)}   \cdot  Z_{(j)}  \cdot  D_{(\alpha)} }{ n_{(i)} \, n_{(j)} }   
 \bigg]_\text{mod 1}
  \ ,
\ea
\ee 
where $\cdot$ denotes intersection of divisors in $X_6$.

\subsection{Examples: $SU(p)_q$ from $Y^{p,q}$}

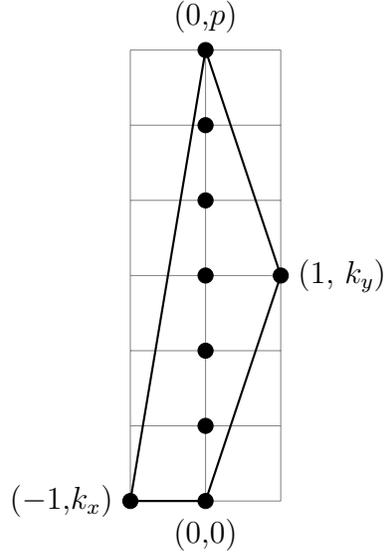
\begin{figure}\centering
\begin{tikzpicture}[x=1cm,y=1cm]
\draw[step=1cm,gray,very thin] (-1,0) grid (1,6);
\draw[ligne] (-1,0)--(0,0)--(1,3)--(0,6)--(-1,0); 
\node[bd] at (-1,0) [label=left:\large{($-1$,$k_x$)}]  {}; 
\node[bd] at (0,0) [label=below:\large{(0,0)}] {};; 
\node[bd] at (1,3) [label=right:\large{(1, $k_y$)}]  {}; 
\node[bd] at (0,6) [label=above:\large{(0,$p$)}] {}; 
\node[bd] at (0,1) {}; 
\node[bd] at (0,2) {}; 
\node[bd] at (0,3) {};
\node[bd] at (0,4) {}; 
\node[bd] at (0,5) {}; 
\end{tikzpicture} 
\caption{The toric diagram for the 5d SCFT realization of  $SU(p)_q$. The example shown is $p=6$, $q= 6-(k_x + k_y) =3$, i.e. $SU(6)_3$, which has  $\mathbb{Z}_3$ 1-form symmetry.  \label{fig:ToricSUpq}}
\end{figure}

This general approach can be exemplified for all toric Calabi-Yau cones, in particular the SCFTs with $SU(p)_q$ IR description, which have from field theory analysis, the anomalies in (\ref{Bcubed}) and (\ref{PL}). 
The Sasaki-Einstein link is given by $Y^{p,q}$, and the Calabi-Yau has simple toric description:
the toric diagram for $SU(p)_q$ is given by figure \ref{fig:ToricSUpq}. For a detailed discussion of this geometry see e.g. \cite{Closset:2018bjz}. The external vertices that determine the toric fan are 
\be
w_0= (0,0)\,, \qquad w_p = (0,p) \,,\qquad w_{x} = (-1, k_x) \,,\qquad w_y = (1, k_y) \,,
\ee
where the CS-level $q$ is determined by 
\be
q=  p-(k_x + k_y) \,.
\ee
There are linear relations among these non-compact divisors $D_{w_i}$, and the instanton $U(1)$ is identified with 
\be
D_{I} = D_{w_x} \,.
\ee
The compact divisors are 
\be
S_a = (0,a) \,,\qquad a= 1, \cdots, p-1 \,. 
\ee
As shown in \cite{Bhardwaj:2020phs} the center symmetry  generator of the gauge theory $SU(p)$ is obtained by taking the linear combination 
\be
Z =  \sum_{a=1}^{p-1} a S_a  \,.
\ee
This compact divisor is also identified with the compact divisor associated to the
generator of ${\rm Tor} \, H^2(L_5;\bZ)$ according to the discussion in section \ref{sec:CS3}.

We will also need an explicit expression for
$p_1(TX_6)=-c_2(TX_6\otimes \bC)$, with $TX_6\otimes \bC$ the
complexification of the tangent bundle of the toric Calabi-Yau
$X_6$. Since $TX_6$ is a complex vector bundle we have
$TX_6\otimes \bC=TX_6\oplus \overline{TX_6}$, so
$c(TX_6\otimes \bC)=c(TX_6)c(\overline{TX_6})$. For a toric variety
$X_6$ with divisors $D_i$ we have \cite{Cox:2011tv}
\begin{equation}
  c(TX_6) = \prod_{i=1}^n (1+D_i)
\end{equation}
so
\begin{equation}
  c(TX_6\otimes\bC) = c(TX_6)c(\overline{TX_6}) = \left(\prod_{i=1}^n (1+D_i)\right)\left(\prod_{i=1}^n (1-D_i)\right) = \prod_{i=1}^n (1-D_i^2)\, .
\end{equation}
and therefore $p_1(TX_6)=\sum_i D_i^2$.

\medskip

With this information at hand it is straightforward to compute the
anomaly coefficients using the formulae \eqref{eq:Omega_formulae},
specialised to the case of one torsion generator of order
${\rm gcd}(p,q)$ and one free generator. We have done these
computations with \textsc{Sage} \cite{sagemath} for $p<20$ and
$|q|\leq p$, and find results compatible with the empirical formulas
\begin{equation} \label{eq:toric_inter}
\ba
Z \cdot Z \cdot Z & = p \, (p-1) \, (p^2 + p\,q - 2 \,q)   
\ ,  \qquad Z \cdot p_1  = 4 \, p \, (p-1)  \ , \cr 
Z \cdot Z \cdot D_I & = -p \, (p-1)  \ ,
\ea
\end{equation}
which we conjecture hold in general. We have also verified in a large
class of examples that we always have $Z \cdot D_I \cdot D_I = 0$, in
accordance to the general claim \eqref{eq:fix_good_basis}, and that
\begin{equation}
\int_{L_5} \diff t_2 \bd \frac{\diff
  p_1(TL_5)}{4} = \bigg[ \frac{Z \cdot p_1}{4 \, {\rm gcd}(p,q)}
\bigg]_\text{mod 1} =0 \ .
\end{equation}
This condition guarantees that the terms in~\eqref{eq:G4X8_result} not
fixed by the mod 24 congruence \eqref{eq:congruence} can indeed be
safely dropped.

Assuming the validity of~\eqref{eq:toric_inter} it is straightforward to verify that
\beq \label{eq:congr}
- \frac 16 \, Z \cdot Z \cdot Z + \frac{1}{24} \, {\rm gcd}(p,q) \, Z \cdot p_1 
= \frac{q \, p \, (p-1) \, (p-2)}{6} -  {\rm gcd}(p,q)^3 (p-1)  \, \frac{P \, (P+1) \, (P-1)}{6}  \ ,
\eeq
where $P = p/{\rm gcd}(p,q)$. Plugging \eqref{eq:toric_inter} in \eqref{eq:Omega_formulae}, and using
\eqref{eq:congr}, we find that the action for the symmetry TFT contains the terms
\beq
\Sym=  \int_{\cW_6} \bigg[ \frac{q \, p \, (p-1) \, (p-2)}{6 \, {\rm gcd}(p,q)^3} \, 
B_2^3
+ \frac{p \, (p-1)}{2\, {\rm gcd}(p,q)^2} \, B_2^2 \, F_I \bigg] \ .
\eeq
This result is in perfect agreement with the field theory results
\eqref{Bcubed} and \eqref{PL}.
It may be worth noting that the result is well-defined, because it is 
invariant under shifts of $B_2$ by ${\rm gcd}(p,q)$ times an arbitrary integral class.
For example,
if we perform the shift $B_2 \rightarrow B_2 + {\rm gcd}(p,q) \, b_2$,
the extra terms generated by the $B_2^3$ term are
\be
\ba
& \frac{q \, p \, (p-1) \, (p-2)}{6 \, {\rm gcd}(p,q)^3} \, 3 \, {\rm gcd}(p,q) \,
\int_{\cW_6}B_2^2 \, b_2 \in \mathbb Z \ , \cr 
\\
& \frac{q \, p \, (p-1) \, (p-2)}{6 \, {\rm gcd}(p,q)^3} \, 3 \, {\rm gcd}(p,q)^2 \,
\int_{\cW_6}B_2 \, b_2^2 \in \mathbb Z \ , \cr
\\
& \frac{q \, p \, (p-1) \, (p-2)}{6 \, {\rm gcd}(p,q)^3} \,  {\rm gcd}(p,q)^3 \,
\int_{\cW_6} b_2^3 \in \mathbb Z \ . 
\ea
\ee
Similar remarks apply to the $B_2^2 \, F_I$ term.


\subsection{Examples: Non-Lagrangian Toric Models}

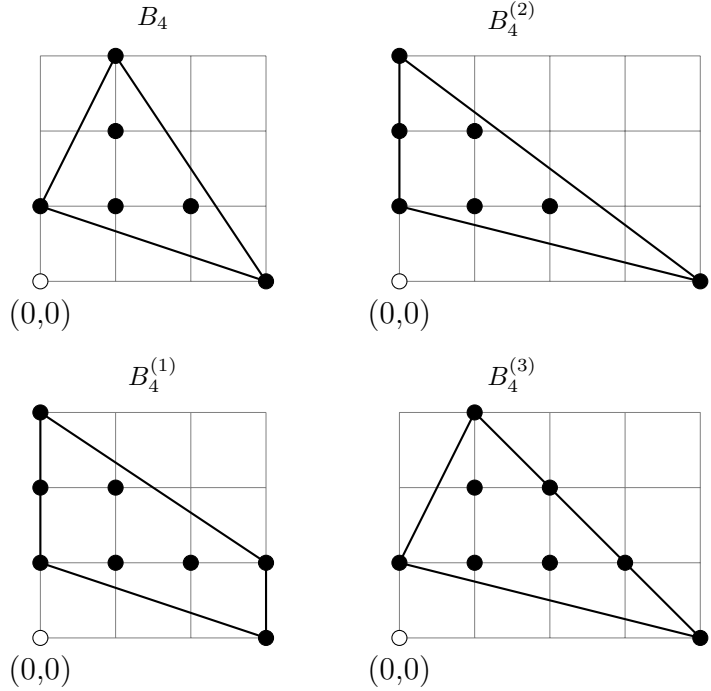
\begin{figure}\centering
\begin{tabular}{cc}
\begin{tikzpicture}[x=1cm,y=1cm]
\draw[step=1cm,gray,very thin] (0,0) grid (3,3);
\node[] at (1.5,3.5) {$B_4$};
\draw[ligne] (3,0)--(1,3)--(0,1)-- (3,0); 
\node[bd] at (3,0) {}; 
\node[bd] at (1,3) {}; 
\node[bd] at (0,1) {};
\node[bd] at (1,1) {};
\node[bd] at (1,2) {};
\node[bd] at (2,1) {};
\node[wd] at (0,0) [label=below:\large{(0,0)}]  {}; 
\end{tikzpicture} 
& \qquad
\begin{tikzpicture}[x=1cm,y=1cm]
\draw[step=1cm,gray,very thin] (0,0) grid (4,3);
\node[] at (1.5,3.5) {$B_4^{(2)}$};
\draw[ligne] (4,0)--(0,3)--(0,2)--(0,1)--(4,0); 
\node[bd] at (4,0) {}; 
\node[bd] at (0,3) {};
\node[bd] at (0,2) {};
\node[bd] at (0,1) {};
\node[bd] at (1,1) {};
\node[bd] at (1,2) {};
\node[bd] at (2,1) {};
\node[wd] at (0,0) [label=below:\large{(0,0)}]  {}; 
\end{tikzpicture} \cr 
\begin{tikzpicture}[x=1cm,y=1cm]
\draw[step=1cm,gray,very thin] (0,0) grid (3,3);
\node[] at (1.5,3.5) {$B_4^{(1)}$};
\draw[ligne] (3,0)--(3,1)--(0,3)--(0,2)--(0,1)--(3,0); 
\node[bd] at (3,0) {}; 
\node[bd] at (3,1) {}; 
\node[bd] at (0,3) {};
\node[bd] at (0,2) {};
\node[bd] at (0,1) {};
\node[bd] at (1,1) {};
\node[bd] at (1,2) {};
\node[bd] at (2,1) {};
\node[wd] at (0,0) [label=below:\large{(0,0)}]  {}; 
\end{tikzpicture} 
&\qquad
\begin{tikzpicture}[x=1cm,y=1cm]
\draw[step=1cm,gray,very thin] (0,0) grid (4,3);
\node[] at (1.5,3.5) {$B_4^{(3)}$};
\draw[ligne] (4,0)--(3,1)--(2,2)--(1,3)--(0,1) -- (4,0); 
\node[bd] at (4,0) {}; 
\node[bd] at (3,1) {};
\node[bd] at (2,2) {};
\node[bd] at (1,3) {};
\node[bd] at (0,1) {};
\node[bd] at (1,1) {};
\node[bd] at (1,2) {};
\node[bd] at (2,1) {};
\node[wd] at (0,0) [label=below:\large{(0,0)}]  {}; 
\end{tikzpicture} 
\end{tabular}
\caption{$B_N$ and $B_N^{(i)}$ non-Lagrangian toric diagrams for $N=4$. \label{fig:BN}}
\end{figure}

Our geometric approach becomes particularly useful when the theories
in question do not have any non-abelian gauge theory description --
i.e. are truly non-Lagrangian.  As an illustration, we consider the
toric models $B_N$ and $B_N^{(i)}$ introduced and studied in
\cite{Eckhard:2020jyr, Morrison:2020ool}, which do not have any
non-abelian gauge theory description in 5d on the Coulomb branch.
They are defined in terms of their toric fan in table \ref{tab:BN},
where also their 1-form symmetry is tabulated. Examples of the $N=4$
models are shown in figure \ref{fig:BN}.  Note that $B_{N=3}$ is the
rank 1 $\mathbb{P}^2$ Seiberg theory, which we show to also have a
non-trivial 1-form symmetry anomaly.

\begin{table}
$$
\begin{array}{|c|c|c|}\hline
\text{Theory} & \Gamma^{(1)} & \text{Toric Fan} \cr \hline\hline
B_N & \mathbb{Z}_{N (N-3)+ 3}&  (N-1, 0,1), (1, N-1, 1), (0,1,1)\cr 
B_N^{(1)} & \mathbb{Z}_{N-1} &   ((N-1,0, 1), (N-1, 1, 1) (0, N-1-k, 1)), \, k= 0, \cdots, N-2  \cr 
B_N^{(2)} & \mathbb{Z}_{N} &  ( (N,0, 1), (0, N-1-k, 1)), \, k= 0, \cdots, N-2  \cr 
B_N^{(3)} & \mathbb{Z}_{N-1} &  ( (0,1, 1),   (N-k, k, 1)), \, k= 0, \cdots, N-1   \cr \hline
\end{array}$$
\caption{Properties of the $B_N$ and $B_N^{(i)}$ non-Lagrangian toric models. \label{tab:BN}}
\end{table}


We can again compute the $B^3$ terms in the SymTFT for $B_N$: 
\be
\Sym^{(B_N)} =  \int_{\mathcal{W}_6} {(N-1)(N-2) \over 6 (N(N-3)+3) }  \, B_2^3 
\ee
For $B_N^{(1)}$ we find 
\be
\Sym^{(B_N^{(1)})} =\int_{\mathcal{W}_6} \frac{(N-3) (N-2)}{6 (N-1)}  \, B_2^3  \,.
\ee
For $B_N^{(2)}$ we find 
\be
\Sym^{(B_N^{(2)})} =\int_{\mathcal{W}_6} \frac{(N-2) (N-1)}{6 N}  \, B_2^3  \,.
\ee
Finally for $B_N^{(3)}$ the $B_2^3$ anomaly vanishes. 

In computing these, we have picked a particular central divisor
associated to the generator of the 1-form symmetry.  e.g. in the case
of $B_N^{(2)}$ theories, we conjecture this to be of the form
\begin{equation}
  Z_{B^{(2)}_N} = (-N+1) \sum_{i=1}^{N-2} v_{1,i} + (-N+2) \sum_{i=1}^{N-3} v_{2,i} + \cdots \,.
\end{equation}
Since there is no non-abelian gauge theory description any central
divisor is in fact equally acceptable, but this choice leads to simple
general formulas.

Finally, let us note that the $B_N^{(1)}$ theories have a $U(1)$
factor in their flavor symmetry groups \cite{Eckhard:2020jyr}. We can
compute the mixed anomaly between this $U(1)$ symmetry and the 1-form
symmetry. To this end, we use the general formula
\eqref{eq:Omega_formulae}, specialised to the case of a single $Z$
central divisor. The non-compact divisor $D$ associated to the $U(1)$
flavor symmetry is identified with the divisor associated to the
vertex with coordinates $(N-1,0)$ in the toric diagram for
$B_N^{(1)}$, see figure \ref{fig:BN}.  We find the following
additional term in the symmetry TFT:
\be
\Sym^{(B_N^{(1)}),\text{mixed}} = \int_{\mathcal{W}_6} \frac{N-2}{2 (N-1)}  \, B_2^2 \, F  \, ,
\ee
where $F$ is the field strength of the background field for the $U(1)$ flavor symmetry.

The application of our approach to this class of non-Lagrangian theories in 5d, demonstrates the flexibility and generality of 
the approach using our differential cohomology extension of dimensional reductions on the link. 
Here we focused on toric models, but any non-compact Calabi-Yau three-fold geometry that has a canonical singularity, i.e. 5d SCFT,  can be 
studied in this way.


\section{5d Anomalies from the Boundary of $(p,q)$ 5-brane Webs}
\label{sec:5branes}

We now consider IIB $(p,q)$ 5-brane webs engineering 5d SCFTs, and in particular we would like to evaluate the IIB supergravity action at the boundary of these webs to compute a 6d bulk action. 
We focus on the part of this action which involves the 1-forms symmetries and corresponds to the anomalies of the 5d SCFTs. 

\subsection{Mixed Anomalies}

In order to describe the boundary geometry, we assume that the topology of the near-horizon limit of the IIB $(p,q)$ 5-brane webs at large $N$ in \cite{DHoker:2017mds} extends generically to all webs. This implies that the boundary is given by $M_4$ that is an $S^2$ fibered over a disc, $\Sigma$, with punctures at the boundary of $\Sigma$ representing the $(p,q)$ 5-brane sources.  The fibered $S^2$ shrinks at the boundary of the disc away from the punctures, and the full space is topologically equivalent to an $S^4$ with punctures.  The non-trivial topological cycles are 3-cycles $\{c_{\ell}\}$,  and the 3-form representatives of the third cohomology of the $M_4$ are denoted by $\{\beta_{\ell} = \nu_{\ell}  \wedge {\rm vol}(S^2)\}$, where $\ell = 1, \ldots, L$ is the total number of semi-infinite 5-brane stacks. $\nu_{\ell}$ corresponds to the angular direction around the punctures and $\beta_{\ell} $ the volume forms for of these 3-cycles.  In particular  $\{c_{\ell}\}$ need to satisfy the following linear relation \cite{Bergman:2018hin}
\begin{equation} \label{eq:linrel}
\sum_{\ell} c_{\ell} = 0, \qquad \sum_{\ell} p_{\ell} c_{\ell}=0, \qquad \sum_{\ell} q_{\ell} c_{\ell}=0\,,
\end{equation}
which are dual to the relations among divisors in eq. (3.24) of \cite{Closset:2018bjz} of the toric Calabi-Yau in the M-theory construction. 

The are also non-trivial 1-cycles with volume forms given by $\{\omega_{\ell} \}$. Locally they can be thought as the hodge dual of the $\beta_{\ell}$,  and they have to satisfy the same constraints \eqref{eq:linrel}. \footnote{Globally the 1-cycles can be thought as segments connecting the puctures,  such that the constraints \eqref{eq:linrel} are satisfied.  F.A. would like to thank Oren Bergman and Christoph Uhlemann for clarifying some of these aspects.}  The intersection pairing reads
\begin{equation}
\int_{M_4} \omega_{\ell} \wedge \beta_{j} = \Omega_{\ell j} \text{Vol}_4 \,.
\end{equation}
We expand $F_3, H_3, C_4$
\be\ba \label{eq:expansions}
& F_3= q^{\ell} \beta_{\ell} + f_2^{\ell} \wedge \omega
_{\ell} + g_3 + \ldots\\
& H_3=p^{\ell} \beta_{\ell}  + h_2^{\ell} \wedge \omega
_{\ell}+ h_3 +\ldots\\
& F_5= f_5+   f_4^{\ell} \wedge \omega_{\ell} +  g_2^{\ell} \wedge \beta_{\ell}   +f_1  \wedge {\rm vol}_4 \,,
\ea \ee
where self-duality of $F_5$ impose that $\ast_6 f_5 = f_1$ and $\ast_6 f_4 = f_2$. In addition there is a field strength for the pair $H_3,F_3$ expanded on $\omega_{\ell}$ corresponding to
\begin{equation}
da_{1}^{\ell} = \sum_j \Omega_{\ell j} (q^{\ell} h_2^{j} -  p^{\ell} f_2^j).
\end{equation}
Not all of $a_{1}^{\ell}$ will be linearly independent, since the 1-cycles as well the 3-cycles, $c_{\ell}$, are also not all independent \eqref{eq:linrel}.  The most relevant aspect is the counting of the massless vector fields which corresponds to the backgrounds of the abelian flavor symmetries, whereas the details of the intersection pairing $\Omega$, as long as it is non-trivial,  will not affect the results in any significant way. 

We study now the reduction of the IIB topological coupling, which we extend to a 11-dimensional coupling to ensure gauge invariance
\begin{equation} \label{eq:11dtop}
S_{11}^{\rm top} = \int F_5\wedge H_3 \wedge F_3 \,.
\end{equation}
The Bianchi identity $dF_5=H_3\wedge F_3$ imposes the following constraints in BF frame $(f_5,f_4)$,
\be\ba \label{eq:bianchiconstBF}
 df_5 = h_3 \wedge f_3   &\ \rightarrow\  f_5 = dc_4+  b_2 \wedge f_3 \\
 df_4^{\ell} = (h_3 \wedge f_2^{\ell} - g_3 \wedge h_2^{\ell})& \  \rightarrow \ f_4^{\ell} =  dc_3^{\ell} + (b_2 f_2^{\ell} - c_2 h_2^{\ell})
\ea \ee
and in the St\"uckelberg  frame $(f_1,f_2)$
\be\ba \label{eq:bianchiconstST}
 df_1 = g_2^{\ell} q_{\ell} - g_2^{\ell} p_{\ell}  &\ \rightarrow \ f_1= dc_0+  b_1^{\ell} q_{\ell} - c_1^{\ell} p_{\ell}  \\
 dg_2^{\ell} = (q^{\ell} h_3 - p^{\ell} g_3) & \ \rightarrow\  g_2^{\ell} =  d{c}_1^{\ell} + (q^{\ell} b_2 - p^{\ell} c_2) \,.
\ea \ee

\paragraph{BF frame.} From the expansion \eqref{eq:expansions} and \eqref{eq:bianchiconstBF}, we obtain various contributions to couplings.
The first interesting coupling is given by the singleton theory, which can be recast in a $SL(2\mathbb{R})$ covariant way as follows,
\begin{equation} \label{eq:singl}
S^{\rm singl}_{7d} =   \int dc_3^{j} \Omega_{j \ell} Q^{\ell I}\wedge \sigma_{IJ}\mathcal{F}^J_3\,,
\end{equation}
where 
\begin{equation}
Q^{\ell I} = \begin{pmatrix}
p_1 & q_1 \\
\vdots & \vdots \\
p_L & q_L \\
\end{pmatrix}, \qquad \mathcal{F}^J_3 = \begin{pmatrix} h_3  \\ g_3\end{pmatrix}, \qquad \sigma_{IJ}= \begin{pmatrix}
0 & 1\\
-1 & 0
\end{pmatrix}\,,
\end{equation}
by suppressing the indices, using matrix multiplication, and defining $C_3=\{c_3^1, \ldots, c_3^L \}$, we can write \eqref{eq:singl} as,
\begin{equation}
S^{\rm singl}_{7d} = \int dC_3 \Omega Q \sigma \mathcal{F} = \int dC_3  \Omega B^{-1} B Q A  A^{-1} \sigma \mathcal{F} =  \int 
d \widetilde{C}_3 Q_{\rm SNF} \widetilde{\mathcal{F}}_3\,,
\end{equation}
where $Q_{\rm SNF} = B Q A$ is the Smith normal form of $Q$ and $A,B$ is the pair of matrices transforming $Q$, moreover we have that $ d\widetilde{C}_3 = (B^{-1})^T \Omega dC_3$ and $ \widetilde{\mathcal{F}}_3 = A^{-1} \sigma \mathcal{F}_3$. The study of boundary conditions and mutual locality of this action determines the 1 or 2-form symmetries of the theory. 

We now study what happens to the vector fields $a_1^{\ell}$ with coupling
\begin{equation}
S_{7d}^{\rm vec}=\int dc_4 \wedge (q^{\ell} \Omega_{\ell j} h_2^{j} -  p^{\ell} \Omega_{\ell j} f_2^{j})\,.
\end{equation}
This can be written in an $SL(2,\mathbb{R})$ covariant way as follows
\begin{equation} \label{eq:BFvec}
S_{7d}^{\rm vec}= \int  dc_4 \wedge  \Omega_{j \ell}  Q^{\ell I} \sigma_{IJ} \mathcal{F}_2^{I j} = \int  dc_4 \wedge \text{Tr}(Q^T\Omega   \mathcal{F}_2^T \sigma) \,.
\end{equation}
where $Q$ and $\sigma$ have been defined in \eqref{eq:singl}
\begin{equation}
\mathcal{F}_2 = \begin{pmatrix} h_2^1  & \ldots &h_2^L \\ f_2^1 & \ldots &f_2^L \end{pmatrix} \,.
\end{equation}
In particular \eqref{eq:BFvec} can be rewritten as follows,
\be \label{eq:stuckvecBF}
\ba
S_{7d}^{\rm vec}&=\int  dc_4 \wedge \text{Tr}( (B^{-1} B Q A A^{-1})^T \Omega  \mathcal{F}_2^T \sigma) = \int  dc_4 \wedge \text{Tr}((A^{-1})^T Q_{\rm SNF}^T  \widetilde{\mathcal{F}}_2^T \sigma)=\\
&= \int  dc_4 \wedge  \text{Tr}(\sigma  \widetilde{\mathcal{F}}_2 Q_{\rm SNF} A^{-1} )\,,
\ea
\ee
where $\widetilde{\mathcal{F}}_2= (B^{-1})^T  \Omega \mathcal{F}_2^T$. This is the dual of the St\"uckelberg mechanism, which makes a combination of the vectors massive, and the massless one satisfies,
\begin{equation}\label{eq:masscond}
 \text{Tr}(\sigma  \widetilde{\mathcal{F}}_2 Q_{\rm SNF} A^{-1} )=0 \,.
\end{equation}
The number of linear independent vectors that are dual to $U(1)$ flavor symmetries is $0$ if $L=3$ and $2(L-3)-1$ otherwise.

Finally, we also have additional couplings, which could potentially lead to anomalies between the $U(1)$ flavor symmetries and the 1-form symmetries, which in $SL(2,\mathbb{Z})$ coviariant form reads
\begin{equation} \label{eq:potanom}
S^{\rm anom}_{7d} =\int   (q^{\ell} b_2 - p^{\ell} c_2) \wedge \Omega_{\ell j} (f_2^{j} h_3 - h_2^j f_3)  = \int \mathcal{C}_2  \sigma Q^T \Omega \mathcal{F}_2^T \sigma \widetilde{\mathcal{F}}_3=\int \widetilde{\mathcal{C}}_2 Q_{\rm SNF}^T \widetilde{\mathcal{F}}_2^T  A \widetilde{\mathcal{F}}_3,  
\end{equation}
where $\mathcal{F}_3 = d\mathcal{C}_2$ and $\widetilde{\mathcal{F}}_3 = d\widetilde{\mathcal{C}}_2$.

\paragraph{St\"uckelberg frame.} The same physical consequences hold also in St\"uckelberg frame. The second Bianchi identity in \eqref{eq:bianchiconstST} implies that
\begin{equation}\label{eq:1formbianchi}
g_2^{\ell}=d\tilde{c}_1^{\ell} + Q \sigma \mathcal{C}_2= d\tilde{c}_1^{\ell} + B^{-1} Q_{\rm SNF} \widetilde{\mathcal{C}}_2 \,.
\end{equation}
The kinetic therm for $g_2^{\ell}$ could make some of the component $\widetilde{\mathcal{C}}_2$ discrete depending on the entries of $Q_{\rm SNF}$.
The first identity in \eqref{eq:bianchiconstST} can be rewritten such that
\begin{equation}
f_1 = dc_0 + \text{Tr}( \sigma \mathcal{F}_2 \Omega Q )  = dc_0 + \text{Tr}(\sigma  \widetilde{\mathcal{F}}_2 Q_{\rm SNF} A^{-1} ) \,,
\end{equation}
leading to the same constraint given by \eqref{eq:stuckvecBF}. Moreover the expansion of \eqref{eq:11dtop}, leads to the same expression \eqref{eq:potanom} 
for the mixed anomaly between the 1-form symmetry and the $U(1)$ flavors. We will see in the example of subsection \ref{sub:examplesIIB}, that this reproduce the anomaly discussed in \cite{BenettiGenolini:2020doj}.

\subsection{'t Hooft Anomaly for the 1-Form Symmetry}
So far we have studied the bulk supergravity action, but the effective 6d anomaly theory can receive contributions from the Chern-Simons terms of the 5-brane sources. To be consistent we add these Chern-Simons actions in the $SL(2,\mathbb{Z})$ covariant form \cite{Bergshoeff:2006gs, Callister:2009sg}, which in general reads,
\begin{equation}
S^{CS}_{6d}  = \sum_{\ell} \int \widetilde{Q}^{\ell I} \sigma_{IJ} \hat{C}^J e^{ Q^{\ell I} \sigma_{IJ}  \hat{F}^{J\ell}} \sqrt{\hat{\mathcal{A}}(R_{\mathcal{T}})/\hat{\mathcal{A}}(R_{\mathcal{N}})}\,,
\end{equation}
where the sum over the 5-branes and on each 5-brane we have $\hat{C}=\hat{C}_0 + \hat{C}_2 + \hat{C}_4 + \hat{C}_6 + \hat{C}_8$, $\hat{\mathcal{A}}$ is the A-roof genus of $R_\mathcal{T},R_\mathcal{N} $ tangent and normal bundle curvatures respectively. The expansion in terms of Pontryagin classes reads
\begin{equation}
\sqrt{\hat{\mathcal{A}}(R_{\mathcal{T}})/\hat{\mathcal{A}}(R_{\mathcal{N}})} = 1 - \frac{1}{48} (p_1(R_{\mathcal{T}})-p_1(R_{\mathcal{N}}))+ \ldots\,.
\end{equation}
Moreover $\widetilde{Q}$ is defined by
\begin{equation}
\widetilde{Q}= \begin{pmatrix}
\tilde{p}_1 & \tilde{q}_1 \\
\vdots & \vdots \\
\tilde{p}_L & \tilde{q}_L \\
\end{pmatrix}, \qquad \widetilde{Q}^{\ell [I} Q^{J],\ell} = \frac{1}{2}\epsilon^{IJ} \quad \forall \ell.
\end{equation}
In addition, we have
\be 
\ba 
& \hat{F}^{\ell}= Q^{\ell I} \sigma_{IJ} (d\hat{a}^{\ell J} +  \mathcal{C}_2^J)\\
& \hat{C}_0 = c_0\\
& \hat{C}_2 = \mathcal{C}_2\\
&\hat{C}_4 = c_4- \frac{1}{2} Q^{\ell I}  \sigma_{IJ} \mathcal{C}_2^J \wedge Q^{\ell I}  \sigma_{IJ} \mathcal{C}_2^J\\
&\hat{C}_6= \mathcal{C}_6 + c_4 \wedge Q^{\ell I}  \sigma_{IJ} \mathcal{C}_2^J+ \frac{1}{6} \widetilde{Q}^{\ell I}  \sigma_{IJ} \mathcal{C}_2^J \wedge Q^{\ell I}  \sigma_{IJ} \mathcal{C}_2^J \wedge Q^{\ell I}  \sigma_{IJ} \mathcal{C}_2^J \\
&\hat{C}_6 = \mathcal{C}_8 + \frac{1}{24} (\widetilde{Q}^{\ell I}  \sigma_{IJ} \mathcal{C}_2^J) \wedge (Q^{\ell I}  \sigma_{IJ} \mathcal{C}_2^J)^3\,,
\ea
\ee 
where  $\mathcal{C}_6=(b_6,c_6)$, and the $\hat{a}^{\ell I}$ correspond to the center of mass mode of the 5-brane stack and all decouple.
We now expand the action and the relevant terms are
\begin{equation} 
S^{CS}_{6d} = \sum_l  \int   \frac{1}{6} \widetilde{Q} \sigma \mathcal{C}_2 \wedge  Q \sigma \mathcal{C}_2 \wedge Q \sigma \mathcal{C}_2 + \frac{1}{48}\widetilde{Q} \sigma \mathcal{C}_2\wedge p_1(R_{\mathcal{T}}) +\ldots \,,
\end{equation}
where  we suppressed all the indices in favour of the matrix product, and we ignored the normal bundle contributions. Moreover, there is no contribution of such kind from the bulk \eqref{eq:11dtop}. In terms of $\widetilde{\mathcal{C}}_2= A^{-1} \sigma \mathcal{C}_2$, which is the diagonal basis where the $(p,q)$ charges take the Smith normal form. We get the contribution to the SymTFT from the CS couplings to be:
\begin{equation} \label{eq:WZaction}
\Sym^{CS}= \sum_l  \int   \frac{1}{6} \widetilde{Q} A \widetilde{\mathcal{C}}_2 \wedge  Q A \widetilde{\mathcal{C}}_2 \wedge Q A \widetilde{\mathcal{C}}_2 + \frac{1}{48}\widetilde{Q} A \widetilde{\mathcal{C}}_2\wedge p_1(R_{\mathcal{T}}) 
\end{equation}
with $A$ defined such that $Q_{\rm SNF}=BQA$. In particular this action reproduces the cubic anomaly introduced in \cite{Gukov:2020btk}, as we will see in the examples of the next subsection. 

\subsection{Examples}\label{sub:examplesIIB}
We now apply these general results to brane-webs that realize 5d SCFTs. 
This complements the geometric analysis in the earlier parts of the paper, in particular the toric Calabi-Yau reductions in M-theory in section \ref{sec:5d}. 
The brane-webs can easily be obtained from the toric diagrams (by passing to a dual graph) and vice versa.

\paragraph{Example: $SU(2)_0$ SCFT.}
The asymptotic 5-branes charges for the $SU(2)_0$ Seiberg theory are 
\begin{equation}
Q=\begin{pmatrix}
1 &1 \\
1 &-1\\
-1&-1\\
-1 &1 \end{pmatrix}
\end{equation}
The Smith normal form reads
\begin{equation}
Q_{\rm SNF}=\begin{pmatrix}
1 & 0\\
0 &2\\
0 &0\\
0 &0\end{pmatrix}\,,
\end{equation}
which together with \eqref{eq:singl} encodes the $\mathbb{Z}_2$ TQFT that upon a suitable choice of boundary condition determines the $\mathbb{Z}_2$ 1-form symmetry of the theory, together with its background field.  
We have 2 independent linear combination of the vector fields, and the condition \eqref{eq:masscond} sets one of them to zero. Then we have a contribution to the anomaly coming from \eqref{eq:potanom}, and the relevant contribution is
\begin{equation}
S^{\rm bulk}_{7d} =2 \int d\tilde{a}_1 \wedge \tilde{c}_2 \wedge \tilde{f}_3 + \ldots  \,,
\end{equation}
where $d\tilde{a}_1= (\tilde{f}^2_2- \tilde{h}^2_2)$ is the background for the instanton $U(1)_I$ of the gauge theory which enhances to $SO(3)$ at the conformal point \cite{Apruzzi:2021vcu, BenettiGenolini:2020doj}. $\tilde{c}_2$ is the background field for the $\mathbb{Z}_2$ 1-form symmetry with $\mathbb{Z}/2$ periods, therefore if we map the field to the one with integral periods, $\tilde{c}_2\rightarrow \tilde{c}_2/2$, and integrate on the 6-dimensional boundary  we get
\begin{equation}
\Sym^{\text{mixed}} =\frac{1}{4}  \int d\tilde{a}_1 \wedge  \tilde{c}_2\wedge \tilde{c}_2 \,.
\end{equation}

We can also compute the anomaly coming from $\eqref{eq:WZaction}$. First of all we use the congruence, that tells us,
\begin{equation} \label{eq:congru}
\tilde{c}_2 p_1(R_{\mathcal{T}})= 4 \tilde{c}_2 \tilde{c}_2 \tilde{c}_2 \quad \text{mod}\quad 24
\end{equation}
with $\tilde{c}_2$ having integer periods.
We also need $\widetilde{Q}$
\begin{equation}
\widetilde{Q}=\begin{pmatrix}
1+x_1 &x_1 \\
-1-x_2 &x_2\\
-1+x_3&x_3\\
1-x_4 &x_4 \end{pmatrix} \,,
\end{equation}
where $x_i$ are general integer parameters. Evaluating \eqref{eq:WZaction} we get,
\begin{equation}
\Sym^{\rm CS}=\frac{(x_1+x_4)}{4} \int \tilde{c}_2^3.
\end{equation}
where $\tilde{c}_2$ is the one with integer periods.
However the freedom of choosing $x_1,x_4$ can be reabsorbed by adding the two terms to the action, which do not change the classical equations of motion of the low-energy anomaly theory,
\begin{equation}
S^{\rm add}_{7d} = \int (d\tilde{g}_2^2 - 2 \tilde{f}_3) (m \tilde{g}_2^2\wedge  \tilde{g}_2^2+m' \frac{1}{48} p_1(R_{\mathcal{T}})) \,,
\end{equation}
where $\tilde{g}_2^2= d\tilde{c}^2_1+ 2 \tilde{c}_2$, moreover $d\tilde{g}_2^2 - 2 \tilde{f}_3$ is no other than  of component of the Bianchi identity \eqref{eq:1formbianchi} defining the $\mathbb{Z}_2$ 1-form symmmetry in St\"uckelberg frame, and $\tilde{g}_2^{\ell} =( B^{-1})^{\ell}_j g_2^j$.  Finally,  since these are Chern-Simons terms $m, m'$ must be integers. By integrating on the 6d boundary and by plugging in the congruence \eqref{eq:congru}, gauge invariance under $\tilde{c}_2\rightarrow \tilde{c}_2 + 2 d\lambda_1$ fixes $3(x_1+x_4)=4m+m'$, which implies
\begin{equation}
\Sym^{\text{cubic}}= \Sym^{\rm CS}+S^{\rm add}_{6d} =0.
\end{equation}

\paragraph{Example: $SU(p)_q$ SCFTs.}
For the general $SU(p)_q$ theory the 5-brane charges and their Smith normal form are 
\begin{equation}
Q=\begin{pmatrix}
-p &1 \\
q &1\\
0&-1\\
(p-q) &-1 \end{pmatrix}
\,,\qquad 
Q_{\rm SNF}=\begin{pmatrix}
1 & 0\\
0 & \text{gcd}(p,q)\\
0 &0\\
0 &0\end{pmatrix} \,,
\end{equation}
which together with \eqref{eq:singl} encodes the $\mathbb{Z}_{\text{gcd}(p,q)}$ 1-form symmetry of the theory, and its background field. 
We have 2 independent linear combinations of vector fields, and the condition \eqref{eq:masscond} sets one of them to zero. Then we have a contribution to the anomaly coming from  \eqref{eq:potanom} that is,
\begin{equation}
S^{\rm bulk}_{7d} =-p\int d\tilde{a}_1 \wedge \tilde{c}_2 \wedge \tilde{f}_3 + \ldots  \,,
\end{equation}
where $d\tilde{a}_1= ( \tilde h^2_2)$ generically corresponds to the background for the instanton symmetry $U(1)_I$ of the gauge theory that enhances to $SO(3)$ at the superconformal point \cite{Apruzzi:2021vcu}.\footnote{We recall that here we have not activated the Stiefel-Whitney class $w_2(SO(3))$ for the non-abelian flavor symmetry. It will be also interesting to compute the mixed anomaly in terms of $w_2$ from a string theory engineering perspective.} $\tilde{c}_2$ is the background field for the $\mathbb{Z}_2$ 1-form symmetry with $\mathbb{Z}/\text{gcd}(p,q)$ periods, therefore if we map the field to the one with integral periods, $\tilde{c}_2\rightarrow \tilde{c}_2/\text{gcd}(p,q)$, and integrate on the 6-dimensional boundary  we get
\begin{equation}
S^{\rm bulk}_{6d} =-\frac{p}{2\text{gcd}(p,q)^2}  \int d\tilde{a}_1 \wedge  \tilde{c}_2\wedge \tilde{c}_2\,.
\end{equation}
This expression is not gauge invariant under $\tilde{c}_2\rightarrow \tilde{c}_2 + \text{gcd}(p,q) d\lambda_1$. We then add the following term which does not change the anomaly theory,
\begin{equation}
S^{\rm add}_{7d}= n (d\tilde{g}_2^2 - \text{gcd}(p,q) \tilde{f}_3)\wedge  d\tilde{a}_1 \wedge \tilde{g}_2^2\,,
\end{equation}
where we recall that $g_2^2= d\tilde{c}^2_1+ \text{gcd}(p,q) \tilde{c}_2$, upon integrating on the 6d boundary, and rescaling $\tilde{c}_2\rightarrow \tilde{c}_2/\text{gcd}(p,q)$, gauge invariance fixes $n$ to be an integer such that $\text{gcd}(p,q) n= -p$, such that
\begin{equation}
\Sym^{\text{mixed}} =\frac{p(p-1)}{2\text{gcd}(p,q)^2}  \int d\tilde{a}_1 \wedge  \tilde{c}_2\wedge \tilde{c}_2\,.
\end{equation}
The $\tilde{c}_2^3$ anomaly comes from evaluating $\eqref{eq:WZaction}$. First of all we use the congruence, that tells us \eqref{eq:congru} and we also need $\widetilde{Q}$, that is
\begin{equation}
\widetilde{Q}=\begin{pmatrix}
1+ p x_1 &x_1 \\
1-px_2 &x_2\\
-1&x_3\\
-1&x_4 \end{pmatrix} \,,
\end{equation}
where $x_i$ are general integer parameters. Evaluating \eqref{eq:WZaction} we get
\begin{equation}
\Sym^{\rm CS}=\left( \frac{q p}{3 \text{gcd}(p,q)^3}+ \frac{(x_1+x_2)}{4}\right) \int \tilde{c}_2\tilde{c}_2\tilde{c}_2\,,
\end{equation}
where $\tilde{c}_2$ is the one with integer periods.
We can now ignore the contribution proportional to $x_1,x_2$. They can be reabsorbed by adding two terms to the action as in the previous example. These new terms do not change the classical equations of motion of the low-energy anomaly theory, since they are proportional to Bianchi identities. They read
\begin{equation}
S^{\rm add}_{7d} = \int  (d\tilde{g}_2^2 - \text{gcd}(p,q)  \tilde{f}_3) \wedge  \left(m \tilde{g}_2^2\wedge  \tilde{g}_2^2+m' \frac{1}{48} p_1(R_{\mathcal{T}})\right)
\end{equation}
where $g_2^2= d\tilde{c}^2_1+ \text{gcd}(p,q)\tilde{c}_2$, and $d\tilde{g}_2^2 - \text{gcd}(p,q) \tilde{f}_3$ is no other than  of component of the Bianchi identity \eqref{eq:1formbianchi} defining the $\mathbb{Z}_2$ 1-form symmmetry in St\"uckelberg frame, and $\tilde{g}_2^{\ell} =( B^{-1})^{\ell}_j g_2^j$. By integrating on the 6d boundary and by plugging in the congruence \eqref{eq:congru}, gauge invariance under $\tilde{c}_2\rightarrow \tilde{c}_2+ \text{gcd}(p,q) d\lambda_1$ fixes $m,m'$ to be integers such that $4m+m'= 2(p-3) \frac{p^2 q}{\text{gcd}(p,q)^3}$, which implies
\begin{equation}
\Sym = \Sym^{\text{CS}}+S^{\rm add}_{6d} =\frac{qp(p-2)(p-1)}{6\text{gcd}(p,q)^3} \int \tilde{c}_2\tilde{c}_2\tilde{c}_2.
\end{equation}

\paragraph{Example: $B_N$ SCFTs.} 
Finally, we apply this method also to the non-Lagrangian theories of type $B_N$ introduced in section \ref{sec:5d} in the toric analysis. 
The asymptotic 5-branes charges of the web and the SNF are 
\begin{equation}
Q=\begin{pmatrix}
N-1 &N-2 \\
-1 &-(N-1)\\
-(N-2)&1\end{pmatrix}
\,,\qquad 
Q_{\rm SNF}=\begin{pmatrix}
1 & 0\\
0 &N(N-3)+3\\
0 &0\\
0 &0\end{pmatrix} \,,
\end{equation}
which together with \eqref{eq:singl} encodes the $\mathbb{Z}_{N(N-3)+3}$ 1-form symmetry of the theory, and its background field. The linear relations \eqref{eq:linrel} fix the cycles to be trivial and the pairing as well. So there is no massless vector, corresponding to the theory having no continuous 0-form symmetry. All we can compute is then the cubic anomaly for the 1-form symmetry. To do so we evaluate \eqref{eq:WZaction} with the congruence \eqref{eq:congru}, where
\begin{equation}
\widetilde{Q}=\begin{pmatrix}
1 &1 \\
x_1 &(N-1)x_1-1\\
-1 - x_2(N-1)&x_2\end{pmatrix} \,,
\end{equation}
where $x_1,x_2$ are integer parameters.
Plugging this and \eqref{eq:congru} into \eqref{eq:WZaction}, we get,
\begin{equation}
\Sym^{\text{CS}}=\left(\frac{x_2 (N(N-3)+3) + 2}{6 (N(N-3)+3) } +\frac{x_2}{12} \right) \int \tilde{c}_2 \tilde{c}_2 \tilde{c}_2 \,.
\end{equation}
Again we can add to the supergravity topological action terms which are proportional to Bianchi identities and do not change the classical equations of motion,
\begin{equation}
S^{\rm add} = \int  (d\tilde{g}_2^2 - (N(N-3)+3) \tilde{f}_3) \wedge  \left(m \tilde{g}_2^2\wedge  \tilde{g}_2^2+m' \frac{1}{48} p_1(R_{\mathcal{T}})\right) \,.
\end{equation}
Again, $m,m'$ are integers that are fixed by gauge invariance under $\tilde{c}_2\rightarrow \tilde{c}_2 +  (N(N-3)+3) d\lambda_1$ once we integrated this additional term on a 7d space with a 6d boundary. $m,m'$ then satisfy
\begin{equation}
4m+m' = 2 (1-x_2) \,.
\end{equation}
This implies that
\begin{equation}
\Sym^{\text{cubic}}= \Sym^{\rm CS}+S^{\rm add}_{6d} =\frac{(N-2)(N-1)}{6(N(N-3)+3)} \int \tilde{c}_2\tilde{c}_2\tilde{c}_2\,.
\end{equation}

\paragraph{Example: $B^{(1)}_N$ SCFTs.} 
The asymptotic 5-branes charges of the web are
\begin{equation}
Q=\begin{pmatrix}
-1 &-(N-1)\\
N-2 &N-1 \\
1 & 0\\
-1&0\\
-1 &0\\
 \vdots & \vdots \end{pmatrix}\,,\qquad 
Q_{\rm SNF}=\begin{pmatrix}
1 & 0\\
0 &N-1\\
0 &0\\
0 &0\\
 \vdots & \vdots
\end{pmatrix}\,,
\end{equation}
where there are $N-2$ $(-1,0)$ 5-branes. 
The SNF together with \eqref{eq:singl} encodes the $\mathbb{Z}_{N-1}$ 1-form symmetry of the theory, and its background field.  To compute the cubic anomaly for the 1-form
symmetry we evaluate \eqref{eq:WZaction} with the congruence \eqref{eq:congru}, where
\begin{equation}
\widetilde{Q}=\begin{pmatrix}
x_1 &x_1(N-1)+1 \\
1&1\\
x_2&1\\
x_3&-1\\
 \vdots & \vdots
\end{pmatrix} \,,
\end{equation}
where $x_1,x_2,x_3, \ldots x_N$ are integer parameters.
Plugging this and \eqref{eq:congru} into \eqref{eq:WZaction}, we get,
\begin{equation}
\Sym^{\text{CS}}=\left(\frac{x_1 (N-1) + 2}{6 (N-1) } +\frac{x_1 +1 }{12} \right) \int \tilde{c}_2\tilde{c}_2 \tilde{c}_2 \,.
\end{equation}
Again we can add to the supergravity topological action terms which are proportional to Bianchi identities and do not change the classical equations of motion,
\begin{equation}
S^{\rm add} = \int  (d\tilde{g}_2^2 - (N-1) \tilde{f}_3) \wedge  \left(m \tilde{g}_2^2\wedge  \tilde{g}_2^2+m' \frac{1}{48} p_1(R_{\mathcal{T}})\right) \,.
\end{equation}
$m,m'$ are integers that are fixed by gauge invariance under $\tilde{c}_2\rightarrow \tilde{c}_2 +  (N-1) d\lambda_1$ once we integrated this additional term on a 7d space with a 6d boundary. $m,m'$ then satisfy
\begin{equation}
4m+m' = 2 p -7 -x_1
\end{equation}
This implies that
\begin{equation}
\Sym^{\text{cubic}}= \Sym^{\rm CS}+S^{\rm add}_{6d} =\frac{(N-3)(N-2)}{6(N-1)} \int \tilde{c}_2\tilde{c}_2\tilde{c}_2.
\end{equation}

In this example we also have a $U(1)$ flavor symmetry. There is indeed a contribution from  \eqref{eq:potanom} that is,
\begin{equation}
S^{\rm bulk}_{7d} =-(N-1)\int d\tilde{a} \wedge \tilde{c}_2 \wedge \tilde{f}_3 + \ldots 
\end{equation}
where $d\tilde{a}_1= ( \tilde{h}^2_2)$ its the background for the $U(1)$ flavor symmetry of the theory, and we display only the interesting contribution. $\tilde{c}_2$ is the background field for the $\mathbb{Z}_2$ 1-form symmetry with $\mathbb{Z}/(N-1)$ periods, therefore if we map the field to the one with integral periods, $\tilde{c}_2\rightarrow \tilde{c}_2/(N-1)$, and integrate on the 6-dimensional boundary  we get
\begin{equation}
S^{\rm bulk}_{6d} =-\frac{1}{2(N-1)}  \int d\tilde{a}_1 \wedge  \tilde{c}_2\wedge \tilde{c}_2.
\end{equation}
This expression is not gauge invariant under $\tilde{c}_2\rightarrow \tilde{c}_2 + (N-1) d\lambda_1$. We then add the following term which does not change the anomaly theory,
\begin{equation}
S^{\rm add}_{7d}= n \int (d\tilde{g}_2^2 - (N-1)\tilde{f}_3)\wedge  d\tilde{a}_1 \wedge \tilde{g}_2^2,
\end{equation}
where we recall that $g_2^2= d\tilde{c}^2_1+ (N-1) \tilde{c}_2$, upon integrating on the 6d boundary, and rescaling $\tilde{c}_2\rightarrow \tilde{c}_2/(N-1)$, gauge invariance fixes $n=-1$, such that
\begin{equation}
\Sym^{\text{mixed}} =\frac{(N-2)}{2(N-1)}  \int d\tilde{a}_1 \wedge  \tilde{c}_2\wedge \tilde{c}_2.
\end{equation}

\paragraph{Example: $B^{(2)}_N$ SCFTs.} 
The asymptotic 5-branes charges and SNF are 
\begin{equation}
Q=\begin{pmatrix}
-1 &-N\\
N-1 &N \\
-1&0\\
-1 &0\\
 \vdots & \vdots \end{pmatrix}\,,\qquad 
 Q_{\rm SNF}=\begin{pmatrix}
1 & 0\\
0 &N\\
0 &0\\
0 &0\\
 \vdots & \vdots
\end{pmatrix} \,,
\end{equation}
where there are $N-2$ $(-1,0)$ 5-branes, and the SNF 
 with \eqref{eq:singl} encodes the $\mathbb{Z}_{N}$ 1-form symmetry of the theory, and its background field. The linear relations \eqref{eq:linrel} fix the cycles to be trivial and the pairing as well. So there is no massless vector, corresponding to the theory having no continuous 0-form symmetry. All we can compute is then the cubic anomaly for the 1-form symmetry. To do so we evaluate \eqref{eq:WZaction} with the congruence \eqref{eq:congru}, where
\begin{equation}
\widetilde{Q}=\begin{pmatrix}
x_1 &x_1N+1 \\
1&1\\
x_2&-1\\
 \vdots & \vdots
\end{pmatrix}\,,
\end{equation} 
where $x_1,x_2,x_3, \ldots x_{N-1}$ are integer parameters.
Plugging this and \eqref{eq:congru} into \eqref{eq:WZaction}, we get
\begin{equation}
\Sym^{\text{CS}}=\left(\frac{x_1 N + 2}{6 N } +\frac{x_1}{12} \right) \int \tilde{c}_2\tilde{c}_2 \tilde{c}_2  \,.
\end{equation}
Again we can add to the supergravity topological action terms which are proportional to Bianchi identities and do not change the classical equations of motion,
\begin{equation}
S^{\rm add} = \int  (d\tilde{g}_2^2 - N\tilde{f}_3) \wedge  \left(m \tilde{g}_2^2\wedge  \tilde{g}_2^2+m' \frac{1}{48} p_1(R_{\mathcal{T}})\right) \,.
\end{equation}
Here $m,m'$ are integers that are fixed by gauge invariance under $\tilde{c}_2\rightarrow \tilde{c}_2 +  N d\lambda_1$ once we integrated this additional term on a 7d space with a 6d boundary. $m,m'$ then satisfy
\begin{equation}
4m+m' =  2 p -7 -x_1 \,.
\end{equation}
This implies that
\begin{equation}
\Sym^{\text{cubic}}= \Sym^{\text{CS}}+S^{\rm add}_{6d} =\frac{(N-1)(N-2)}{6N} \int \tilde{c}_2\tilde{c}_2\tilde{c}_2 \,.
\end{equation}

We can also compute the 't Hooft anomaly for the $\mathbb{Z}_N$ 1-form symmetry form symmetry of the  $B^{(3)}_N$. The $\Sym^{\text{CS}}$ has trivial $\frac{1}{N}$ contribution sufficiently implies that $\Sym^{\text{cubic}}=0$. 

These examples illustrate that the approach using the webs and the geometry agree nicely and reproduce the same anomalies also in non-Lagrangian theories. 
In the context of the webs it is possible to describe theories, which are not described by toric geometries (but so-called generalized toric polygons (GTPs) \cite{Benini:2009gi,vanBeest:2020kou, VanBeest:2020kxw}). One can still compute from the web-data using a simple combinatorial prescription the 1-form symmetry \cite{Bhardwaj:2020phs}. It would be very interesting to generalize the above analysis to webs which are dual to such GTPs.


\section{Little String Theory Anomaly from Linear Dilaton Holography}
\label{sec:LST}

The final application is more holographic in spirit, but follows the same philosophy as the other parts of our analysis. 
We derive he anomaly of the LST from a holoraphic point of view -- again using an expansion on the boundary of the space. 

$N$ NS5 branes in IIB are believed to be described by a Little String Theory \cite{Intriligator:1997pq}, that is a non-gravitational theory with non-local stringy excitation \cite{Aharony:1999ks}. The low-energy limit of this 6d theory is given by $(1,1)$ SYM in 6 dimensions. Many little string theories have 7d bulk gravity dual solution with linear dilaton behaviour \cite{Maldacena:1997cg}. In this case the dual is given by 
(string frame)
\begin{equation} \label{eq:NS5LSTmet}
ds^2= dx_6^2 + N (d \rho ^2 + d\Omega_3), \qquad e^{\phi} = e^{\phi_0 - \rho},  \qquad H= N \text{vol}_3\,,
\end{equation}
where $dx_6^2$ is an abbreviation for a flat 6-dimensional Minkowski metric, $d\Omega_3$ is the metric on a round 3-sphere and vol$_3$ its volume, whereas $\phi_0$ is a constant. 

Given this background we can now implement a reduction  of the IIB  topological coupling
\begin{equation}
S_{\text{top}}=\int F_5 \wedge H_3 \wedge F_3\,,
\end{equation}
where we expand the fluxes on the cycle of the internal $S^3$ geometry,
\begin{align}
&F_5 =  f_5 + f_2 \wedge \text{vol}_3\\
&F_3 = f_3 \\
&H_3 = h_3 +N \text{vol}_3 \, .
\end{align}
Because of self-duality of $F_5$, satisfy $f_5= \ast_7 f_2$, where $\ast_7$ is the hodge dual in the seven-dimensional space spanned by Minkowski$_6$ and the $\rho$ coordinate. The Bianchi identity $dF_3=dH_3=0$ and $dF_5=H_3\wedge F_3$ imply that
\begin{align} 
df_3=0  &\quad \rightarrow\quad h_3 = db_2 \\
dh_3 = 0 &\quad \rightarrow\quad  f_3= dc_2\\
df_2 = N f_3 &\quad \rightarrow\quad  f_2 = dc_1 + N c_2 \label{eq:BianchiLST}\\
df_5 = N h_3\wedge f_3 &\quad \rightarrow\quad  f_5 = dc_4 - N c_2 \wedge db_2\,.
\end{align}

Since self-duality of $F_5$ imposes $f_5= \ast_7 f_2$ we manifestly have two frames.
\paragraph{BF-frame.} In this frame we choose to keep $f_5$, and therefore what we get is the following coupling,
\begin{equation}
S_{{\text{top}}_{8d}} = N \int dc_4 \wedge  dc_2 - N \int db_2 \wedge c_2 \wedge  dc_2 \,.
\end{equation}
The first one give rise to the singleton theory, which upon choice of boundary condition, fixes whether the boundary theory has a $\mathbb{Z}_N$ 1-form symmetry with background $c_2$, such that $Ndc_2=0$, or a dual 3-form symmetry, or a mixture \cite{Witten:1998wy, Aharony:1998qu, Gross:1998gk, Belov:2004ht, Maldacena:2001ss, Freed:2006yc, Freed:2006ya, Morrison:2020ool, Albertini:2020mdx, GarciaEtxebarria:2019caf}. The background $c_2$ corresponds to the center 1-form symmetry of the 6d $SU(N)$ gauge theory low-energy limit  of the LST. The second coupling is the holographic realization of part of the mixed anomaly between the continuous 1-form symmetry associated with the current $J=\ast_6 \text{Tr} (F\wedge F)$ of the low-energy 6d $SU(N)$ gauge theory \cite{Cordova:2020tij}, whose background is $b_2$, and $c_2$, the background of the discrete $\mathbb{Z}_N$ 1-form symmetry corresponding to the center of the $SU(N)$, with periods  such that $\oint c_2 \in \frac{\mathbb{Z}}{N}$.
\paragraph{St\"uckelberg-frame.} In this frame the discrete $\mathbb{Z}_N$ 1-form symmetry is realised via the St\"uckelberg mechanism that is induced by the kinetic term for $f_2=dc_1 + N c_2$. This implies that $c_1$ can be  entirely gauged away, and $c_2$ is the background for a $\mathbb{Z}_N$ 1-form symmetry. The surviving topological coupling is then,
\begin{equation} \label{eq:partmixanLST}
S_{\text{top}_{8d}} =- \int db_2 \wedge f_2 \wedge  dc_2 \,,
\end{equation}
which will give rise to part of the mixed anomaly among the 2 different 1-forms symmetries of the theory. In order to evaluate the full mixed anomaly we need to take into account the invariance under large gauge transformation for the $b_2$,
\begin{equation} \label{eq:b2LGT}
b_2 \rightarrow b_2 + \Lambda_2, \qquad \oint \Lambda_2 \in \mathbb{Z}
\end{equation} 
for any integer value of $N$. The expression \eqref{eq:partmixanLST} is not invariant under these large gauge transformation on a space with a non-trivial boundary. Indeed the term, 
\begin{equation} 
S_{\text{top}_{7d}}  =- \frac{N}{2}\int db_2 \wedge c_2 \wedge  c_2 
\end{equation}
is not invariant under \eqref{eq:b2LGT}. The general principle is that any effective action describing the reduction of the IIB action on $S^3$ and in particular the correct classical equations of motion, is valid. We can indeed add the following term, 
\begin{equation} 
S_{\text{top}_{8d}}  =-   N \int  db_2 \wedge f_2 \wedge  dc_2  - m \int db_2 \wedge f_2\wedge (N f_3 - df_2)  \,,
\end{equation}
where adding the second term does not change the theory and the classical configuration, since $(N f_3 - df_2)$ is vanishing due to the Bianchi identity \eqref{eq:BianchiLST}. We can now pick this action and integrate on a space with a 7d boundary to obtain the SymTFT
\begin{equation} 
\Sym  =- \frac{N}{2} \int db_2 \wedge c_2  \wedge  c_2  + m \frac{N^2}{2} \int db_2 \wedge c_2 \wedge  c_2\,.
\end{equation}
where we forget about total derivatives.
We need to require that this action is gauge invariant under \eqref{eq:b2LGT} and that the coefficient is well defined modulo 1. This fixes $m=1$, so that the final result for the anomaly reads,
\be
\Sym = \frac{N-1}{2N} \int db_2 \wedge \tilde{c}_2  \wedge  \tilde{c}_2 \,,
\end{equation}
where $\frac{\tilde{c}_2}{N}  = c_2$. 

\section{Conclusions and Outlook}

The main object of study in this paper is the Symmetry Topological
Field Theory (SymTFT). 
This is an object that encodes the choice of symmetries given the local dynamics, and the anomalies of these symmetries.

Our main result is that this SymTFT can be computed from
string/M-theory by a reduction on the boundary of the compactifaction
spacetime -- this can be in a purely geometric engineering setup, as
in the M-theory analysis in sections \ref{sec:5d} and \ref{sec:7d},
but also brane-setups fall into this framework, as in section~\ref{sec:5branes}.

In the geometric engineering setup we made a case that this requires a
refined notion of dimensional reduction, recasting the topological
terms in the supergravity action in terms of differential cohomology
classes. This differential cohomology approach is {\it essential} in
order to accommodate torsion (co)-cycles and their contributions.  The
latter is pertinent when the symmetries under consideration are
discrete, such as is commonly the case in higher-form symmetries.

There are various obvious applications to other geometric engineering
setups. For instance, within the M-theory setting that we have
discussed in this paper the natural extension is to consider reduction
to 4d on $G_2$-holonomy manifolds, such as the ones proposed by Bryant
and Salamon \cite{bryant1989}, and generalizations thereof, which
model the confining-deconfining transition of SYM theories. The reduction
on Calabi-Yau four- and five-folds should also result in interesting
anomalies in 3d and 1d.  Specializing to the case of elliptic
Calabi-Yau $n$-folds, the results in M-theory have an uplift to
F-theory, and thus anomalies in the context of even-dimensional QFTs,
like 6d SCFTs (for a review see \cite{Heckman:2018jxk}), 4d
$\mathcal{N}=1$ SQFTs (for a review see \cite{Weigand:2018rez}) and 2d
(0,2) theories \cite{Schafer-Nameki:2016cfr}. More generally, type IIB
compactifications can yield supersymmetric gauge theories, which can
have 1-form symmetries
\cite{GarciaEtxebarria:2019caf,DelZotto:2020esg, Bhardwaj:2021pfz,
  Bhardwaj:2021zrt,Hosseini:2021ged}.

In addition to the existence of 1-form (and higher-form) symmetries and resulting anomalies, supersymmetric QFTs can have higher-group symmetries. E.g. in 6d SCFTs \cite{Apruzzi:2021mlh} and LSTs \cite{Cordova:2020tij, DelZotto:2020sop}, 5d SCFTs \cite{Apruzzi:2021vcu}, and 4d class S \cite{Bhardwaj:2021wif}.  Such higher-groups (continuous or discrete) are dual to mixed anomalies, and thus should have an imprint in the supergravity link reduction. 
Beyond this multitude of purely geometric constructions there are brane-systems, and of course the more familiar setting of holography. 
We hope to return to many of these applications in the future. 

\subsection*{Acknowledgements}

We thank Ibou Bah, Vladimir Bashmakov, Marieke van Beest, Oren
Bergman, Lakshya Bhardwaj, Cyril Closset, Michele Del Zotto, Dewi
Gould, Max Hubner, Dave Morrison and Yi-Nan Wang for helpful
discussions. We also thank Dan Freed for allowing us to reproduce his
characterisation of the symmetry theory, which in part motivated this
work. FA, FB and SSN are supported in part by the European Union's
Horizon 2020 Framework: ERC grant 682608. STFC Consolidated Grant
ST/T000864/1 supports in part FB and SSN.  I.G.E.\ is partially
supported by STFC grant ST/T000708/1 and by the Simons Foundation
collaboration grant 888990 on Global Categorical Symmetries. S.S.H. is
funded by the STFC grant ST/T506035/1. SSN acknowledges support by the
 Simons Collaboration grant  "Special Holonomy in Geometry, Analyis, and Physics".


\addtocontents{toc}{\protect\setcounter{tocdepth}{1}}

\appendix

\section{Integral Quantization of $G_4$-flux}
\label{app:DC}

In order to diagnose whether $G_4$ has integral or half-integral periods
on a four-cycle $\cC_4$, we can  compute the integral 
 of the fourth
Stiefel-Whitney class of the tangent bundle $T \cM_{11}$ on $\cC_4$ \cite{Witten:1996md},
\beq
\int_{\cC_4} G_4 = \frac 12 \, \int_{\cC_4} w_4(T\cM_{11}) \mod 1  \ ,
\eeq
where the pullback  to $\cC_4$ is implicit.
In this work, we consider 11d spacetimes that are a direct product,
$\cM_{11} = \cW_{11-n} \times L_n$,
where $n=3$ and $L_3 = S^3/\Gamma$ (with $\Gamma$ and ADE subgroup of $SU(2)$),
or $n=5$ and $L_5$ a smooth Sasaki-Einstein manifold.

The total Stiefel-Whitney class splits as $w(T\cM_{11}) = w(T\cW_{11-n}) \smile w(T L_n)$.
Possible contributions to $w_4(T\cM_{11})$ are therefore of the form
$w_{4-i}(T\cW_{11-n}) \smile w_i(TL_n)$, $i =0,1,2,3,4$.
We observe that all the spaces $L_n$ in this work 
are the base of a Calabi-Yau cone.
In particular, each $L_n$ is orientable and Spin,
and therefore $w_1(TL_n) = 0 = w_2(TL_n)$.
We also assume that external spacetime is orientable and Spin,
so that $w_1(T\cW_{11-d}) = 0 = w_2(T\cW_{11-n})$.
We conclude that terms of the form $w_{4-i}(T\cW_{11-n}) \smile w_i(TL_n)$ with
 $i=1,2,3$ cannot give any non-zero   contributions to $w_4(T\cM_{11})$.
The remaining potential contributions thus have $i=0$ or $i = 4$.
To kill the contribution with $i=0$ we assume that external spacetime
satisfies $w_4(T\cW_{11-n}) = 0$.
What remains to be checked is whether $w_4(TL_n)$ is trivial.
Clearly, we have $w_4(TL_3) = 0$ for dimensional reasons.\footnote{For completeness,
let us point out that also $w_3(S^3/\Gamma) =0$. This follows 
directly from \eqref{eq:wu}, recalling that $w_1$ and $w_2$ are zero because
$S^3/\Gamma$ is orientable and spin.
}
Next, we prove that $w_4(TL_5) =0$ for any smooth Sasaki-Einstein five-manifold $L_5$.
In fact, we can prove a stronger statement: $w_{i>0}(TL_5) = 0$.

Let us adopt the shorthand notation $w_i= w_i(TL_5)$.
We have already observed $w_1 = 0 = w_2$. We also have (in
general, from the Wu formula)
\beq \label{eq:wu}
  \Sq1(w_2) = w_3 + w_1\smile w_2 \ ,
\eeq
which implies $w_3=\Sq1(0)=0$ in our case. We also have that the Wu
class
\be 
  \nu_4  = w_1^4 + w_2^2 + w_1\smile w_3 + w_4
\ee
necessarily vanishes on a 5-manifold for degree reasons (as it
represents $\Sq4$ acting on $H^1(L_5;\bZ_2)$, which vanishes by
general properties of the Steenrod squares), so we conclude
$w_4=0$. Finally, again from the Wu formula
\beq
  \Sq2(w_3) = w_2\smile w_3 + w_1\smile w_4 + w_5 \ ,
\eeq
which implies, given $w_1=w_2=w_3=0$, that $w_5=0$.

Let us conclude with further comments on 
$p_1 = p_1(TL_5)$.
Note that on a Spin manifold we
have \cite{Wu1954}
\beq
  \mathfrak{P}(w_2) = \rho_4(p_1) + \theta_2(w_1\Sq1(w_2) + w_4) \ ,
\eeq
so, since $w_1=w_2=w_4=0$, we learn from the analysis above that
$\rho_4(p_1)=0$, or equivalently that $p_1$ vanishes mod 4.

\section{Type IIA Analysis}
\label{app:IIA}

We have seen that the differential cohomological approach is very
effective in computing the SymTFTs starting from M-theory. 
To cross-check and to complement this with more familiar
methods in this appendix we compute the same SymTFTs using the
IIA background, which is given by not only geometry but also RR-flux. 
This setups is however more canonical in that there are no torsion cycles. 
The results will agree (to the
extent that the more limited IIA analysis below is applicable, not
every M-theory background that we analyse above has a well-behaved IIA
limit), but this more conventional approach will be a countercheck and contrast to the much more 
efficient 
 differential cohomology approach that we are
proposing.

\subsection{Type IIA Background for 7d SYM}

The Einstein metric on $S^3/\mathbb Z_N$ that descends from the quotient
of the round unit-radius metric on $S^3$ can be written as
\beq \label{3dmetric}
ds^2(S^3/\mathbb Z_N) = \frac 14 \, (d\theta^2 + \sin^2 \theta \, d\phi^2) + \frac{1}{N^2} \, D\psi^2 \ , \qquad
D\psi = d\psi - \frac N2 \cos \theta \, d\phi \ .
\eeq
The coordinate $\theta$ has range $[0,\pi]$ and the angles $\phi$, $\psi$ have
period $2\pi$. 
We identify the Hopf fiber $S^1_\psi$ in \eqref{3dmetric}
with the M-theory circle. It then follows that in type IIA the link geometry
consists simply of a round $S^2$ (described  by the coordinates $\theta$, $\phi$ in \eqref{3dmetric})
threaded by $N$ units of $F_2$ R-R 2-form flux.
Let us stress that in the type IIA setup there are no
torsional cohomology classes. As a result, the reduction can be performed
according to the standard paradigm, including only the modes
associated to harmonic forms in the internal space.
In the present situation, the latter is $S^2$ and the only non-trivial harmonic
forms are 1 and
\beq \label{omega2_def}
\omega_2 = \frac{1}{4\pi} \, \sin \theta \, d\theta \wedge d\phi \ , \qquad \int_{S^2} \omega_2 = 1 \ .
\eeq
The Chern-Simons topological terms in the Type IIA effective action can be written as an 11-form
\beq \label{IIA_I11}
S_{\rm 10d,top}  = 2\pi \int_{\cM_{10}} I_{10}^{(0)} \ , \qquad
dI_{10}^{(0)} = I_{11} = - \frac 12 \, {H_3}\wedge 
{F_4} \wedge {F_4}- {H_3} \wedge X_8
\ .
\eeq
The 8-form $X_8$ is as in M-theory, but constructed with
Pontryagin classes of the tangent bundle to 10d spacetime.
The 3-form $H_3$ is the field strength of the NS-NS 2-form field,
while $F_4$ is the field strength of the R-R 3-form field. We will not consider in this section $X_8$.
The Bianchi identities for $H_3$,  $F_4$  read\footnote{Our normalizations
for the NS-NS 2-form $B_2$, 
the R-R 1-form $C_1$, and the R-R 3-form $C_3$
are such that their minimal couplings to a fundamental string with worldsheet $\cW_2$,
a D0-brane with worldline $\cW_1$, and a D2-brane with worldvolume $\cW_3$
 introduce factors $e^{2\pi i \int_{\cW_2} B_2}$, $e^{2\pi i \int_{\cW_1} C_1}$,
 $e^{2\pi i \int_{\cW_3} C_3}$ in the
path integral, respectively.
In these conventions
$H_3 = dB_2$,
$F_2 = dC_1$,
$F_4 = dC_3- C_1
\wedge dB_2$.
  }
\beq \label{IIA_Bianchi}
dH_3 = 0 \ , \qquad
{d F_4} = - {H_3}
\wedge {F_2}  \ ,
\eeq
where $F_2$ is the   field strength of the RR 1-form field,
obeying $dF_2 = 0$.

As already noted above, we have a non-zero $F_2$ flux threading
$S^2$
\beq \label{S2_flux}
{F_2 }= N \, \omega_2 \ . 
\eeq
The expansions of the field strengths $H_3$, $F_4$ take the form
\beq \label{IIA_on_S2}
{H_3} = { \widetilde H_3}
+  {   f_1} \wedge \omega_2 \ , \qquad
{F_4} = - {\gamma_4}
- {\widetilde F_2} \wedge \omega_2 \ .
\eeq
The quantities $\widetilde H_3$, $f_1$, $\gamma_4$, $\widetilde F_2$
are external field strengths, whose Bianchi identities follow
from \eqref{IIA_Bianchi}, \eqref{S2_flux},
\beq
d\widetilde H_3= 0 \ , \qquad
df_1 = 0 \ , \qquad
d\gamma_4 = 0 \ , \qquad
d\widetilde F_2 = N \, \widetilde H_3 \ . 
\eeq
Upon plugging \eqref{IIA_on_S2} into $I_{11}$ in \eqref{IIA_I11} and fiber integrating along $S^2$, we obtain 
the following 9-form expression in the external field strengths,
\beq 
I_9 = - { \widetilde F_2 } \wedge {\widetilde H_3} \wedge {\gamma_4} \,.
\eeq
The key subtlety in the following is to note that we have the freedom 
of adding an integer multiple of the Bianchi identity to this anomaly
\be
I_9^{\text{extra}} = (N \widetilde{H}_3 - d \widetilde{F}_2 ) \widetilde{F}_2 \gamma_4  = N \widetilde{H}_3  \widetilde{F}_2 \gamma_4 - d(\widetilde{F}_2 \widetilde{F}_2 \gamma_4) \,.
\ee
Adding these contributions 
\be
I_9^{\text{improved}} =( N-1) \widetilde{H}_3  \widetilde{F}_2 \gamma_4 - d(\widetilde{F}_2 \widetilde{F}_2 \gamma_4) 
\ee
Applying an anti-derivative, and dropping total derivatives, we get 
\be
I_8^{(0)} = {1\over 2} N(N-1) \tilde{F}_2 \tilde {F}_2 \gamma_4 = {(N-1)\over  2 N } B_2 B_2 \gamma_4 \,,
\ee
where $\tilde{F}_2 =  B_2/N$, which agrees with the result in the main text. We should note that this analysis, though based on a more standard reduction process, is very limited to situations, where there is a IIA reduction of the M-theory background.

\subsection{Mixed Anomaly for M-theory on $C(Y^{p,0})$ from Type IIA}

Like in the 7d case we can also repeat the computation of the SymTFT from a IIA point of view for the $SU(p)_0$ SCFTs in 5d . 
Again the circle reduction from M-theory to IIA implies a background flux in the geometry. 
We take the relation between the 11d supergravity fields
and the Type IIA fields (in the string frame)
\begin{align}
ds^2_{11} &= e^{- \frac 23 \, \Phi} \, ds^2_{10} + e^{\frac 43 \, \Phi} \, (dy + C_1)^2 \ , \qquad y \sim y + 2 \pi R \ , \nn \\
C_3^{\rm 11d} & = C_3 + B_2 \wedge dy  \ .
\end{align}
These relations are written in the supergravity normalization.
We work with integral fluxes
\beq
d \widetilde F_4 = - H_3 \, F_2 \ , \qquad
\mathcal I_{11} = - \frac 12 \, H_3 \,  \widetilde F_4 \,  \widetilde F_4 \ .
\eeq
We reduce of $X_4 = S^2 \times S^2$. The volume forms on the two $S^2$ factors
are denoted $\omega_{2,a=1}$, $\omega_{2,a=2}$. The intersection
pairing in this basis is
\beq
I_{ab} =\int_{X_4} \omega_{2a} \, \omega_{2b} = 
\begin{pmatrix}
0 & 1 \\
1 & 0
\end{pmatrix} \ .
\eeq
The IIA field strengths are given by
\beq
F_2= n^a \, \omega_{2a} \ , \qquad
H_3 = \widetilde H_3 + f_1^a \, \omega_{2a} \ , \qquad
\widetilde F_4 = \widetilde u_0 \, V_4 + F_2^a \, \omega_{2a} + \gamma_4 \ .
\eeq
We have $\int_{X_4} V_4 = 1$. The background flux quanta $n^a$ in the basis used so far are
\beq \label{flux_vec_before}
n^a = \begin{pmatrix}
p \\ p
\end{pmatrix}  \ .
\eeq
The non-trivial Bianchi identities after reduction on $X_4$ are
\beq
dF_2^a = - n^a \, \widetilde H^3 \ , \qquad d\widetilde u_0 = - I_{ab} \, n^a \, f_1^b \ .
\eeq
We compute
\beq\label{general}
I_7 = - \int_{X_4} \mathcal I_{11} = \widetilde u_0 \, \widetilde H_3 \, \gamma_4
+ I_{ab} \, F_2^a \, f_1^b \, \gamma_4 
+ \frac 12 \, I_{ab} \, F_2^a \, F_2^b \, \widetilde H_3 \ .
\eeq

Note that we have 
\be
d F_2 ^{a} = - p \tilde{H}^3 
\ee
however there is one combination that is a massless $U(1)$ gauge field 
\be
F_{I} = F_2^1 - F_2^2 \,,
\ee
and another that is a torsion $p$-form $F_2^1$.
We perform this change of basis, which is a lattice automorphism
with determinant $+1$, 
where
\be \label{flux_vec_after}
M^a{}_b =  \begin{pmatrix}
1 & - 1 \\
1 & 0 
\end{pmatrix} 
\,,\qquad 
n'^a = \begin{pmatrix}
0 \\ p
\end{pmatrix} \  , \qquad
I'_{ab} = \begin{pmatrix}
0 & -1 \\
-1 & 2
\end{pmatrix} \ .
\eeq
In this new basis, we can write
\be
I_7  =
- \int_{X_4} \mathcal I_{11} = \widetilde u_0 \, \widetilde H_3 \, \gamma_4
+ I'_{ab} \, (F_{I}, F_2^1)^a \, f_1' {}^b \, \gamma_4 
-  F_{I} F_2^1  \widetilde{H}_3 + F_2^1 F_2^1 \widetilde{H}_3 \,,
\ee
with the non-trivial Bianchi identities 
\be
d   F_2^1 = - p \, \widetilde H_3 \ , \qquad
d\widetilde u_0 = p \, f_1'^1 - 2 \,  p \, f_1'^2 \,.
\ee
We introduce gauge potentials $B_2$, $c_3$ for $\widetilde H_3$, $\gamma_4$, respectively,
\be
\widetilde H_3 = d\tilde{B}_2 \ , \qquad \gamma_4 = dc_3 \ .
\ee
Let us focus on the case with $\gamma_4=f_1=0$.
Again as in 7d there is an ambiguity that we can add a multiple of the Bianchi identity
\be
I_7^{\text{improved}}  =
 -  F_{I} F_2^1  \widetilde{H}_3 + F_2^1 F_2^1 \widetilde{H}_3  + m (d F_2^1 + p \widetilde{H}_3) F_2^1 F_2^1 
+ n (d F_2^1 + p \widetilde{H}_3) F_2^1 F_I 
\ee
An antiderivative of $I_7$ again rescaling $\tilde{B}_2 = B_2/p$ we find 
\be \label{full_I60}
I_6^{(0)}  =-{ (1+ m p) \over 3 p^2} B_2^3 - { np-1 \over 2 p} B_2^2 F_I \,.
\ee
We are free to choose $n$ to make the term well-defined under gauge transformations. The mixed anomaly between the instanton $U(1)_I$ and the 1-form symmetry is invariant for $n=1$ 
\be
I_6^{0-1} = - {p-1 \over 2p} B_2^2 F_I.
\ee
Similarly the $B^3$ anomaly has an ambiguity, however as we know from the M-theory computation this also acquires 
contributions from $C_3 X_8$. The match with the M-theory computation is more subtle and might require a detailed analysis of the higher derivative couplings studied in \cite{Liu:2013dna}. We leave this for the future and focus our endeavours on the differential cohomology approach.


\bibliographystyle{JHEP}
\bibliography{refs}

\end{document}